\documentclass[prl,twocolumn,10pt,longbibliography,amsmath,amssymb,aps,
]{revtex4-2}

\usepackage{graphicx}
\usepackage{dcolumn}
\usepackage{bm}

\usepackage{multirow}
\usepackage{xcolor}
\usepackage{ulem}
\usepackage{cancel}
\usepackage{epsfig,psfrag}
\usepackage{amsmath}
\usepackage{amssymb}
\usepackage{amsbsy}
\usepackage{amsthm}
\usepackage{amsfonts}
\usepackage{wasysym}
\usepackage{bbm}
\usepackage{tabularx}
\usepackage{euscript}
\usepackage{enumerate}
\usepackage{amsfonts}
\usepackage{exscale}
\usepackage{bbold}
\usepackage{float}
\usepackage{slashed}
\usepackage{dsfont}
\usepackage{csquotes}
\usepackage{booktabs}
\usepackage{footmisc}
\usepackage{tikz}

\usepackage{dsfont}
\usepackage{comment}

\usepackage[]{hyperref}

\def\bea{\begin{eqnarray}}
\def\eea{\end{eqnarray}}

\def\ba{\begin{array}}
	\def\ea{\end{array}}

\def\sgn{\text{sgn}}

\def\abs#1{\left|#1\right|}

\def\sgn{{\rm sgn}}

\def\be{\begin{equation}}       \def\ee{\end{equation}}
\def\bea{\begin{eqnarray}}      \def\eea{\end{eqnarray}}
\def\ba{\begin{array}}
	\def\ea{\end{array}}
\def\bnum{\begin{enumerate} }
	\def\enum{\end{enumerate}}

\def\=>{\Rightarrow}
\def\>{\rightarrow}

\def\eye2{Fathbb{I}}

\graphicspath{{Figures/}}

\begin{document}

\vspace*{-2.5cm}
\hfill \textbf{YITP-25-101}

\title{Growth of block diagonal operators and symmetry-resolved Krylov complexity} 
\preprint{Report Number XYZ-2025-01}
\author{Pawel Caputa$^{1,2,3}$}
\author{Giuseppe Di Giulio$^1$} \email{Mail:
giuseppe.di-giulio@fysik.su.se}
\author{Tran Quang Loc$^1$}
\affiliation{$^1$The Oscar Klein Centre and Department of Physics, Stockholm University, AlbaNova, 106 91 Stockholm, Sweden
}
\affiliation{$^2$Yukawa Institute for Theoretical Physics, Kyoto University, Kitashirakawa Oiwakecho, Sakyo-ku, Kyoto 606-8502, Japan
}
\affiliation{$^3$Faculty of Physics, University of Warsaw, Pasteura 5, 02-093 Warsaw, Poland
}

\date{\today}

\begin{abstract}

This work addresses how the growth of invariant operators is influenced by their underlying symmetry structure. For this purpose, we introduce the symmetry-resolved Krylov complexity, which captures the time evolution of each block into which an operator, invariant under a given symmetry, can be decomposed.
We find that, at early times, the complexity of the full operator is equal to the average of the symmetry-resolved contributions. At later times, however, the interplay among different charge sectors becomes more intricate.
In general, the symmetry-resolved Krylov complexity depends on the charge sector, although in some cases this dependence disappears, leading to a form of Krylov complexity equipartition.
Our analysis lays the groundwork for a broader application of symmetry structures in the study of Krylov space complexities with implications for thermalization and universality in many-body quantum systems.
\end{abstract}

\maketitle

\paragraph*{Introduction.---}

 Restricting the analysis of many-body quantum systems with global symmetries to a fixed-charge sector can reveal structure and correlations that remain hidden when all symmetry sectors are treated collectively \cite{LaFlorencie2014,Goldstein:2017bua,Xavier:2018kqb, Lukin19}.
Symmetry-resolved entanglement entropies are bipartite entanglement measures that account for a fixed value of a conserved charge restricted to a subregion.
In critical systems, it has been found they capture more universal features than the total entanglement entropy \cite{Goldstein:2017bua,Xavier:2018kqb,Murciano:2020vgh,Bonsignori:2020laa,Capizzi:2020jed,Estienne:2020txv,Zhao:2020qmn,Horvath:2021fks,Capizzi:2021kys,Calabrese:2021wvi,Ghasemi:2022jxg,Foligno:2022ltu,Capizzi:2023bpr,Fossati:2023zyz,Kusuki:2023bsp}.
Furthermore, whenever we neglect microscopic details of many-body systems, symmetry-resolved entanglement entropies have been shown to be independent of the charge sector, a property known as {\it entanglement equipartition} \cite{Xavier:2018kqb,Murciano:2019wdl,Bonsignori:2019naz,Calabrese:2020tci,Ares:2022hdh,Magan:2021myk,Weisenberger:2021eby,Zhao:2022wnp,DiGiulio:2022jjd,Northe:2023khz,Benedetti:2024dku}.

These developments have motivated a broader use of symmetries—and their breaking \cite{Casini:2019kex,Casini:2020rgj,Casini:2021zgr,Ares:2022koq,Ares:2025onj}—as tools to probe many-body systems from a quantum information perspective (see e.g. review \cite{Castro-Alvaredo:2024azg}). In this spirit, several physical quantities have been decomposed into symmetry sectors, and their symmetry-resolved counterparts have found diverse applications \cite{Cornfeld:2018wbg,Murciano:2021djk,Capizzi:2021zga,Chen:2021nma,Chen:2021pls,Ares:2022gjb,Chen:2022gyy,Rath:2022qif,Parez:2022sgc,Parez:2022xur,Murciano:2023ofp,DiGiulio:2023nvz,Yan:2024rcl,Nie:2021ond}.
In this work, we introduce symmetry-resolved Krylov complexity to the list.

Originally proposed as a diagnostic tool for quantum chaos \cite{Parker:2018yvk}, Krylov complexity quantifies the growth of operators undergoing time evolution. This idea was later extended to quantum states that spread through the Hilbert space during evolution, leading to the notion of spread complexity \cite{Balasubramanian:2022tpr}. These complexity measures have already found applications in quantum chaos \cite{Parker:2018yvk,Roberts:2018mnp,Rabinovici:2022beu,Balasubramanian:2022dnj,Erdmenger:2023wjg,Loc:2024oen}, topological phases of matter \cite{Caputa:2022eye,Caputa:2022yju}, quantum field theory \cite{Dymarsky:2021bjq,Avdoshkin:2022xuw,Camargo:2022rnt,Caputa:2021ori,Caputa:2023vyr,Malvimat:2024vhr,Caputa:2025dep}, open quantum systems, and non-unitary dynamics \cite{Liu:2022god,Bhattacharya:2022gbz,Medina-Guerra:2025wxg,Medina-Guerra:2025rwa} (see review \cite{Nandy:2024htc}).
Moreover, operator and state complexities are also believed to capture aspects of gravity and black hole physics in Anti-de Sitter spacetimes \cite{Rabinovici:2023yex,Balasubramanian:2024lqk,Miyaji:2025yvm,Caputa:2024sux}, as described by AdS/CFT \cite{Maldacena:1997re}.

In this work, we consider systems with a conserved charge and Hermitian operators that are invariant under this symmetry. These operators exhibit a block-diagonal structure across the charge sectors of the theory. We define the symmetry-resolved Krylov complexities as the Krylov complexities of these block operators.
%
This analysis allows us to understand how different symmetry sectors contribute to the Krylov complexity of the full operator, and whether the total Krylov complexity can be meaningfully decomposed into symmetry-resolved components. We find that such a decomposition holds at early times, whereas at later times the charge sectors correlate intricately in contributing to the total complexity.
In this context, it is insightful to study the dependence of symmetry-resolved Krylov complexity on the charge and investigate whether an equipartition emerges in certain regimes. We find that, in general, equipartition of Krylov complexity does not always occur; however, we present an example in which it does appear.

These results suggest that symmetry resolution can provide a fine-grained lens on operator growth, especially relevant for systems where symmetries constrain dynamics, such as those exhibiting prethermalization \cite{Berges:2004ce,Kollar11,Langen:2016vdb}, deep thermalization \cite{Choi:2021npc,Cotler:2021pbc,Chang:2024cic}, or quantum many-body scars \cite{Turner:2018kjz,Choi:2019wqq}. 
%
More broadly, symmetry-resolved Krylov complexity is a promising tool for understanding the spread of information in many-body systems where symmetries and the consequent presence of charge sectors are associated with universal properties.


\paragraph*{Setup and tools.---}
Consider a system evolving under a time-independent Hamiltonian $H$. The Heisenberg evolution of a Hermitian operator $O(0)$ at a fixed time $t=0$ reads
\begin{equation}
    O(t)=e^{{\rm i} H t} O(0) e^{-{\rm i} H t}=\sum_{n=0}^\infty\frac{({\rm i} t)^n}{n!}\tilde{O}_n\,,
    \label{eq:Heisem_time_ev}
\end{equation}
where $\tilde{O}_0=O(0)$, $\tilde{O}_1=[H,O(0)]$, $\tilde{O}_2=[H,[H,O(0)]]$ and, as $n$ increases, the number of nested commutators in $\tilde{O}_n$ grows correspondingly. The goal is then to quantify this growth in time by defining the notion of operator's size, i.e. complexity. To this end, we endow the operator algebra describing our theory with a Hilbert space structure \cite{LanczosBook,Lanczos:1950zz}. We define the vectors $\{|\tilde{O}_n)\,,\;n=0,1,2,\dots\}$ representing the operators $\tilde{O}_n$ in this Hilbert space, and we introduce an inner product. The choice of the latter is detailed in the Appendix. 
Given the inner product, we can orthonormalize the set $\{|\tilde{O}_n)\,,\;n=0,1,2,\dots\}$ through the Gram-Schmidt procedure and obtain the Krylov basis $\{|K_n)\,,\;n=0,1,2,\dots,\mathcal{K}-1\}$. The number $\mathcal{K}$ of elements of the Krylov basis is bounded by the dimension of operator Hilbert space.
We call $\phi_n(t)$ the amplitudes of the operator $O(t)$ on the Krylov basis. These amplitudes evolve in time satisfying a Schrödinger equation with tridiagonal Hamiltonian, whose entries are the Lanczos coefficients $b_n$ \cite{LanczosBook,Lanczos:1950zz}.
%
The choice of representing $O(t)$ on the operator Hilbert space so that $(O(t)\vert O(t))=1$ implies $\sum_{n=0}^{\mathcal{K}-1}\vert\phi_n(t)\vert^2=1$. 
This allows us to represent the operator evolution by a particle, described by a wavefunction $\phi_n(t)$, hopping on a 1D chain (Krylov chain). As time progresses, the particle moves away from the origin, and its average position provides a useful measure of the operator's complexity as a function of time, the {\it Krylov complexity} \cite{Parker:2018yvk}
 \begin{equation}
\label{eq:total_C}
  C_K(t)= \sum_{n=0}^{\mathcal K-1} n \vert\phi_{n}(t)\vert^2\,.
\end{equation}
 In the Appendix we review how to compute $C_K(t)$ in practice. 

In this manuscript, we focus on systems endowed with a global symmetry generated by the charge $Q$, such that $[H,Q]=0$. The presence of a conserved charge $Q$ implies that the Hilbert space $\mathcal{H}$ organizes into superselection sectors, $\mathcal{H}=\bigoplus_{q}\mathcal{H}_q$, which are labeled by $q$, the eigenvalues of $Q$, corresponding to the irreducible representations of the symmetry group. We refer to the superselection sectors as {\it charge sectors}.
Although our construction applies more generally, we focus on the case of  U(1) symmetries, for which the direct sum above runs over the integers $q\in\mathbb{Z}$.
We now consider an invariant operator, i.e. $[O(0),Q]= [O(t),Q]=0$, and assume that it does not commute with $H$, so that its time evolution is non-trivial. Due to the commutation with $Q$, the operator $O$ acquires a block diagonal structure
\begin{equation}
\label{eq:Odecomposition}
 O(t)=\sum_{q\in\mathbb{Z}} O_q(t)\,,\quad O_q(t)=\Pi_q O(t)=e^{- {\rm i} H t}\Pi_q O(0) e^{ {\rm i} H t}\,.
\end{equation}
Each of the blocks $O_q$ is obtained by acting on $O$ with the projector $\Pi_q$ associated with the subspace $\mathcal{H}_q$ and, therefore, belongs to the charge sector labeled by the integer $q$.
The main question we address in this work is how the different charge sectors contribute to the Krylov complexity of an invariant operator. Thus, we first introduce a quantifier of Krylov complexity per charge sector. Considering the block operator $O_q$ in \eqref{eq:Odecomposition}, we can simply apply the same Krylov procedure as for the full operator. Namely, introduce a block operator $|O_q(t))$ and expand it on the fixed-charge Krylov basis $\{|K^{(q)}_n)\,,\;n=0,1,2,\dots,\mathcal{K}_q-1\}$.
The number $\mathcal K_q$ of elements in the Krylov basis for $O_q(t)$ satisfies $\mathcal K_q\leq \mathcal{K}$.
Since $[H,Q]=0$, the Hamiltonian decomposes $H=\sum_q H_q$, and the Krylov basis is obtained by the orthonormalization of nested commutators of $H_q$ with $O_q(0)$.  We call $\phi^{(q)}_{n}$ the amplitudes resulting from the expansion of $|O_q(t))$ on the Krylov basis. The dynamics of $\phi^{(q)}_{n}$ is given by a tridiagonal Hamitlonian, whose entries are the fixed-charge Lanczos coefficients $b^{(q)}_{n}$.
This is sufficient to define the {\it symmetry-resolved Krylov complexity}
\begin{equation}
\label{eq:fixedcharge_C}
  C^{(q)}_K(t)= \sum_{n=0}^{\mathcal K_q-1} n \left\vert\phi^{(q)}_{n}(t)\right\vert^2\,,
\end{equation}
which quantifies the growth of the block operator associated with the charge sector $q$. Importantly, \eqref{eq:fixedcharge_C} accounts only for a single block, neglecting the existence of the others. 
By definition, we can compute $C^{(q)}_K(t)$ starting from the block $O_q(t)$ and extracting the amplitudes $\phi^{(q)}_{n}(t)$.
For this purpose, we observe that $\vert O(t))$ and $\vert O_q(t))$ are related by
\begin{equation}
\label{eq:O decomposition_vect}
\vert O(t))=\sum_{q\in\mathbb{Z}} \sqrt{p_q}\vert O_q(t))\,.
\end{equation}
The presence of the factors $\sqrt{p_q}$ is due to the assumption that both $\vert O(t))$ and $\vert O_q(t))$ are normalized to one. 
This also implies that $\sum_q p_q =1$, which provides the interpretation of $p_q$ as the probability associated with the charge sector $q$.
The most efficient way to access $\phi^{(q)}_{n}(t)$ to compute \eqref{eq:fixedcharge_C} is described in the Appendix. 
The goal of the next sections is to understand if and how \eqref{eq:fixedcharge_C} captures the information encoded in the total Krylov complexity \eqref{eq:total_C} and what can we learn from its dependence on $q$.

We emphasize that the symmetry resolution of entanglement measures presents a sharp difference compared to the one of Krylov complexity. In the former, the reduced density matrices are decomposed into charge sectors associated with the conserved charge {\it restricted to a subsystem}, while, in the latter, the charge sectors are defined with respect to the {\it full} conserved charge.
%

\paragraph*{General results.---}

To discuss the interplay between the growths of the operator $O$ and of its fixed-charge blocks, we study how the elements of the Krylov basis $\{|K_n)\}$ can be written in terms of the fixed-charge Krylov vectors $\{|K^{(q)}_n)\}$. This analysis is detailed in the Appendix; here, we report the main findings.
Setting $t=0$ in \eqref{eq:O decomposition_vect} and defining $\vert K_0)=\vert O(0))$ and $\vert K^{(q)}_0)=\vert O_q(0))$, we conclude that $\vert K_0)$ can be written as a linear combination of Krylov vectors with the same label $n=0$ but in different charge sectors.
This is not the case for any value of $n$; when $n\geq 2$, $\vert K_n)$ cannot be written anymore in terms of $\vert K^{(q)}_n)$, and vectors with $m\neq n$ arise in the linear combination. This fact prevents us from writing $\phi_{n}(t)$ in terms of $\phi^{(q)}_{n}(t)$ in the different charge sectors, which implies that the total complexity $C_K(t)$ in \eqref{eq:total_C} does not have a simple relation in terms of the symmetry-resolved complexities. If a relation between $C_K(t)$ and the $C^{(q)}_K(t)$s can be found, one has to proceed case-by-case, as we show for the example of a general block-diagonal $4\times 4$ operator. 
Nevertheless, to compare the total complexity with the cumulative information extracted from the individual charge sub-sectors \eqref{eq:fixedcharge_C}, we first define an average complexity over the sectors as
\begin{equation}
\label{eq:weighted_Complexity}
    \bar{C}(t)=\sum_{q\in\mathbb{Z}}p_q C^{(q)}_K(t)\,,
\end{equation}
where the Krylov complexity of each block operator is weighted with the corresponding probability $p_q$. We point that a different attempt to average complexity contributions in ensembles of initial states was discussed in \cite{Craps:2024suj}.
It is then natural to ask how much information from the total complexity \eqref{eq:total_C} is captured by the average \eqref{eq:weighted_Complexity}, and more broadly, how these two quantities are related.
As shown in the Appendix, by expanding $C_K(t)$ and $\bar{C}(t)$ for early times up to order $t^4$, we obtain
\begin{equation}
\label{eq:differenceCpCK}
  C_K(t)-\bar{C}(t)= \left(
  \sum_{q\in\mathbb{Z}} p_q\left(b^{(q)}_{1}\right)^4-b^4_1\right)\frac{t^4}{2} +O(t^6)  \,,
\end{equation}
The expression \eqref{eq:differenceCpCK} shows that, at leading order in the expansion of early times, $\bar{C}(t)$ behaves exactly as
$C_K(t)$ and, at higher orders, it starts to deviate. In other words, $\bar{C}(t)$ reproduces well the early time growth of the Krylov complexity of the full operator. The coefficient of $t^4$ is the variance of $\left(b^{(q)}_{1}\right)^2$ computed over the probability distribution $p_q$ and, therefore, is positive. We conjecture that the difference in \eqref{eq:differenceCpCK} is positive for any time, and we show that this holds in all the examples in the manuscript.
This inequality could be justified as follows. The average Krylov complexity $\bar{C}(t)$ combines the information on the charge sectors without taking into account possible mixing during the construction of the Krylov basis $\{|K_n)\}$. This mixing of the sectors can enhance the total complexity $C_K(t)$, which, therefore, is expected to be larger than $\bar{C}(t)$. Thus, $C_K(t)-\bar{C}(t)$ measures how much the charge sectors are correlated in the construction of the Krylov basis. We hope to come back to a general proof of $C_K(t)-\bar{C}(t)\geq 0$ in future work.

Finally, we provide an interpretation of the early-time behaviour \eqref{eq:differenceCpCK} in terms of the effective dynamics on the Krylov chain. At early times, the dynamics of the full operator can be represented by a particle for each charge sector moving on the Krylov chain, whose average positions (weighted with $p_q$) determine the total Krylov complexity. At later times, the fixed-charge particles do not move on the Krylov chain independently; due to the possible mixing of the charge sectors in determining the Krylov basis, the dynamics are best represented by a single degree of freedom (in which we can imagine all the fixed-charge particles coalesce), whose position on the chain gives the total Krylov complexity.

\paragraph*{Examples.---}
In the final part of the paper, we report three examples where we compute the symmetry-resolved Krylov complexities, highlighting the properties discussed so far.
\\
\noindent \underline{Two-spins:}
It is insightful to start from a simple model of two spins $1/2$, described by two sets of Pauli matrices $X_{\textrm{S}},\,Y_{\textrm{S}}\,,Z_{\textrm{S}}$ with ${\textrm{S}}\in\{{\textrm{L}},{\textrm{R}}\}$, interacting through the Hamiltonian $H_{\textrm{spin}} =J(X_{\textrm{L}}X_{\textrm{R}}+Y_{\textrm{L}}Y_{\textrm{R}}+Z_{\textrm{L}}Z_{\textrm{R}})$. This Hamiltonian commutes with the total magnetization in the $Z$-direction $Q_{\textrm{spin}}=Z_{\textrm{L}}+Z_{\textrm{R}}$, which is, therefore, conserved. The operator $Q_{\textrm{spin}}$ has eigenvalues $q=0,\pm 1$, which label the three charge sectors of this system.
\\
The left spin operator $Z_{\textrm{L}}$ (we could choose, equivalently, $Z_{\textrm{R}}$) commutes with $Q_{\textrm{spin}}$, but does not with $H_{\textrm{spin}}$. Thus, it exhibits a non-trivial time evolution and a block-diagonal structure
\begin{equation}
\label{eq:spinL_evolved}
    Z_{\textrm{L}}(t)=
    \begin{pmatrix}
       1 & 0 & 0 & 0\\
        0& \cos(J t) & {\rm i}\sin(J t) &0 \\
        0& - {\rm i}\sin(J t) & -\cos(J t)& 0\\
       0 & 0 & 0 & -1
    \end{pmatrix}\,,
\end{equation}
where the first and the last $1\times 1$ blocks correspond to $q=1$ and $q=-1$ respectively, while the $2\times 2$ block to $q=0$.
Due to the small dimensionality of the Hilbert space, the Krylov basis and complexities can be obtained explicitly for the full $4\times 4$ operator and the three blocks.
Referring to the Appendix for the details, we find $C_K(t)=1-\cos(Jt)$, $C^{(q=0)}_K(t)=\sin^2(Jt)$
and $C^{(q=\pm1)}_K(t)=0$. The fact that the symmetry-resolved Krylov complexity in the sectors with $q=\pm1$ is vanishing is consistent with the entries of the $1\times 1$ blocks not evolving in time. Even if in a simple way, the symmetry resolved complexities depend on the charge, not showing equipartition. We also identify the probabilities \eqref{eq:O decomposition_vect}, $p_{\pm 1}=\frac{1}{4}$ and $p_{0}=\frac{1}{2}$, that lead, from \eqref{eq:weighted_Complexity}, to $\bar{C}(t)=\sin^2(Jt)/2$. As expected, the early time growth coincides with the one of $C_K(t)$ above. Computing their difference, we have $C_K(t)-\bar{C}(t)=2\sin^4(Jt/2)$, which is positive not only at order $t^4$ but at any time. This is in agreement with our expectations. Interestingly, comparing $C_K $ and $C^{(q=0)}_K $, it is straightforward to observe that, for some values of $t$, the symmetry-resolved component is larger than the total Krylov complexity. This makes the positivity of $C_K(t)-\bar{C}(t)$ at any time even less trivial from the mathematical point of view.
\\
\noindent \underline{General 4-dimensional operator:}
\begin{figure}[t]
	
	\centering
\includegraphics[width=1.1\linewidth]{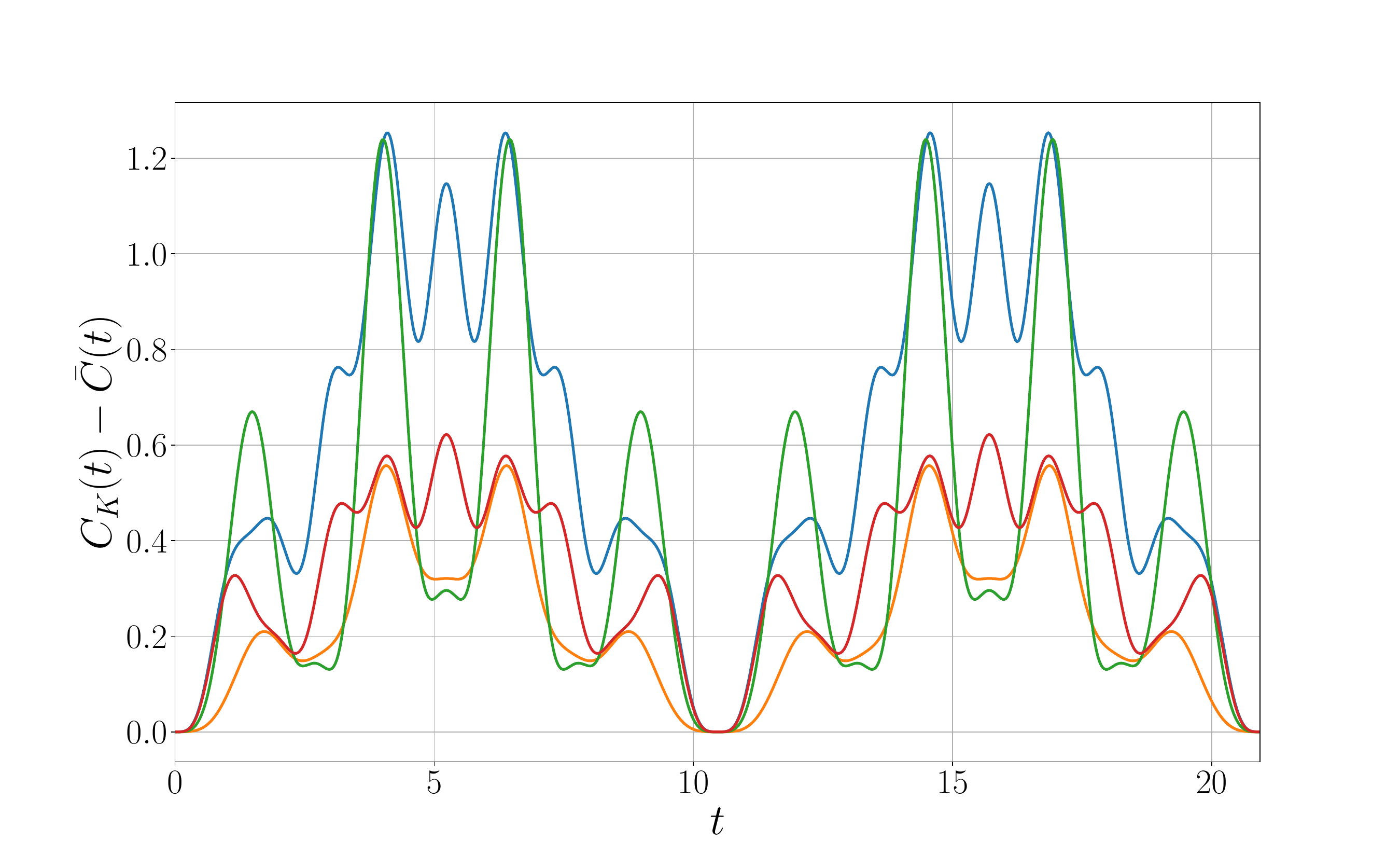}
	\caption{Difference between the Krylov complexity of the operator \eqref{eq:operator4times4} and the average \eqref{eq:weighted_Complexity} over the (two) sectors. All the curves are obtained for $E_1=8,\,E_2=5\,,E_3=3.4\,,E_4=1$. Different colors correspond to distinct choices of the parameters $A_{ii}$ ($i=1,2,3,4$), $A_{12}$, $A_{34}$, $B_{12}$, $B_{34}$ (see Table I in the Appendix). In all the instances shown here, $C_K(t)-\bar{C}(t)\geq 0$, supporting our surmise.}
	\label{fig:4x4Example}
\end{figure}
The inequality $C_K(t)-\bar{C}(t)\geq 0$ in the previous example might be an artifact of the simplicity of the system. To improve on this, we consider the most general $4\times 4$ Hermitian operator with two $2\times 2$ blocks along the diagonal
\begin{equation}
\label{eq:operator4times4}
   O_4=
    \begin{pmatrix}
       A_{11} & A_{12}+{\rm i}B_{12} & 0 & 0\\
        A_{12}-{\rm i}B_{12}&  A_{22} &  0 &0 \\
        0& 0 &  A_{33}& A_{34}+{\rm i}B_{34}\\
       0 & 0 & A_{34}-{\rm i}B_{34} &  A_{44}
    \end{pmatrix}\,,
\end{equation}
where $A_{ii}\, (i=1,2,3,4),\, A_{12},\,A_{34},\,B_{12},\, B_{34}\in\mathbb{R}$. We assume that the block structure is induced by a conserved charge, which, in the diagonal basis, reads $Q_4={\rm diag}\left(q_+,q_+,q_-,q_-\right)$.
The Hamiltonian governing the (non-trivial) time evolution of $O_4$ is diagonal in the same basis and is given by
$H_4={\rm diag}\left(E_1,E_2,E_3,E_4\right)$.

The two blocks of $O_4$ evolve separately in time. The Krylov complexity of a general Hermitian operator in a two-dimensional Hilbert space has been computed in \cite{Caputa:2024vrn}, and we can use those results to determine the two symmetry-resolved Krylov complexities in this example. 
As shown in the Appendix, we can also extract the probabilities associated with the two blocks. Combining them with the symmetry-resolved Krylov complexities as in \eqref{eq:weighted_Complexity}, we obtain $\bar{C}(t)$.
The computation of the Krylov complexity of the $4\times 4$ operator is more involved. In the Appendix, we describe a method to obtain analytic expressions. 
It is insightful to plot the difference $C_K(t)-\bar{C}(t)$ as a function of time; we do it in Fig.\,\ref{fig:4x4Example} for various choices of parameters. Consistently with our expectations, all the curves are positive. In the Appendix, we provide further evidence to support this fact.
As already verified for the previous example, this instance also demonstrates that the total Krylov complexity can be smaller than one of the two symmetry-resolved complexities for some time values. This is illustrated, for some parameter values of the model, in a dedicated figure in the Appendix.

\noindent \underline{Harmonic oscillators:}
To include infinite-dimensional Hilbert spaces in our analysis and move towards the study of many-body quantum systems, let us consider a model of coupled complex harmonic oscillators described by $N$ (non-Hermitian) position operators $\Phi_j$ and conjugate momenta $\Pi_j$ such that $[\Phi_i,\Pi_j]={\rm i}\delta_{ij}$. If we assume that the Hamiltonian of the model
can be rewritten as the sum of two quadratic Hamiltonians of independent real oscillators (see the Appendix), we can diagonalize it as
\cite{Serafini17book}
\begin{equation}
\label{eq:HamcomplexHO_diag}
H_{\textrm{\tiny HO}}=\sum_{k=1}^N\omega_k(b_k^\dagger b_k+c_k^\dagger c_k+1) \,,
\end{equation}
where we have introduced two sets of creation operators $b^\dagger_k$ (particles) and $c^\dagger_k$ (antiparticles), with $k=1,2,\dots,N$. 
The modes are labeled by the integer $k$ and characterized by the dispersion relation $\omega_k$, whose details depend on the specific Hamiltonian with which we start, but whose form does not affect our results. This system has a conserved charge that, in the Fock basis, reads $Q_{\textrm{\tiny HO}}=\sum_{k=1}^N(b_k^\dagger b_k-c_k^\dagger c_k)$ and generates a U(1) symmetry. The charge $Q_{\textrm{\tiny HO}}$ is interpreted as the difference between the total number of particles and antiparticles and has integer eigenvalues; thus, the charge sectors are labeled by $q\in \mathbb{Z}$.
To study an operator commuting with $Q_{\textrm{\tiny HO}}$ but not with $H_{\textrm{\tiny HO}}$ (retaining a non-trivial dynamics), we consider
\begin{equation}
\label{eq:good_operator_CHO}
 O_{\textrm{\tiny HO}} =\prod_{k=1}^N \left(b_kc_k+b_k^\dagger c_k^\dagger\right)\,.
\end{equation}
It is already insightful to focus on the single-mode case, i.e. $N=1$ (we call $\omega\equiv\omega_1$).
In the Appendix, we report the computation of the Krylov complexity,
finding that the only non-vanishing Lanczos coefficient is $b_1=2\omega$ and correspondingly, $C_K(t)=\sin^2(2\omega t)$.

Since $[Q_{\textrm{\tiny HO}},O_{\textrm{\tiny HO}}]=0$, $O_{\textrm{\tiny HO}}$ has a block-diagonal structure with infinitely many blocks labeled by all the integer numbers. 
In the Appendix, we find that all the fixed-charge Lanczos coefficients are the same as the one determining the Krylov chain dynamics of the full operator. Thus, also all the symmetry-resolved complexities have the same expression as the total one
\begin{equation}
\label{eq:equi_HC1mode}
   C^{(q)}_K(t)= C_K(t)=\sin^2(2\omega t) \,,
\qquad\forall\, q\in \mathbb{Z}\,.
\end{equation}
The independence of the charge in $C^{(q)}_K(t)$ implies that the operator in \eqref{eq:good_operator_CHO} exhibits the equipartition of the Krylov complexity.
Despite the equipartition, the blocks of $O_{\textrm{\tiny HO}}$ have a non-trivial probability distribution $p_q$ associated with them, as shown in the Appendix. 
Due to \eqref{eq:equi_HC1mode}, $C_K(t)-\bar{C}(t)= 0$ in this case.
In extending the analysis to the many-body operator \eqref{eq:good_operator_CHO} with $N>1$, we find that the amount of non-vanishing Lanczos coefficients grows with $N$ and, thus, we are not able to determine a general expression for the Krylov complexity analytically. However, in the Appendix, we observe that the Lanczos coefficients associated with the evolution of $O_{\textrm{\tiny HO}}$ and the fixed-charge ones are the same, and the latter are independent of $q$ also for generic $N$. Thus, we conclude that the equipartition of the Krylov complexity is also present for the general many-body operator in
\eqref{eq:good_operator_CHO}. We further analytically compute the early-time growth of the Krylov complexity for generic $N$ and its full expression when $N=2$.

Finally, we elaborate on the occurrence of the equipartition of Krylov complexity. In general, we expect this is due to the specific choice \eqref{eq:good_operator_CHO}.
Focusing for simplicity on $N=1$, we observe that this operator creates and destroys a particle and an antiparticle. Thus, it connects Fock states with an energy difference $\Delta E/\omega-1=\pm 2$. Supported by the analysis in the Appendix, we show that in this example equipartition of Krylov complexity is related to the locality in the energy basis of the considered operator. We hope to test this hypothesis in more general scenarios in future works.


\paragraph*{Conclusions and Outlook.---}
In this work, we introduced the symmetry-resolved Krylov complexity to probe the contribution of different charge sectors to the growth of operators invariant under a given symmetry. This approach highlights the intricacy of the relationship between the total complexity $C_K(t)$ of the operator and the complexities $C^{(q)}_K(t)$ of its fixed-charge blocks.
We find that, although a general relation between $C_K(t)$ and the $C^{(q)}_K(t)$'s is complicated and system-dependent, the initial growth of the total complexity is perfectly captured by the average $\bar{C}(t)$ in \eqref{eq:weighted_Complexity} of the symmetry-resolved complexities over the charge sectors. We interpret this as an effective early-time dynamics on the Krylov chain characterized by several charged particles hopping independently and then coalescing in a unique degree of freedom at late times. 
In other words, at early times, we can precisely identify the individual charge contributions, as if the fixed-charge Krylov particles have a manifest effect on the total Krylov complexity through the average $\bar{C}(t)$. On the other hand, at late times, given that $\bar{C}(t)$ deviates from $C_K(t)$,  this picture of charged Krylov particles individually contributing to the total complexity breaks down. However, the charge sectors still play a collective role in determining $C_K(t)$, despite not being described by hopping particles. 
It would be interesting to understand how the time scale at which the behaviour changes depends on the details of the system considered.
Beyond the initial times, we conjecture that $C_K(t) - \bar{C}(t)$ is non-negative and captures the correlations among the charge sectors in determining the optimal basis to describe the operator growth.
These correlations are not evident from other quantities, as the autocorrelation functions. Indeed, the total autocorrelation function is given by the average over the charge sectors of the resolved contributions (see the Appendix), hiding the mixing among the sectors. For this reason, $C_K- \bar{C}$, provides a helpful tool for improving our understanding of Krylov complexity as a measure of operator spread.
This is supported by the examples considered in the manuscript. For a complete picture, 
it will be important to understand if the complexity of the full operator captures both quantum and classical correlations between its charge sectors, as it happens for the symmetry-resolved entanglement entropies \cite{Lukin:2019tkq}. Finally, we find that the U(1)-invariant operator \eqref{eq:good_operator_CHO} involving complex harmonic oscillators exhibits the equipartition of Krylov complexity among the charge sectors. 

We remark that the absence of a general relationship between Krylov complexity and symmetry-resolved Krylov complexity should not be regarded as an incompleteness. Indeed, the counterintuitive fact that, when studying the complexity of simpler blocks of a given operator, we cannot directly reconstruct the total complexity serves as evidence that Krylov complexity probes the quantum dynamics of operators at a deep level, also accounting for the aforementioned correlations between charge sectors.

This work paves the way to understand how the symmetry structure of operators in theories endowed with global symmetries determines their growth across time evolution.
There are several directions to further explore in this respect. Here, we have focused on the operator growth and its complexity. However, symmetry resolution could also be used to study the spread of quantum states \cite{Caputa:2025ozd}. Indeed, when not eigenstates of a conserved charge, a state may have components in different charge sectors, i.e. distinct charge eigenspaces. The study of state complexity can be unified with the approach presented in this work, by starting from an expansion like \eqref{eq:O decomposition_vect} for the evolving quantum state (not generally the associated with some operator). In addition, due to charge conservation, focusing on a charge sector means restricting to certain energy windows in the spectrum of our system and this should be related to micro-canonical inner products considered in \cite{Kar:2021nbm}. 
Moreover, the symmetry resolution of entanglement has shown that entanglement measures in fixed charge sectors may lead to the emergence of more universal properties of the theory compared to what is found in the unresolved analysis. Paralleling this remark, we expect the same to happen for Krylov complexities in critical systems and their continuum limit given by CFTs.
For this purpose, investigating the Krylov chain dynamics induced by Hamiltonians symmetric under Lie groups from the perspective of symmetry resolution could be insightful \cite{Dymarsky:2019elm,Caputa:2021sib}. In addition, it would be very interesting to analyze other complexity measures in quantum many-body systems and quantum field theories (see \cite{Baiguera:2025dkc} for a recent review) from the perspective of symmetry resolution and understand if examples of equipartition of complexity can be found for these new probes. This would be an important step towards a more detailed characterization of existing complexity measures and their role in many-body physics.

In this work, we have focused only on invariant operators, characterized by a block-diagonal structure, which naturally leads to the question of how the total Krylov complexity is affected by that of the blocks. Including charged operators in our analysis relates to recent investigations using entanglement asymmetry to detect symmetry breaking in reduced density matrices \cite{Ares:2022koq}. Indeed, the symmetry breaking implies a reduced density matrix that does not commute with the charge. Finding a complexity-based proxy of symmetry-breaking would bring us closer to understanding how the presence or absence of symmetries influences the operator growth.
Finally, the operator entanglement provides a way to characterize the complexity of an operator in terms of its approximability by matrix product operators \cite{Zanardi:2001zza,Prosen:2007gfp,Dubail:2016xht}. It would be interesting to compare the dynamics of symmetry-resolved operator entanglement \cite{Rath:2022qif,Murciano:2023ofp} and Krylov complexity in view of studying universal properties of many-body quantum systems.

\section{Acknowledgments}
\noindent We are grateful to  Javier Magan, Sara Murciano and Dimitrios Patramanis for discussions. This work is supported by the ERC Consolidator grant (number: 101125449/acronym: QComplexity). Views and opinions expressed are however those of the authors only and do not necessarily reflect those of the European Union or the European Research Council. Neither the European Union nor the granting authority can be held responsible for them.


\onecolumngrid
\newpage

\section{Appendix}

In this appendix, we provide computational details on the following aspects discussed in the main text:

\begin{itemize}

    \item Lanczos algorithm for a generic Hermitian operator
    \item Lanczos algorithm for the evolution of fixed-charge blocks in which invariant operators decompose.
    \item Resolution of Krylov complexity in a two-spin system.
    \item Krylov complexity of $4\times 4$ block-diagonal operators and its symmetry resolution.
    \item Equipartition of Krylov complexity in harmonic oscillator systems.
\end{itemize}

\subsection{Review of the Lanczos algorithm}

We first briefly review the iterative procedure to determine the Krylov basis associated with the dynamics of the operator $O$. We simply refer to this procedure as the Lanczos algorithm. To represent the operator $O(t)$ on a Hilbert space, we need to introduce an inner product. We choose the Wightman inner product
\begin{equation}
\label{eq:innerproduct}
(A|B)= \langle e^{\beta H  /2} A^\dagger e^{-\beta H  /2} B\rangle_\beta=\frac{\textrm{Tr}\left(e^{-\beta H /2} A^\dagger e^{-\beta H  /2} B\right)}{\textrm{Tr}\left(e^{-\beta H  } \right)}\,,
\end{equation}
parametrized by the inverse temperature of the system $\beta=1/T$. 
After introducing $\vert O(t))$, the Krylov basis is constructed through the recursion relations
\cite{LanczosBook,Lanczos:1950zz}
\begin{equation}
\label{eq:Lanczos}
  \vert A_{n})= \mathcal{L}\vert K_{n-1})-b_{n-1}\vert K_{n-2})\,,
  \qquad
  \vert K_n)=b_n^{-1}\vert A_n) \,,
\qquad
b_n=\sqrt{(A_n\vert A_n)}
   \,,
\end{equation}
with the initial conditions
\begin{equation}
\label{eq:initialCond_Lanczos}
   \vert K_0)=\vert O(0)) \,,
\qquad
b_0=0
   \,.
\end{equation}
The operator $\mathcal{L}$ in \eqref{eq:Lanczos} is called Liouvillian and its action is defined as
\begin{equation}
\label{eq:Liovillian}
  \mathcal{L}\vert O(0)) = \big\vert [H, O(0)]\big)\,.
\end{equation}
Since we require $(O(t)\vert O(t))=1$, when plugged in \eqref{eq:innerproduct}, $O(t)$ has to be appropriately normalized.
The Lanczos algorithm leads to the $\mathcal{K}$ elements of the Krylov basis, on which we expand $\vert O(t))$ as
\begin{equation}
\label{eq:Krylov expansion}
  \vert O(t))=\sum_{n=0}^{\mathcal{K}-1}  {\rm i}^n \phi_n(t) \vert K_n)\,.
\end{equation}
 The number $\mathcal{K}$ is bounded by the dimension of the operator Hilbert space, and it is often smaller than that as the Gram-Schmidt procedure stops before the Krylov basis spans the full operator Hilbert space.
The last formula in \eqref{eq:Lanczos} shows that, by carrying out the full Lanczos algorithm, we find the Lanczos coefficients that we can use to determine the amplitudes $\phi_n(t)$ via the Schroedinger equation
\begin{equation}
\label{eq:LanczosEvolution}
\partial_t \phi_n(t)=b_n \phi_{n-1}(t)-b_{n+1} \phi_{n+1}(t) \,.   
\end{equation} 
The orthonormalization procedure above can be difficult in Hilbert spaces with large dimensionality. Thus, it is helpful to extract the Lanczos coefficients differently. For that, we introduce the autocorrelation function
\begin{equation}
R(t)=(O(t) |O(0))=
\frac{\langle e^{\beta H  /2} O(t) e^{-\beta H  /2} O(0)\rangle_\beta}{\langle e^{\beta H  /2} O(0) e^{-\beta H  /2} O(0)\rangle_\beta}=
\frac{\textrm{Tr}\left(e^{-\beta H  /2} O(t) e^{-\beta H /2} O(0)\right)}{\textrm{Tr}\left(e^{-\beta H /2} O(0) e^{-\beta H  /2} O(0)\right)}
\,.
\label{eq:autocorrelationSM}
\end{equation}
 The Lanczos coefficients are determined through the moment recursion method \cite{LanczosBook}, which consists of using recursive relations between the moments of the autocorrelation function 
\begin{equation}
\label{eq:moments_mun_def}
\mu_n=(-{\rm i})^n\frac{d^n R(t)}{dt^n}\bigg\vert_{t=0}\,,
\end{equation}
and the $b_n$s.
If $O(t)$ is Hermitian, the autocorrelation is real and an even function of time, and, therefore, only the even moments are non-vanishing.
For instance, the first two Lanczos coefficients are obtained from $\mu_2=b_1^2$ and $\mu_4=b_1^2(b_1^2+b_2^2)$, and we refer the interested reader to \cite{LanczosBook,Parker:2018yvk} for more details on how to obtain the ones for larger values of $n$.


\subsection{Lanczos coefficients of fixed-charge blocks}

To compute the symmetry-resolved Krylov complexity, we need the fixed-charge amplitudes $\phi^{(q)}_{n}(t)$ obtained by expanding the block operator $\vert O_q(t))$ in the fixed-charge Krylov basis. Combining this expansion with \eqref{eq:Krylov expansion}, we find
\begin{equation}
\label{eq:Krylov basis_O_v2}
  \vert O(t))=\sum_{q\in\mathbb{Z}}\sqrt{p_q}\sum_{n=1}^{\mathcal K_q-1}  {\rm i}^n \phi^{(q)}_{n}(t) \vert K^{(q)}_{n})\,, 
\end{equation}
which implies that, due to the lack of relation between $\vert K_n)$ and the $\vert K^{(q)}_n)$s for fixed values of $n$ (see the following section), the full complexity is not easily expressed in terms of the symmetry-resolved ones.

As discussed in the main text, the fixed-charge amplitudes are determined starting from $\vert O_q(t))$ and proceeding as done to compute $\phi_{n}(t)$. Following the discussion above, it is convenient to consider the autocorrelation function of $O_q(t)$
\begin{equation}
\label{eq:projected_amplitude}
R_q(t)\equiv(O_q(t)\vert O_q(0))=
\frac{\langle \Pi_q e^{\beta H  /2} O(t)^\dagger e^{-\beta H  /2} O(0)\rangle_\beta}{\langle \Pi_q e^{\beta H  /2} O(0)^\dagger e^{-\beta H  /2} O(0)\rangle_\beta}
=
\frac{\textrm{Tr}\left(\Pi_q e^{-\beta H  /2} O(t) e^{-\beta H  /2} O(0)\right)}{\textrm{Tr}\left(\Pi_q e^{-\beta H  /2} O(0) e^{-\beta H  /2} O(0)\right)}\,,
\end{equation}
that we denote symmetry-resolved autocorrelation function. The second step in \eqref{eq:projected_amplitude} is performed by taking into account that the projectors $\Pi_q$ are idempotent and commute with $O$ and $H$. Also here, the normalization $(O_q(t)\vert O_q(t))=1$ requires rescaling the operators $O_q(t)$ when plugged into \eqref{eq:innerproduct} to compute scalar products. 
Once $R_q(t)$ is obtained, we can apply the moment recursion methods mentioned above to compute the fixed-charge Lanczos coefficients $b^{(q)}_n$ from the moments of $R_q(t)$, whose definition parallels \eqref{eq:moments_mun_def}.

Using the inner product \eqref{eq:innerproduct} and the orthogonality of the projectors, i.e. $\Pi_q\Pi_{q'}=\delta_{qq'}\Pi_{q}$, we can also compute the charge sector probability distribution in \eqref{eq:Odecomposition}. It reads 
\begin{equation}
    p_q=\left|(O_q(0) |O(0))\right|^2=\frac{\textrm{Tr}\left(\Pi_q e^{-\beta H  /2} O(0) e^{-\beta H  /2} O(0)\right)}{\textrm{Tr}\left(e^{-\beta H  /2} O(0) e^{-\beta H  /2} O(0)\right)}
    \label{eq:pq_SM}
    \,.
\end{equation}
Using \eqref{eq:autocorrelationSM}, \eqref{eq:projected_amplitude} and \eqref{eq:pq_SM}, we can prove that
\begin{equation}
\label{eq:resummation amplitude}
R(t)=\sum_q p_q R_q(t)\,,
\end{equation}
which provides a charge decomposition of the autocorrelation function in its symmetry-resolved components.

If the block operator $O_q$ is known explicitly, the computation of $R_q$ is generally straightforward. However, there are instances where accessing block operators might be difficult, as could happen for operators acting on infinite-dimensional Hilbert spaces. To bypass this problem, we take inspiration from the toolkit to study symmetry-resolved entanglement entropies \cite{Goldstein:2017bua}. 
Since the charge sectors are accessed via a projector $\Pi_q=\Pi_q^2$ (see \eqref{eq:Odecomposition}), we can write it  in its Fourier representation
\begin{equation}
\label{eq:FourierProjector}
\Pi_q=\int_{-\pi}^\pi\frac{d \alpha}{2\pi}e^{{ \rm i}\alpha(Q-q)} \,.
\end{equation}
Plugging this relation into \eqref{eq:projected_amplitude}, we find
\begin{equation}
\label{eq:FourierTransf_amplitude}
R_q(t)=\frac{\int_{-\pi}^\pi\frac{d \alpha}{2\pi}e^{-{ \rm i}\alpha q}
\textrm{Tr}\left(e^{{ \rm i}\alpha Q} e^{-\beta H  /2} O(t) e^{-\beta H  /2} O(0)\right)}{\int_{-\pi}^\pi\frac{d \alpha}{2\pi}e^{-{ \rm i}\alpha q}
\textrm{Tr}\left(e^{{ \rm i}\alpha Q} e^{-\beta H  /2} O(0) e^{-\beta H  /2} O(0)\right)}\,.
\end{equation}
Crucially, computing the right-hand side of \eqref{eq:FourierTransf_amplitude} does not require to know the block operator $O_q$. Instead, this is effectively replaced by inserting an Aharonv-Bohm flux in the correlation function of the full operator $O$ and computing its Fourier transform. The formula \eqref{eq:FourierTransf_amplitude} is used to obtain the results for the last examples discussed in the main text.
The Fourier representation \eqref{eq:FourierProjector} can also be exploited to derive a useful formula for the probability distribution \eqref{eq:pq_SM}
\begin{equation}
\label{eq:pq_SM_Fourier}
  p_q= \int_{-\pi}^\pi\frac{d \alpha}{2\pi}e^{-{ \rm i}\alpha q}\frac{\textrm{Tr}\left(e^{{ \rm i}\alpha Q}e^{-\beta H  /2} O(0) e^{-\beta H  /2} O(0)\right)}{\textrm{Tr}\left(e^{-\beta H  /2} O(0) e^{-\beta H  /2} O(0)\right)}\,.
\end{equation}
The expression \eqref{eq:pq_SM_Fourier} is used in the Appendix to compute the non-trivial probability distribution of the charge sectors in which $O_{\textrm{\tiny HO}}$ in \eqref{eq:good_operator_CHO} decomposes.

\subsection{Lanczos algorithm and charge sectors}
\label{subsec:fixchargeLanczos}

In this appendix, we describe how to adapt the Lanczos algorithm to study the dynamics of fixed-charge blocks of operators invariant under a given symmetry.
If the block structure of $O$ is due to a conserved charge $Q$, we have that $[H,Q]=0$, implying that the Hamiltonian $H$ has the same block-diagonal structure as $O$, i.e. $H=\sum_q H_q$. This implies that
$ [H_q,Q_{q'}]\propto\delta_{qq'}$,
and therefore $ [H,O_q]=[H_q,O_q]$.
As a consequence, the Liouvillian operator defined by \eqref{eq:Liovillian} also decomposes as
\begin{equation}
    \mathcal{L}=\sum_q\mathcal{L}_q\,,
\qquad
    \mathcal{L}_{q'} \vert O_q(0))=\delta_{qq'}\mathcal{L}_{q}\vert O_q(0))=\delta_{qq'}\vert [H_q,O_q(0)])
    \,.
\end{equation}
Since every $\mathcal{L}_{q}$ acts non-trivially on the block operators with the same label $q$, the Gram-Schmidt orthonormalization can be carried out sector by sector. This leads to the fixed-charge Krylov basis defined by the following recursive relations
\begin{equation}
\label{eq:LanczosAlg_fixedq}
   \vert A^{(q)}_{n})= \mathcal{L}_q\vert K^{(q)}_{n-1})-b_{n-1}^{(q)}\vert K^{(q)}_{n-2})\,,
  \qquad
   \vert K^{(q)}_{n})=\left(b_n^{(q)}\right)^{-1}\vert A^{(q)}_{n}) \,,
   \end{equation}
with $b_n^{(q)}\equiv\sqrt{(A^{(q)}_{n}\vert A^{(q)}_{n})}$ and the initial condition  $\vert K^{(q)}_{0})=\vert O_q(0))$.

To understand how the fixed-charge Krylov basis vectors $\vert K^{(q)}_{n})$ are related to the total ones $\vert K_{n})$, we explicitly compute $\vert K^{(q)}_{0})$, $\vert K^{(q)}_{1})$, $\vert K^{(q)}_{2})$ and we compare them with the first Krylov vectors $\vert K_1)$ and $\vert K_2)$  of the full unresolved evolution. For convenience, we report these vectors here
\begin{equation}
\label{eq:K1}
   \vert K_1)=\frac{\mathcal{L}\vert K_0)}{\sqrt{(K_0\vert\mathcal{L}^2\vert K_0)}} \,,
\qquad
    \vert K_2)=\frac{\mathcal{L}\vert K_1)-\sqrt{(K_0\vert\mathcal{L}^2\vert K_0)}\vert K_0)}{\sqrt{(K_1\vert\mathcal{L}^2\vert K_1)-(K_0\vert\mathcal{L}^2\vert K_0)}} \,.
\end{equation}

\noindent $\boldsymbol{n=0}$:
Using the initial condition and \eqref{eq:O decomposition_vect} for $t=0$, we obtain
\begin{equation}
\label{eq:K0_decompos}
| K_{0})=\sum_q \sqrt{p_q} | K^{(q)}_{0})\,.
\end{equation}
\noindent $\boldsymbol{n=1}$: 
From \eqref{eq:LanczosAlg_fixedq}, we obtain
\begin{equation}
   \vert K_{1}^{(q)})=\frac{\mathcal{L}_q\vert K_{0}^{(q)})}{\sqrt{(K_{0}^{(q)}\vert\mathcal{L}_q^2\vert K_{0}^{(q)})}} \,,
\end{equation}
which, from \eqref{eq:K1} and \eqref{eq:O decomposition_vect}, can be used to decompose $\vert K_1)$ in \eqref{eq:K1} as
\begin{equation}
\label{eq:K1_decompos}
  \vert K_{1})=\sum_q \sqrt{p_q}\sqrt{\frac{(K^{(q)}_{0}\vert\mathcal{L}_q^2\vert K^{(q)}_{0})}{(K_{0}\vert\mathcal{L}^2\vert K_{0})}}\vert K^{(q)}_{1})  \,.
\end{equation}
Notice that $\vert K_{1})$ can be decomposed in terms of the $\vert K^{(q)}_{1})$ only, without involving vectors of the fixed-charge Krylov basis with different label $n$.
\\

\noindent $\boldsymbol{n=2}$: Using \eqref{eq:LanczosAlg_fixedq}, we straghtforwardly obtain 
\\
\begin{equation}
   \vert K_{2}^{(q)})=\frac{\mathcal{L}_q\vert K_{1}^{(q)})-\sqrt{(K_{0}^{(q)}\vert\mathcal{L}_q^2\vert K_{0}^{(q)})}\vert K_{0}^{(q)})}{\sqrt{(K_{1}^{(q)}\vert\mathcal{L}_q^2\vert K_{1}^{(q)})-(K_{0}^{(q)}\vert\mathcal{L}_q^2\vert K_{0}^{(q)})}} \,.
\end{equation}
The vector  $\vert K_{2})$ in \eqref{eq:K1} can be decomposed as
\begin{equation}
\label{eq:K2_decomposition}
 \vert K_{2})= \sum_q \sqrt{p_q}\sqrt{\frac{(K_{0}^{(q)}\vert\mathcal{L}_q^2\vert K_{0}^{(q)})}{(K_{0}\vert\mathcal{L}^2\vert K_{0})}} \frac{\mathcal{L}_q\vert K_{1}^{(q)})-\frac{(K_{0}\vert\mathcal{L}^2\vert K_{0})}{\sqrt{(K_{0}^{(q)}\vert\mathcal{L}_q^2\vert K_{0}^{(q)})}}\vert K_{0}^{(q)})}{\sqrt{(K_{1}\vert\mathcal{L}^2\vert K_{1})-(K_{0}\vert\mathcal{L}^2\vert K_{0})}}  \,.
\end{equation}
However, since 
\begin{equation}
\label{eq:K2_decomposition_v2}
   \mathcal{L}\vert K^{(q)}_{1})-\frac{(K_{0}\vert\mathcal{L}^2\vert K_{0})}{\sqrt{(K_{0}^{(q)}\vert\mathcal{L}_q^2\vert K_{0}^{(q)})}}\vert K_{0}^{(q)})\,\, \cancel{\propto}\,\,\vert K_{2}^{(q)})\,,
\end{equation}
 we cannot decompose $\vert K_{2})$ only in terms of the fixed-charge Krylov vectors with the same label $n=2$.
 This property also holds for $n>2$ and implies that the total complexity $C_K(t)$ in \eqref{eq:total_C} does not have a simple relation in terms of the symmetry-resolved complexities. If a relation between $C_K(t)$ and the $C^{(q)}_K(t)$s can be found, it has to be worked case-by-case, as we show later in this Appendix for the example of a general block diagonal $4\times 4$ operator.

The explicit construction \eqref{eq:LanczosAlg_fixedq} allows us to prove that Krylov vectors associated with different charge sectors are orthogonal. Indeed, using the inner product in \eqref{eq:innerproduct} and the fact that the projectors commute with $O(0)$ and the Hamiltonian, we find
\begin{equation}
(K_{0}^{(q)}\vert K_{0}^{(q')})\propto \langle  e^{\beta H  /2}\Pi_q O(0) e^{-\beta H  /2} \Pi_{q'} O(0)\rangle_\beta \propto \delta_{qq'}\,,
\end{equation}
for vectors with $n=0$. This relation is extended to
$(K_{n}^{(q)}\vert K_{m}^{(q')})\propto \delta_{qq'}$ by noticing that $\Pi_q$ commutes also with the nested commutators of $O(0)$ with the Hamiltonian.

To compute the fixed-charge Lanczos coefficients, we have to carry out the full Lanczos algorithm in \eqref{eq:LanczosAlg_fixedq} or, alternatively, compute the moments of the symmetry-resolved autocorrelation function \eqref{eq:projected_amplitude}. Expanding in series \eqref{eq:resummation amplitude}, we find a tower of relations between the moments of $R_q(t)$ and those of $R(t)$
\begin{equation}
\label{eq:mun_chargedecoposition}
  \mu_{n}= \sum_{q\in \mathbb{Z}} p_q \mu^{(q)}_{n}   \,.
\end{equation}
Due to these formulas, we can write the first Lanczos coefficient of the full operator dynamics in terms of the first fixed-charge Lanczos coefficients, namely $b_1^2    =\sum_q p_q \left(b^{(q)}_{1}\right)^2$.
The nonlinearity of the relations between $b^{(q)}_{n}$ and $\mu^{(q)}_{n}$ does not allow to find further simple decompositions of the Lanczos coefficients $b_{n}$ in terms of their fixed-charge components.  Such relations can be worked out case-by-case, but they are not linear in the probability distribution $p_{q}$ and they become difficult to write in a closed form. We can use these decompositions to derive the formula \eqref{eq:differenceCpCK}. The early time growth of the full Krylov complexity reads \cite{Fan:2022xaa}
\begin{equation}
  C_K(t)=\mu_2 t^2+\left(\frac{\mu_4}{6}-\frac{\mu^2_2}{2}\right) t^4+O(t^6)  \,,
\end{equation}
and the same holds for $C^{(q)}_K(t)$ by replacing $\mu_{n}$ with $\mu^{(q)}_{n}$. From \eqref{eq:mun_chargedecoposition} and the first two relations between $b^{(q)}_{n}$ and $\mu^{(q)}_{n}$, we \eqref{eq:differenceCpCK}.

\subsection{Details on the two-spin example}


The first example considered in the main text involves two spins $1/2$ evolving through the Hamiltonian
\begin{equation}
    \label{eq:spinHamiltonian}
H_{\textrm{spin}}=J(X_{\textrm{L}}X_{\textrm{R}}+Y_{\textrm{L}}Y_{\textrm{R}}+Z_{\textrm{L}}Z_{\textrm{R}})\,,
\end{equation}
where the spin variables are written in terms of the Pauli matrices $X_{\textrm{S}},\,Y_{\textrm{S}}\,,Z_{\textrm{S}}$ with ${\textrm{S}}\in\{{\textrm{L}},{\textrm{R}}\}$.
Here, $J$ is the spin coupling.
This Hamiltonian commutes with the total magnetization of the two spins along the $Z$-direction, i.e.
\begin{equation}
Q_{\textrm{spin}}=Z_{\textrm{L}}+Z_{\textrm{R}}\,.
\end{equation}
We focus on a single spin operator $Z_{\textrm{L}}$, which commutes with $Q_{\textrm{spin}}$ but not with $H_{\textrm{spin}}$, having, therefore, a non-trivial time evolution. The Heisenberg evolution of $Z_{\textrm{L}}$ is given in \eqref{eq:spinL_evolved} and exhibits the expected block-diagonal structure.  

Using the inner product in \eqref{eq:innerproduct} with $\beta=0$, we compute the autocorrelation function of $Z_{\textrm{L}}(t)$ and, from its moments in \eqref{eq:moments_mun_def}, we find that there are only two non-vanishing Lanzcos coefficients determining  dynamics of $Z_{\textrm{L}}(t)$ on the Krylov chain
\begin{equation}
\label{eq:LanczoscoeffspinL}
    b_1=b_2=J/\sqrt{2}\,.
\end{equation}
This implies that the Krylov space is generated by three independent vectors. When we expand the operator on this basis, we obtain the amplitudes \begin{equation}
\label{eq:phi0_2spin}
\phi_{0}(t) =\cos^2(Jt/2)\,,\qquad
 \phi_{1}(t)  =\frac{\sin(Jt)}{\sqrt{2}}\,,\qquad
  \phi_{2}(t) =\sin^2(Jt/2)\,. 
\end{equation}
From \eqref{eq:phi0_2spin}, we can compute the Krylov complexity $C_K(t)=1-\cos(Jt)$, as reported in the main text.

Let us move to the Krylov complexities of the three blocks, i.e. the symmetry-resolved complexities.
Since the two blocks associated with $q=\pm 1$ do not evolve in time, their Krylov complexity is zero. On the other hand, the $2\times 2$ block has non-trivial time evolution and non-trivial complexity.
As explained above, we can extract the Lanczos coefficients from the moments of the autocorrelation function $R_{q=0}(t)$, finding only one of them being non-vanishing, i.e. $b_{1}^{(q=0)}=J$. This corresponds to a two-dimensional Krylov space and fixed-charge amplitudes for the block operator expanded on the Krylov basis given by
\begin{equation}
\label{eq:phi0_2spin_sectors}
\phi^{(q=0)}_{0}(t) =\cos(Jt)\,,\qquad
 \phi^{(q=0)}_{1}(t)  =\sin(Jt)\,. 
\end{equation}
Using these amplitudes and \eqref{eq:fixedcharge_C}, the symmetry-resolved Krylov complexity with $q=0$ reads $C^{(q=0)}_K(t)=\sin^2(Jt)$. From \eqref{eq:pq_SM}, we can compute the probabilities associated with the blocks of $Z_{\textrm{L}}(t)$, finding $p_{\pm 1}=\frac{1}{4}$ and $p_{0}=\frac{1}{2}$. Combining these with the symmetry-resolved Krylov complexities as in \eqref{eq:weighted_Complexity}, we find $\bar{C}(t)=\sin^2(Jt)/2$, and we compute the difference $C_K(t)-\bar{C}(t)$ reported in the main text.
It is insightful to notice that, in this example, the total Krylov complexity can be expressed in terms of the only non-vanishing symmetry resolved component or of the average $\bar{C}(t)$. It reads $C_K(t)=1-\sqrt{1-C^{(q=0)}_K(t)}=1-\sqrt{1-2\bar{C}(t)}$, which is a simple, yet non-linear relation. 

\subsection{Krylov complexity symmetry-resolution of a block diagonal $4 \times 4$ operator}

In this appendix, we provide the details for computing the Krylov complexity (and its symmetry resolution) of a $4 \times 4$ operator with two $2 \times 2$ blocks along its diagonal.

\subsubsection{Krylov complexity of the $2 \times 2$ blocks}

Let us first consider an arbitrary system defined on a two-dimensional Hilbert space and described by a Hamiltonian which, in the energy basis, is represented as
\begin{equation}
\label{eq:Hamiltonian2x2}
H_2 = \begin{pmatrix}
E_1 & 0 \\
0 & E_2
\end{pmatrix}\,.
\end{equation}
The initial state operator at time $t=0$ is taken to be
the most general $2 \times 2$ Hermitian matrix $O_2$
\begin{equation}
O_2 = \begin{pmatrix}
A_{11} & A_{12} + {\rm i} B_{12} \\
A_{12} - {\rm i} B_{12} & A_{22}
\end{pmatrix}\,,
\end{equation}
where $A_{11},\, A_{22},\,A_{12},\,B_{12}\in\mathbb{R}$.
For convenience, we define the squared sums of the diagonal and off-diagonal components of $O_2$ as
\begin{equation}
M = A_{12}^2 + B_{12}^2\,, \quad T = A_{11}^2 + A_{22}^2\,.\label{RT}
\end{equation}
The Krylov complexity of the time evolution of the operator $O_2$ with respect to the Hamiltonian $H_2$ \label{eq:Hamiltonian2x2} (see \eqref{eq:Heisem_time_ev})
 has been computed in \cite{Caputa:2024vrn} and reads
\begin{equation}
\label{eq:2x2KComplexity_SM}
C^{(2\times 2)}_K(t) = \frac{2 M}{T + 2 M} 
\left\{
\sin^2 (\Delta E t) 
+ \frac{8 T}{T + 2 M} \sin^4 \left( \frac{\Delta E}{2} t \right)
\right\}\,,
\end{equation}
where $T$ and $M$ are given in \eqref{RT} and $\Delta E=E_1-E_2$.

In the next section, we use \eqref{eq:2x2KComplexity_SM} as symmetry-resolved Krylov complexity of the $4\times 4$ operator $O_4$ \eqref{eq:operator4times4}. For notational convenience, we label the two $2\times 2$ blocks along its diagonal as $O_+$ and $O_-$, i.e.
\begin{equation}
\label{eq:O4_SM}
O_4 =
\begin{pmatrix}
O_+ & 0 \\
0 & O_-
\end{pmatrix}.
\end{equation}
The blocks $O_\pm$ have Krylov complexities $C^{(\pm)}_K(t)$ written as in \eqref{eq:2x2KComplexity_SM} in terms of the energy differences $\Delta E_+=E_1-E_2$ and $\Delta E_-=E_3-E_4$ and the quantities defined in \eqref{RT} specified for the two sets of entries of the blocks. We denote these latter auxiliary quantities for the upper (lower) block as $M_+$ and $T_+$ ($M_-$  and $T_-$).
As  $O_\pm$ are interpreted as labeling the charge sectors induced by a conserved charge, we think of $C^{(\pm)}_K(t)$ as symmetry-resolved Krylov complexities of $O_4$.

\subsubsection{Krylov complexity of a $4 \times 4$ operator}

We move now to discuss the total Krylov complexity of $O_4$ evolving via the Hamiltonian $H_4={\rm diag}\left(E_1,E_2,E_3,E_4\right)$. When represented as a vector in a Hilbert space, the evolution of $O_4$ is realized by acting with the Liouvillian in \eqref{eq:Liovillian} associated with $H_4$. We denote it as $\mathcal{L}_4$. The Hilbert space considered here is equipped with the inner product \eqref{eq:innerproduct} with $\beta=0$.
Since $\mathcal{L}_4$ is Hermitian, we can diagonalize it as
\begin{equation}
    \mathcal{L}_4 = V D V^{-1}\label{phi0t}\,,
\end{equation}
where $D$ is a diagonal matrix whose entries are the real eigenvalues $\lambda_j$ of $\mathcal{L}_4$, and the columns of $V$ are the corresponding eigenvectors. 
When we represent the evolving operator in a Hilbert space, the autocorrelation function can be seen as a return amplitude. This, in turn, is equal to the complex conjugate of the amplitude of $\vert O_4(t)) $ on the first Krylov vector. We have \cite{Parker:2018yvk}
\begin{equation}
    R(t)=\phi^*_0(t) = 
\sum_k \abs{V_{0k}}^2
e^{{\rm i}\lambda_k t}\,, 
\qquad
V_{0k}\equiv(O_4(0)\vert V\vert k)\,,
\label{eq:O0t}
\end{equation}
where $\vert k)$ is the eigenvector of $\mathcal{L}_4$ associated with $\lambda_k$.
It is convenient to introduce the time-dependent moments as
\begin{equation}
    \mu_n(t) = (-{\rm i})^n\frac{\partial_t^n \phi_0(t)}{\phi_0(t)} 
    = \frac{\sum_k \abs{V_{0k}}^2 \lambda_k^n e^{{\rm i}\lambda_k t}}{\sum_k \abs{V_{0k}}^2 e^{{\rm i}\lambda_k t}} \,\label{mu_n}\,,
\end{equation}
which, when $t=0$, reduce to the moments \eqref{eq:moments_mun_def}.
By applying the moments recursion method on the autocorrelation function \eqref{eq:O0t}, we find that there are only four non-vanishing Lanczos coefficients. 
Now we can use \eqref{eq:LanczosEvolution} to express the other four non-vanishing amplitudes $\phi_n(t)$, with $n=1,2,3,4$, in terms of the Lanczos coefficients and $\phi_0(t)$ and its derivatives. We obtain 
\begin{align}
& \phi_1(t)= - \frac{1}{b_1} \partial_t \phi_0(t) \,, 
\qquad\qquad \qquad \qquad \qquad \;\;\,\, 
\phi_2(t) =  \frac{b_1^2 + (\partial_t)^2}{b_1 b_2} \phi_0(t) \,, 
\\
&
\phi_3(t) = - \frac{(b_1^2 + b_2^2)(\partial_t) + (\partial_t)^3}{b_1 b_2 b_3} \phi_0(t) \,,
\qquad\qquad
\phi_4(t) = \frac{b_1^2 b_3^2 + (b_1^2 + b_2^2 + b_3^2)(\partial_t)^2 + (\partial_t)^4}{b_1 b_2 b_3 b_4} \phi_0(t) \,.
\end{align}
From \eqref{mu_n}, we re-express these amplitudes in terms of time-dependent moments as
\begin{align}
&\phi_1(t) = -\frac{1}{b_1} {\rm i} \mu_1(t) \phi_0(t)\,,
\qquad\qquad \qquad \qquad \qquad \;\;\, 
\phi_2(t) = 
\frac{b_1^2 - \mu_2(t)}{b_1 b_2} \phi_0(t)\,,\\
&\phi_3(t) =-
\frac{\left(b_1^2 + b_2^2\right){\rm i} \mu_1(t)  -{\rm i} \mu_3(t)}{b_1 b_2 b_3} \phi_0(t)\,, \qquad\qquad
\phi_4(t) =
\frac{b_1^2 b_3^2 - \left(b_1^2 + b_2^2 + b_3^2\right)\mu_2(t) + \mu_4(t)}{b_1 b_2 b_3 b_4} \phi_0(t)\,.
\end{align}
At this point, we have all the ingredients to explicitly compute the amplitudes $\phi_n(t)$, given that we know the time-dependent moments from \eqref{mu_n} and the Lanczos coefficients from the initial moments $\mu_n(0)$.
This allows us to compute the Krylov complexity of $O_4(t)$, which reads
\begin{align}
\label{eq:CK4x4_SM}
&C_K(t) = \sum_{n=0}^4 n\, |\phi_n(t)|^2 =g_4(t) |\phi_0(t)|^2\,,
\end{align}
where we have defined 
\begin{align}
\label{eq:g4_def}
g_4(t) \equiv
\left| \frac{\mu_1(t)}{b_1} \right|^2 +
2  \left| \frac{b_1^2 - \mu_2(t)}{b_1 b_2} \right|^2 +
3  \left| \frac{(b_1^2 + b_2^2)\mu_1(t) - \mu_3(t)}{b_1 b_2 b_3} \right|^2 +
4  \left| \frac{b_1^2 b_3^2 - (b_1^2 + b_2^2 + b_3^2)\mu_2(t) + \mu_4(t)}{b_1 b_2 b_3 b_4} \right|^2\,.
\end{align}
Although we already know the Krylov complexity of the two $2\times 2$ time-dependent matrices (see \eqref{eq:2x2KComplexity_SM}), it is convenient to adapt the approach shown here to the $2\times 2$ blocks $O_\pm$.  Starting from the knowledge of the two amplitudes $\phi_0^{(\pm)}(t)$ (or, equivalently, the two autocorrelation functions $R_\pm(t)$), we can extract the only non-vanishing Lanczos coefficients $b_1^{(\pm)}$ and $b_2^{(\pm)}$. Following what we did before, we compute the other two non-vanishing amplitudes for each block operator, finding
\begin{equation}
    \phi^{(\pm)}_1(t)= - \frac{1}{b^{(\pm)}_1} \partial_t \phi^{(\pm)}_0(t) \,, 
\qquad\qquad \, 
\phi^{(\pm)}_2(t) =  \frac{\left(b^{(\pm)}_1\right)^2 + \partial_t^2}{b^{(\pm)}_1 b^{(\pm)}_2} \phi^{(\pm)}_0(t)\,.
\end{equation}
Adapting the definition \eqref{mu_n} to $\mu_n^{(\pm)}(t)$ defined in terms of $\phi_0^{(\pm)}(t)$, we can write the Krylov complexity of $O_\pm$ as
\begin{equation}
\label{eq:Cplusminus}
   C_K^{{(\pm)}}(t) = \sum_{n=0}^2 n\, |\phi^{(\pm)}_n(t)|^2 =f^{(\pm)}_2(t) |\phi^{(\pm)}_0(t)|^2\,,  \qquad
   f_{2}^{(\pm)}(t) \equiv  \left| \frac{\mu_1^{(\pm)}(t)}{b_{1}^{(\pm)}} \right|^2 +
2  \left| \frac{\left(b^{(\pm)}_1\right)^2 - \mu_{2}^{(\pm)}(t)}{b_{1}^{(\pm)} b_{2}^{(\pm)}} \right|^2\,. 
\end{equation}
As expected, these expressions become \eqref{eq:2x2KComplexity_SM} when we write the Lanczos coefficients and the time-dependent moments in terms of the entries of  $O_\pm$.

\begin{figure}[t!]
	\centering
	\includegraphics[width=\linewidth]{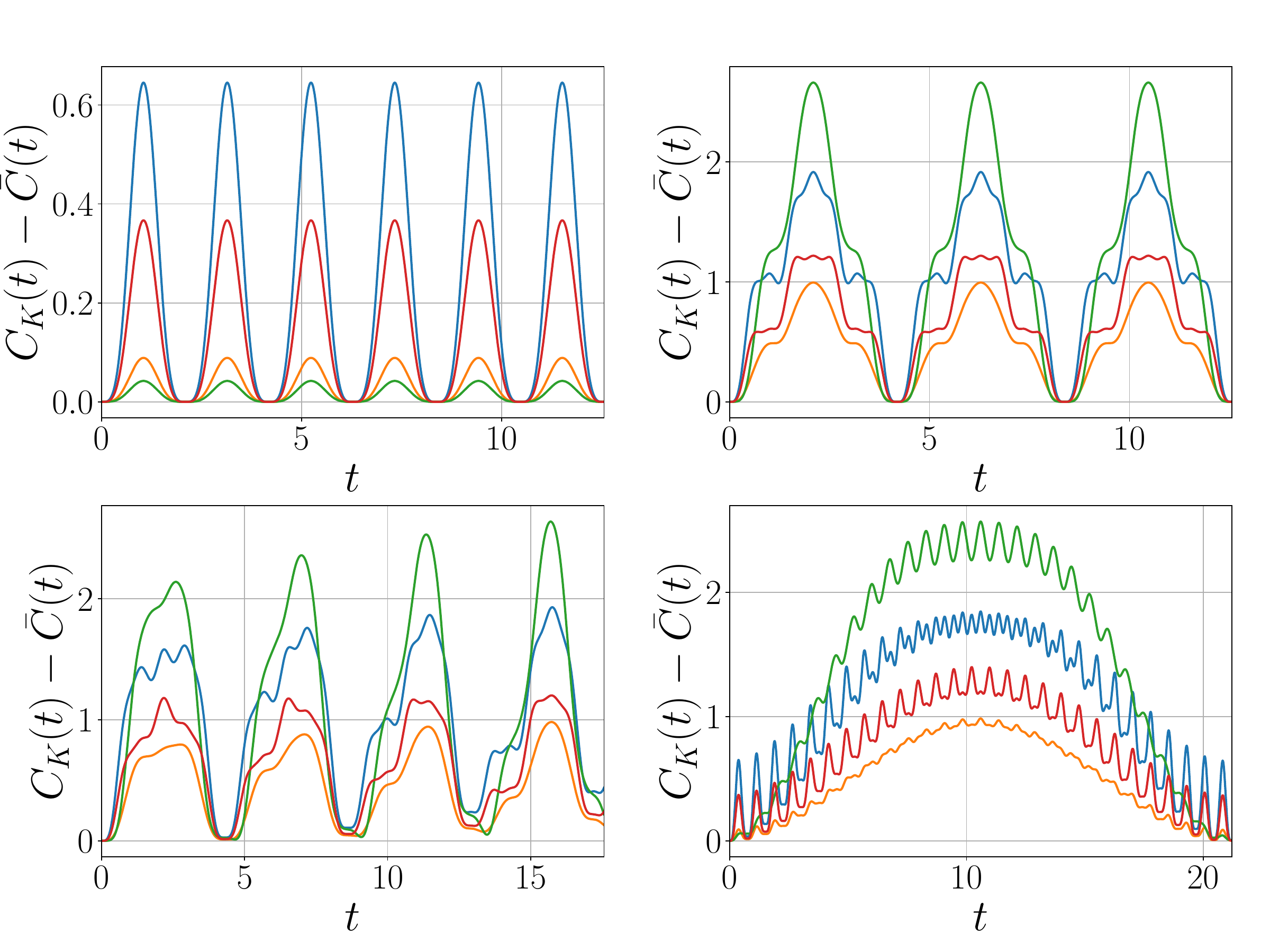}
	\caption{Difference between the Krylov complexity of the operator \eqref{eq:operator4times4} and the average \eqref{eq:weighted_Complexity} over the (two) sectors. Different colors correspond to distinct choices of the parameters $A_{ii}$ ($i=1,2,3,4$), $A_{12}$, $A_{34}$, $B_{12}$, $B_{34}$ (see Table I). In all the instances shown here, $C_K(t)-\bar{C}(t)\geq 0$, supporting our surmise.
     Curves in different panels are obtained for different sets of energies: top-left $E_1=8.00,\,E_2=5.00\,,E_3=1.00\,,E_4=1.00$, top-right $E_1=8.00,\,E_2=2.00\,,E_3=2.50\,,E_4=1.00$, bottom-left $E_1=7.10,\,E_2=2.72\,,E_3=3.14\,,E_4=1.73$, bottom-right $E_1=10.50,\,E_2=2.20\,,E_3=3.00\,,E_4=3.30.$}
	\label{fig:4x4Example_App}
\end{figure}

\begin{figure}[t!]
	\centering
\includegraphics[width=.5\textwidth]{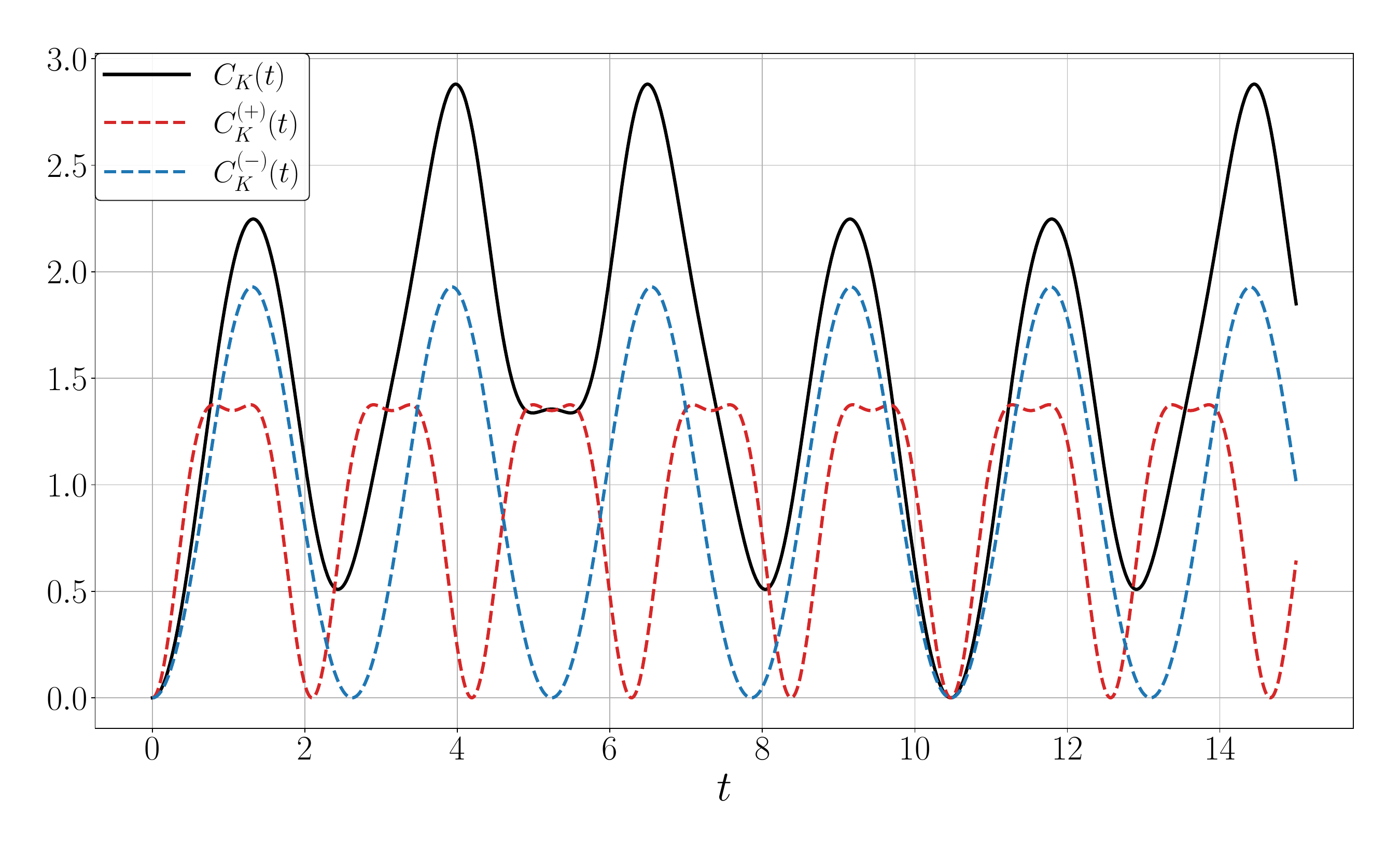}
	\caption{Krylov complexity \eqref{c4} and symmetry resolved complexities \eqref{eq:Cplusminus} for the two charge sectors plotted as functions of time. The curves have been obtained for $E_1=8.0,\,E_2=5.0\,,E_3=3.4\,,E_4=1.0$ and $A_{ij}$ and $B_{ij}$ in the first line of Table I. We observe that, for some values of time, the symmetry-resolved Krylov complexity $C_K^{(+)}(t)$ is larger than the Krylov complexity of the total $4\times 4$ operator.}
	\label{fig:CplusminusCK}
\end{figure}

When we regard the operators $O_\pm$ as the two blocks along the diagonal of $O_4$ (see $\eqref{eq:O4_SM}$), we can decompose the amplitude $\phi_0$ in terms of $\phi_0^{(\pm)}$, namely
\begin{align}
\label{eq:phi0fromplusminus}
\phi_0(t) = p_+\, \phi_0^{(+)}(t) + p_-\, \phi_0^{(-)}(t)\,.
\end{align}
This fact comes from taking the complex conjugate of the decomposition  \eqref{eq:resummation amplitude}.
The probabilities $p_\pm$ can be written in terms of the parameters of the entries of $O_\pm$ as
\begin{equation}
\label{eq:pplusminus}
p_\pm \equiv \frac{T_\pm + 2 M_\pm}{T_+ + 2 M_+ + T_- + 2 M_-}\,,
\end{equation}
where
$T_\pm$ and $M_\pm$ are defined below \eqref{eq:O4_SM}. Since the autocorrelation function in \eqref{eq:O0t} is real for Hermitian operators, $\phi_0(t)$, $\phi^{(\pm)}_0(t)$ are real. Using the fact that these amplitudes are real, from \eqref{eq:phi0fromplusminus}, we can write the square modulus of $\phi_0(t)$ as
\begin{align}
|\phi_0(t)|^2 = 
p_+^2\, \left(\phi^{(+)}_0(t)\right)^2 
+ p_-^2\, \left(\phi^{(-)}_0(t)\right)^2
+ 2p_+ p_- 
\phi^{(+)}_0(t)\, \phi^{(-)}_0(t) 
\label{normO44}\,.
\end{align}
Plugging this expression into \eqref{eq:CK4x4_SM} and using \eqref{eq:Cplusminus}, we obtain
\begin{align}
C_K(t) 
&= g_4(t) \left(
p^2_+ \frac{C_K^{(+)}(t)}{f^{(+)}_2(t)}
+
p^2_- \frac{C_K^{(-)}(t)}{f^{(-)}_2(t)}
+ 2p_+ p_- \phi^{(+)}_0(t)\phi^{(-)}_0(t)
\right)
\\
&
=
g_4(t) \left(
p_+ \sqrt{\frac{C_K^{(+)}(t)}{f^{(+)}_2(t)}}
+ 
s_\pm(t) p_- \sqrt{\frac{C_K^{(-)}(t)}{f^{(-)}_2(t)}}
\right)^2
\,,\label{c4}
\end{align}
where $g_4(t)$ is given in \eqref{eq:g4_def}, $f_2^{(\pm)}(t)$ in \eqref{eq:Cplusminus} and $p_\pm$ in \eqref{eq:pplusminus} and we have defined $s_\pm(t)\equiv\sgn \left [\phi^{(+)}_0(t)\phi^{(-)}_0(t)\right]$. Let us comment on this result.
This expression highlights the intricacy of reconstructing full complexity from the contributions of the blocks. The weights from the different charge sectors occur through the squares of the probabilities, and an interference term $\phi^{(+)}_0(t)\phi^{(-)}_0(t) $ mixing the distinct sectors also emerges. 
We can use the analytical expressions \eqref{c4}, \eqref{eq:Cplusminus} and \eqref{eq:pplusminus} to compute $\bar{C}(t)$ in \eqref{eq:weighted_Complexity} and study the behaviour of $C_K(t) - \bar{C}(t)$. This is reported in Fig.\,\ref{fig:4x4Example} and Fig.\,\ref{fig:4x4Example_App}. Curves in different panels are drawn for different sets of energies $E_i$ (specified in the captions), while curves in the same panel differ by the choice of the entries of the initial operator in \eqref{eq:operator4times4} (see Table I).  We observe that $C_K(t) - \bar{C}(t)$ is positive in all the cases considered. Despite not being a proof, these curves support our expectation on the sign of $C_K(t) - \bar{C}(t)$.
The fact that difference $C_K(t) - \bar{C}(t)$ is positive for any time is mathematically even less trivial if we notice that one of the symmetry resolved Krylov complexity in \eqref{eq:Cplusminus}, i.e. $C_K^{(+)}(t)$, can be either larger or smaller than $C_K(t)$ in \eqref{c4}, depending on $t$. This is shown in Fig.\,\ref{fig:CplusminusCK}, where we report these three quantities as a function of time for certain values of parameters of the model. The plot shows that there is no fixed order among these complexities, as it changes with $t$.

\begin{table}[t]
\centering
\begin{tabular}{c|cccccccc}
\toprule
\textbf{Curve} & $A_{11}$ & $A_{22}$ & $A_{33}$ & $A_{44}$ & $A_{12}$ & $B_{12}$ & $A_{34}$ & $B_{34}$ \\
\midrule
\tikz{\draw[thick, color={rgb,255:red,31;green,119;blue,180}] (0,0) -- (0.6,0);}   & 0.7 & 0.2 & 2.1 & 1.9 & 0.4 & 0.9 & 1.5 & 0.7 \\
\tikz{\draw[thick, color={rgb,255:red,255;green,127;blue,14}] (0,0) -- (0.6,0);}    & 2.2 & 1.5 & 4.7 & 0.4 & 0.7 & -0.8 & 0.9 & -1.3 \\
\tikz{\draw[thick, color={rgb,255:red,44;green,160;blue,44}] (0,0) -- (0.6,0);}    & 0.3 & 1.9 & 1.1 & 1.8 & -0.6 & 0.2 & 2.4 & 0.6 \\
\tikz{\draw[thick, color={rgb,255:red,214;green,39;blue,40}] (0,0) -- (0.6,0);}  & 0.5 & 0.3 & 4.9 & 2.2 & 1.0 & 0.3 & -1.7 & 0.8 \\
\bottomrule
\end{tabular}
\label{Table}
\caption{The table shows the values of the parameters of the operator in \eqref{eq:operator4times4} chosen to draw the curves in Fig.\,\ref{fig:4x4Example} and Fig.\,\ref{fig:4x4Example_App}.}
\end{table}

We hope that the form \eqref{c4} of the Krylov complexity may offer a valuable framework for studying more general block-diagonal operators defined in higher-dimensional Hilbert spaces. In particular, it could help verify the expectation that $C_K(t) - \bar{C}(t)$ is non-negative while also shedding light on how correlations across different charge sectors contribute to this behavior.

\subsection{Krylov complexity of complex harmonic oscillators}
In this final appendix, we detail the computations for the Krylov complexity and its resolution for U(1)-invariant operators defined in systems of coupled complex harmonic oscillators.
\subsubsection{Example of model of coupled oscillators}
As discussed in the main text, any Hamiltonian of non-Hermitian canonically commuting operators $\Phi_j$ and $\Pi_j$ that can be rewritten as the sum of two independent real oscillator Hamiltonians is diagonalized in the form \eqref{eq:HamcomplexHO_diag}.
Here, for concreteness, we provide an example of a Hamiltonian with these properties. Consider the one-dimensional chain
\begin{equation}
\label{eq:HC_ex_SM}
    H_{\textrm{\tiny HO}}=\sum_{j=1}^N\left(\Pi_j^\dagger\Pi_j+\omega^2\Phi_j^\dagger\Phi_j+(\Phi^\dagger_{j+1}-\Phi^\dagger_j)(\Phi_{j+1}-\Phi_j)\right)\,,
\end{equation}
where periodic boundary conditions $\Phi_{j+N}=\Phi_{j}$ are imposed. We introduce two sets of Hermitian canonically commuting operators
$\Phi_j=(\varphi_j+{\rm i}\tilde \varphi_j)/\sqrt{2}$, $\Pi_j=(p_j+{\rm i}\tilde p_j)/\sqrt{2}$ so that $[\varphi_j,p_l]=[\tilde\varphi_j,\tilde p_l]={\rm i}\delta_{jl}$ and all the other commutators are vanishing. In these new variables, \eqref{eq:HC_ex_SM} becomes
\begin{equation}
    H_{\textrm{\tiny HO}}=\sum_{j=1}^N\left(\frac{1}{2}p_j^2 +\frac{1}{2}\omega^2\varphi_j^2+\frac{1}{2}(\varphi_{j+1}-\varphi_j)^2\right)+\sum_{j=1}^N\left(\frac{1}{2}\tilde p_j^2+\frac{1}{2}\omega^2\tilde\varphi_j^2+\frac{1}{2}(\tilde\varphi_{j+1}-\tilde\varphi_j)^2\right)\,.
\end{equation}
We have written $ H_{\textrm{\tiny HO}}$ as a sum of two real oscillator Hamiltonians with the same frequency. Thus, we can recast it in the diagonal form \eqref{eq:HamcomplexHO_diag} after having introduced two sets of creation and annihilation operators. In this example, we know the explicit expression of the dispersion relation, which reads \cite{Serafini17book}
\begin{equation}
    \omega_k=\sqrt{\omega^2+4\sin^2\left(\frac{\pi k}{N}\right)}\,, \qquad k=1,\dots,N.
\end{equation}
However, this expression does not affect our analysis, and we can maintain it general in what follows.

\subsubsection{Krylov complexity of the operator \eqref{eq:good_operator_CHO} with $N=1$ and its symmetry resolution}
Here, we compute the Krylov complexity of the full operator $O_{\textrm{\tiny HO}}(t)$ in \eqref{eq:good_operator_CHO} and its components resolved with respect to the U(1) symmetry generated by $Q_{\textrm{\tiny HO}}$ given in the main text. For this purpose, we need to compute the autocorrelation functions
$R(t)$ and $R_q(t)$. 
For evaluating $R(t)$, we observe that
\begin{equation}
\label{eq:forR_OHO}
    \textrm{Tr}\left(e^{-H \beta /2} O_{\textrm{\tiny HO}}(t) e^{-H \beta /2} O_{\textrm{\tiny HO}}(0)\right)
=\frac{\cos(2\omega t)}{8\sinh^4(\beta \omega/2) }\,.
\end{equation}
Notice that here we are considering general value of $\beta$ in the inner product \eqref{eq:innerproduct}. 
The return amplitude is obtained by dividing \eqref{eq:forR_OHO} by its value at initial time $t = 0$. We obtain $R(t)=\cos(2\omega t)$. Applying the moment recursion method to this autocorrelation function, we find, as described in the main text, $b_1=2\omega$ and all the other Lanczos coefficients equal to zero. This leads to $C_K(t)=\sin^2(2\omega t)$.

Moving to the symmetry resolution, computing $R_q(t)$ is not straightforward, as we do not have direct access to the fixed-charge blocks. Indeed, the full Hilbert space of the system and the fixed-charge Hilbert spaces are infinite-dimensional. To overcome this difficulty, we resort to the formula \eqref{eq:FourierTransf_amplitude}. Thus, 
we first compute the trace above with the insertion of a flux operator $e^{{\rm i} \alpha Q_{\textrm{\tiny HO}}}$, obtaining
\begin{equation}
\label{eq:fluxS_OCHO}
  \textrm{Tr}\left(e^{{ \rm i}\alpha Q_{\textrm{\tiny HO}}}e^{-H \beta /2} O_{\textrm{\tiny HO}}(t) e^{-H \beta /2} O_{\textrm{\tiny HO}}(0)\right)= 
  \frac{\cos(2\omega t)}{2[\cos(\alpha)-\cosh(\beta\omega )]^2}\,.
\end{equation}
We see that the dependence on time and the one on $\alpha$ factorize, which implies that, after the Fourier transform in $\alpha$, we find $\cos(2\omega t)$ multiplied by a non-trivial function of $q$. Dividing this outcome by itself evaluated at $t=0$, the dependence on $q$ drops out and we find $R_q(t)=\cos(2\omega t)$ for any integer $q$. This implies that the fixed-charge Lanczos coefficients are the same as the ones for the dynamics of the full operator, and as a consequence, the symmetry-resolved complexity is $C^{(q)}_K(t)=C_K(t)=\sin^2(2\omega t)$ in any charge sector. The fact that $C^{(q)}_K(t)$ is independent of $q$ realizes the equipartition of Krylov complexity. Before commenting on the possible reason behind the equipartition, we observe that, using \eqref{eq:pq_SM}, the probability associated with each sector reads
\begin{equation}
\label{eq:prob_OCHO}
  p_q=e^{-\vert q\vert \omega \beta }(\vert q\vert+\coth(\beta \omega))\tanh^2(\beta \omega/2)  \,,
\end{equation} 
whose normalization implies that $C_K(t)-\bar{C}(t)= 0$ in this case. The distribution \eqref{eq:prob_OCHO} shows that the various charge sectors of the operator $O_{\textrm{\tiny HO}}$ have different properties. This makes the occurrence of equipartition of Krylov complexity even less trivial.



\subsubsection{Equipartition and its breaking}

To investigate possible reasons for the occurrence of the equipartition of Krylov complexity, we consider a generic operator invariant under the U(1) symmetry generated by $Q_{\textrm{\tiny HO}}$.
We preliminarily characterize the fixed-charge Hilbert spaces in which the total Hilbert space decomposes.
The Hilbert space of the $q$-th sector (with $q$ integer)
is generated by the basis $\lbrace\vert n,q\rangle\,,\, n-\vert q \vert /2\geq 0\rbrace$, where the integer or half-integer $n$ is related to the energy eigenvalues $E$ of $H_{\textrm{\tiny HO}}$ as $n = \frac{E}{2\omega} - \frac{1}{2}$.
Using these bases, a generic operator commuting with $Q_{\textrm{\tiny HO}}$ (but not necessarily with $H_{\textrm{\tiny HO}}$) can be written as
\begin{equation}
    \label{eq:chargeenergy decomposition O}O=\sum_{q\in\mathbb{Z}}O_q\,,
    \qquad
    O_q=\sum_{n,n'=\vert q \vert /2 }^\infty O^{(q)}_{n,n'} \vert n,q\rangle\langle n',q\vert\,,
\end{equation}
which automatically provides an expression for the fixed-charge block $O_q$.

The representation in \eqref{eq:chargeenergy decomposition O} allows us to provide an alternative formula for the symmetry-resolved return amplitude $R_q(t)$  associated with each block operator. 
Plugging it into the definition of the inner product $(O_q(t)\vert O_q(0))$, using the orthonormality of the bases $\lbrace\vert n,q\rangle\rbrace$ and the orthogonality among different charge sectors, we find
\begin{equation}
\label{eq:resolvedS_OCHO_newbasis}
R_q(t)=\frac{\sum_{n,n'=\vert q\vert/2}^\infty\vert O^{(q)}_{n,n'}\vert^2e^{-\beta\omega(n+n'+1)}e^{-2{\it i}\omega t(n'-n)}}{\sum_{n,n'=\vert q\vert/2}^\infty\vert O^{(q)}_{n,n'}\vert^2e^{-\beta\omega(n+n'+1)}}\,.
\end{equation}
Moreover, using again the inner product \eqref{eq:innerproduct}, we could also conveniently write the probability $p_q=\left|(O_q(0) |O(0))\right|^2$ associated with the corresponding charge sector as
\begin{equation}
\label{eq:p_OCHO_newbasis}
p_q=\frac{\sum_{n,n'=\vert q\vert/2}^\infty\vert O^{(q)}_{n,n'}\vert^2e^{-\beta\omega(n+n'+1)}}{\sum_{q=-\infty}^\infty\sum_{n,n'=\vert q\vert/2}^\infty\vert O^{(q)}_{n,n'}\vert^2e^{-\beta\omega(n+n'+1)}}\,.
\end{equation}
We can rewrite the operator $O_{\textrm{\tiny HO}}$ in the basis \eqref{eq:chargeenergy decomposition O}, finding the components
\begin{equation}
\label{eq:good_operator_nonHerm_component}
(O^{(q)}_{\textrm{\tiny HO}})_{n,n'}=\sqrt{n^2-\frac{q^2}{4}}\delta_{n,n'+1}+\sqrt{n'^2-\frac{q^2}{4}}\delta_{n',n+1}\,,
\end{equation}
and their squared norm
\begin{equation}
\label{eq:good_operator_nonHerm_component_sq}
\vert(O^{(q)}_{\textrm{\tiny HO}})_{n,n'}\vert^2=\left(n^2-\frac{q^2}{4}\right)\delta_{n,n'+1}+\left(n'^2-\frac{q^2}{4}\right)\delta_{n',n+1}\,.
\end{equation}
We can use this expression in \eqref{eq:resolvedS_OCHO_newbasis} to confirm our result for the symmetry-resolved return amplitude.
Moreover, we could straightforwardly generalize our findings for the operator given by $\tilde O^{(q)}_{n,n'}=o_n^{(q)}\delta_{n,n'+a}+o_{n'}^{(q)}\delta_{n',n+a}$ with $a\in\mathbb{Z}$ and $o_n^{(q)}\in\mathbb{R}$. Importantly, the operator with components $\tilde O^{(q)}_{n,n'}$ is the most general example of an operator local in the energy space, as it connects only Fock states with a fixed energy difference $\Delta E/\omega-1=\pm 2a$. Its symmetry-resolved autocorrelation function $R_q(t)$ reads
\begin{equation}
\label{eq:resolvedS_OCHO_newbasis_general}
R_q(t)=\cos(2a\omega t)\,,
\end{equation}
and it coincides with the total autocorrelation function $R(t)$. This implies that, for a general operator local in the energy space, we always find equipartition of Krylov complexity.
If we instead consider an operator $\hat{O}$ which connects Fock states in such a way that $\Delta E/\omega-1=2a,2b$ ($a\neq b\in\mathbb{Z}$), i.e.
$\hat O^{(q)}_{n,n'}=A_n^{(q)}\delta_{n,n'+a}+A_{n'}^{(q)}\delta_{n',n+a}+B_n^{(q)}\delta_{n,n'+b}+B_{n'}^{(q)}\delta_{n',n+b}$, a similar computation shows that the corresponding symmetry-resolved autocorrelation function acquires a dependence of $q$ which leads to a breaking of the equipartition of Krylov complexity.
This shows that, as soon as we do not consider operators local in the energy space, the equipartition of Krylov complexity breaks down. In the example of complex harmonic oscillators, locality in energy space appears crucial for equipartition. It would be interesting to investigate whether this connection manifests in other systems.

\subsubsection{Equipartition of Krylov complexity in the many-body case}

We finally generalize our computation
to the operator \eqref{eq:good_operator_CHO} defined for a system of $N>1$ coupled harmonic oscillators. 
Since the considered operator has the factorized form
$O_{\textrm{\tiny HO}}=\prod_{k=1}^N O^{(k)}$ with each single-mode operator defined by $O^{(k)}=c_kb_k+c_k^\dagger b_k^\dagger$, the corresponding time-evolved state reads
$\vert O_{\textrm{\tiny HO}}(t))=\otimes_{k=1}^N\vert O^{(k)}(t))$.
 Due to this factorization, the total autocorrelation function also factorizes into a product of the autocorrelation functions for each oscillator mode
\begin{equation}
    R(t)=\prod_{k=1}^N (O^{(k)}(t)\vert O^{(k)}(0))=\prod_{k=1}^N \cos(2\omega_k t)\,,
    \label{eq:autocorr_manymode}
\end{equation}
where, in the last step, we have used the result for the single oscillator derived in the previous sections.

In this many-body case, the fixed-charge blocks in which $O_{\textrm{\tiny HO}}$ decomposes are even more intricate. Since the sectors are labeled by the integer values of the charge, we can still use \eqref{eq:FourierTransf_amplitude} to compute the symmetry-resolved autocorrelation functions. We obtain
\begin{equation}
\label{eq:SRautocorr_manybody}
    R_q(t)=\frac{\int_{-\pi}^\pi\frac{d \alpha}{2\pi}e^{-{ \rm i}\alpha q}\prod_{k=1}^N
\textrm{Tr}\left(e^{{ \rm i}\alpha Q_{\textrm{\tiny HO}}} e^{-\beta H  /2} O^{(k)}(t) e^{-\beta H  /2} O^{(k)}(0)\right)}{\int_{-\pi}^\pi\frac{d \alpha}{2\pi}e^{-{ \rm i}\alpha q}
\prod_{k=1}^N\textrm{Tr}\left(e^{{ \rm i}\alpha Q_{\textrm{\tiny HO}}} e^{-\beta H  /2} O^{(k)}(0) e^{-\beta H  /2} O^{(k)}(0)\right)}\,,
\end{equation}
where the traces in the integrand can be computed and read
\begin{equation}
\label{eq:SRautocorr_manybody_mode}
    \textrm{Tr}\left(e^{{ \rm i}\alpha Q_{\textrm{\tiny HO}}} e^{-\beta H  /2} O^{(k)}(t) e^{-\beta H  /2} O^{(k)}(0)\right)=  \frac{\cos(2\omega_k t)}{2[\cos(\alpha)-\cosh(\beta\omega_k )]^2}\,.
\end{equation}
In \eqref{eq:SRautocorr_manybody}, the mode factorization occurs before the Fourier transform and, therefore, the symmetry-resolved autocorrelation does not factorize.
  This hampers the direct evaluation of the integrals. However, using \eqref{eq:SRautocorr_manybody_mode}, we find again that the dependence on time and $\alpha$ factorizes. Thus, the time dependence at the numerator of \eqref{eq:SRautocorr_manybody} can be brought out from the integral, and the two complicated integrals at the numerator and denominator cancel.
As a result, $R_q(t)=R(t)$ for any $q\in\mathbb{Z}$, which, as discussed before, leads to the equipartition of Krylov complexity. This example demonstrates that equipartition is not a feature of the single-body model, but can also occur in many-body scenarios.

To compute the Krylov complexity and its resolution for a generic value of $N$, we have to apply the moment recursion method to  \eqref{eq:autocorr_manymode}. Doing so, we find that the number of non-vanishing Lanczos coefficients (and corresponding amplitudes $\phi_n(t)$) grows with $N$ and these quantities are hard to express in a closed form. This does not allow us to provide a formula for the Krylov complexity of $O_{\textrm{\tiny HO}}(t)$ for any $N$. 
For gaining some analytical understanding, we can compute the early-time growth of $C_K(t)$ at the first two non-vanishing orders, obtaining
\begin{equation}
C_K(t)=C^{(q)}_K(t)=4t^2\sum_{k=1}^N\omega_k^2-\frac{16}{3}t^4\sum_{k=1}^N\omega_k^4+O(t^6)\,.
\end{equation}
Here we notice that the oscillator modes do not couple with each other at early times. 
To verify whether this feature survives at higher orders, we can compute the exact expression of the Krylov complexity for the operator with $N = 2$. Here, the non-vanishing Lanczos coefficients are only three, and this allows us to find the formula
\begin{equation}
   C_K(t)=C^{(q)}_K(t)=\frac{3}{2}- \frac{(\omega_1+\omega_2)^2\cos(4t(\omega_1-\omega_2))+(\omega_1-\omega_2)^2\cos(4t(\omega_1+\omega_2))+4\omega_1^2\cos(4t\omega_2)+4\omega_2^2\cos(4t\omega_1)}{4(\omega_1^2+\omega_2^2)}\,.
\end{equation}
Expanding this result for small times, we see that it is enough to reach the order $O(t^6)$ to find products between $\omega_1$ and $\omega_2$, corresponding to a coupling between the oscillator modes.

\bibliography{Refs}

\begin{thebibliography}{96}%
\makeatletter
\providecommand \@ifxundefined [1]{%
 \@ifx{#1\undefined}
}%
\providecommand \@ifnum [1]{%
 \ifnum #1\expandafter \@firstoftwo
 \else \expandafter \@secondoftwo
 \fi
}%
\providecommand \@ifx [1]{%
 \ifx #1\expandafter \@firstoftwo
 \else \expandafter \@secondoftwo
 \fi
}%
\providecommand \natexlab [1]{#1}%
\providecommand \enquote  [1]{``#1''}%
\providecommand \bibnamefont  [1]{#1}%
\providecommand \bibfnamefont [1]{#1}%
\providecommand \citenamefont [1]{#1}%
\providecommand \href@noop [0]{\@secondoftwo}%
\providecommand \href [0]{\begingroup \@sanitize@url \@href}%
\providecommand \@href[1]{\@@startlink{#1}\@@href}%
\providecommand \@@href[1]{\endgroup#1\@@endlink}%
\providecommand \@sanitize@url [0]{\catcode `\\12\catcode `\$12\catcode
  `\&12\catcode `\#12\catcode `\^12\catcode `\_12\catcode `\%12\relax}%
\providecommand \@@startlink[1]{}%
\providecommand \@@endlink[0]{}%
\providecommand \url  [0]{\begingroup\@sanitize@url \@url }%
\providecommand \@url [1]{\endgroup\@href {#1}{\urlprefix }}%
\providecommand \urlprefix  [0]{URL }%
\providecommand \Eprint [0]{\href }%
\providecommand \doibase [0]{https://doi.org/}%
\providecommand \selectlanguage [0]{\@gobble}%
\providecommand \bibinfo  [0]{\@secondoftwo}%
\providecommand \bibfield  [0]{\@secondoftwo}%
\providecommand \translation [1]{[#1]}%
\providecommand \BibitemOpen [0]{}%
\providecommand \bibitemStop [0]{}%
\providecommand \bibitemNoStop [0]{.\EOS\space}%
\providecommand \EOS [0]{\spacefactor3000\relax}%
\providecommand \BibitemShut  [1]{\csname bibitem#1\endcsname}%
\let\auto@bib@innerbib\@empty
\bibitem [{\citenamefont {LaFlorencie}\ and\ \citenamefont
  {Rachel}(2014)}]{LaFlorencie2014}%
  \BibitemOpen
  \bibfield  {author} {\bibinfo {author} {\bibfnamefont {N.}~\bibnamefont
  {LaFlorencie}}\ and\ \bibinfo {author} {\bibfnamefont {S.}~\bibnamefont
  {Rachel}},\ }\bibfield  {title} {\bibinfo {title} {{Spin-resolved
  entanglement spectroscopy of critical spin chains and Luttinger liquids}},\
  }\href {https://doi.org/10.1088/1742-5468/2014/11/P11013} {\bibfield
  {journal} {\bibinfo  {journal} {J. Stat. Mech.}\ }\textbf {\bibinfo {volume}
  {2014}},\ \bibinfo {pages} {P11013} (\bibinfo {year} {2014})}\BibitemShut
  {NoStop}%
\bibitem [{\citenamefont {Goldstein}\ and\ \citenamefont
  {Sela}(2018)}]{Goldstein:2017bua}%
  \BibitemOpen
  \bibfield  {author} {\bibinfo {author} {\bibfnamefont {M.}~\bibnamefont
  {Goldstein}}\ and\ \bibinfo {author} {\bibfnamefont {E.}~\bibnamefont
  {Sela}},\ }\bibfield  {title} {\bibinfo {title} {{Symmetry-resolved
  entanglement in many-body systems}},\ }\href
  {https://doi.org/10.1103/PhysRevLett.120.200602} {\bibfield  {journal}
  {\bibinfo  {journal} {Phys. Rev. Lett.}\ }\textbf {\bibinfo {volume} {120}},\
  \bibinfo {pages} {200602} (\bibinfo {year} {2018})},\ \Eprint
  {https://arxiv.org/abs/1711.09418} {arXiv:1711.09418 [cond-mat.stat-mech]}
  \BibitemShut {NoStop}%
\bibitem [{\citenamefont {Xavier}\ \emph {et~al.}(2018)\citenamefont {Xavier},
  \citenamefont {Alcaraz},\ and\ \citenamefont {Sierra}}]{Xavier:2018kqb}%
  \BibitemOpen
  \bibfield  {author} {\bibinfo {author} {\bibfnamefont {J.~C.}\ \bibnamefont
  {Xavier}}, \bibinfo {author} {\bibfnamefont {F.~C.}\ \bibnamefont
  {Alcaraz}},\ and\ \bibinfo {author} {\bibfnamefont {G.}~\bibnamefont
  {Sierra}},\ }\bibfield  {title} {\bibinfo {title} {{Equipartition of the
  entanglement entropy}},\ }\href {https://doi.org/10.1103/PhysRevB.98.041106}
  {\bibfield  {journal} {\bibinfo  {journal} {Phys. Rev. B}\ }\textbf {\bibinfo
  {volume} {98}},\ \bibinfo {pages} {041106} (\bibinfo {year} {2018})},\
  \Eprint {https://arxiv.org/abs/1804.06357} {arXiv:1804.06357
  [cond-mat.stat-mech]} \BibitemShut {NoStop}%
\bibitem [{\citenamefont {Lukin}\ \emph
  {et~al.}(2019{\natexlab{a}})\citenamefont {Lukin}, \citenamefont {Rispoli},
  \citenamefont {Schittko}, \citenamefont {Tai}, \citenamefont {Kaufman},
  \citenamefont {Choi}, \citenamefont {Khemani}, \citenamefont {Léonard},\
  and\ \citenamefont {Greiner}}]{Lukin19}%
  \BibitemOpen
  \bibfield  {author} {\bibinfo {author} {\bibfnamefont {A.}~\bibnamefont
  {Lukin}}, \bibinfo {author} {\bibfnamefont {M.}~\bibnamefont {Rispoli}},
  \bibinfo {author} {\bibfnamefont {R.}~\bibnamefont {Schittko}}, \bibinfo
  {author} {\bibfnamefont {M.~E.}\ \bibnamefont {Tai}}, \bibinfo {author}
  {\bibfnamefont {A.~M.}\ \bibnamefont {Kaufman}}, \bibinfo {author}
  {\bibfnamefont {S.}~\bibnamefont {Choi}}, \bibinfo {author} {\bibfnamefont
  {V.}~\bibnamefont {Khemani}}, \bibinfo {author} {\bibfnamefont
  {J.}~\bibnamefont {Léonard}},\ and\ \bibinfo {author} {\bibfnamefont
  {M.}~\bibnamefont {Greiner}},\ }\bibfield  {title} {\bibinfo {title}
  {{Probing entanglement in a many-body localized system}},\ }\href
  {https://doi.org/10.1126/science.aau0818} {\bibfield  {journal} {\bibinfo
  {journal} {Science}\ }\textbf {\bibinfo {volume} {364}},\ \bibinfo {pages}
  {256} (\bibinfo {year} {2019}{\natexlab{a}})}\BibitemShut {NoStop}%
\bibitem [{\citenamefont {Murciano}\ \emph
  {et~al.}(2020{\natexlab{a}})\citenamefont {Murciano}, \citenamefont
  {Di~Giulio},\ and\ \citenamefont {Calabrese}}]{Murciano:2020vgh}%
  \BibitemOpen
  \bibfield  {author} {\bibinfo {author} {\bibfnamefont {S.}~\bibnamefont
  {Murciano}}, \bibinfo {author} {\bibfnamefont {G.}~\bibnamefont
  {Di~Giulio}},\ and\ \bibinfo {author} {\bibfnamefont {P.}~\bibnamefont
  {Calabrese}},\ }\bibfield  {title} {\bibinfo {title} {{Entanglement and
  symmetry resolution in two dimensional free quantum field theories}},\ }\href
  {https://doi.org/10.1007/JHEP08(2020)073} {\bibfield  {journal} {\bibinfo
  {journal} {JHEP}\ }\textbf {\bibinfo {volume} {08}},\ \bibinfo {pages}
  {073}}\BibitemShut {NoStop}%
\bibitem [{\citenamefont {Bonsignori}\ and\ \citenamefont
  {Calabrese}(2021)}]{Bonsignori:2020laa}%
  \BibitemOpen
  \bibfield  {author} {\bibinfo {author} {\bibfnamefont {R.}~\bibnamefont
  {Bonsignori}}\ and\ \bibinfo {author} {\bibfnamefont {P.}~\bibnamefont
  {Calabrese}},\ }\bibfield  {title} {\bibinfo {title} {{Boundary effects on
  symmetry resolved entanglement}},\ }\href
  {https://doi.org/10.1088/1751-8121/abcc3a} {\bibfield  {journal} {\bibinfo
  {journal} {J. Phys. A}\ }\textbf {\bibinfo {volume} {54}},\ \bibinfo {pages}
  {015005} (\bibinfo {year} {2021})},\ \Eprint
  {https://arxiv.org/abs/2009.08508} {arXiv:2009.08508 [cond-mat.stat-mech]}
  \BibitemShut {NoStop}%
\bibitem [{\citenamefont {Capizzi}\ \emph {et~al.}(2020)\citenamefont
  {Capizzi}, \citenamefont {Ruggiero},\ and\ \citenamefont
  {Calabrese}}]{Capizzi:2020jed}%
  \BibitemOpen
  \bibfield  {author} {\bibinfo {author} {\bibfnamefont {L.}~\bibnamefont
  {Capizzi}}, \bibinfo {author} {\bibfnamefont {P.}~\bibnamefont {Ruggiero}},\
  and\ \bibinfo {author} {\bibfnamefont {P.}~\bibnamefont {Calabrese}},\
  }\bibfield  {title} {\bibinfo {title} {{Symmetry resolved entanglement
  entropy of excited states in a CFT}},\ }\href
  {https://doi.org/10.1088/1742-5468/ab96b6} {\bibfield  {journal} {\bibinfo
  {journal} {J. Stat. Mech.}\ }\textbf {\bibinfo {volume} {2007}},\ \bibinfo
  {pages} {073101} (\bibinfo {year} {2020})},\ \Eprint
  {https://arxiv.org/abs/2003.04670} {arXiv:2003.04670 [cond-mat.stat-mech]}
  \BibitemShut {NoStop}%
\bibitem [{\citenamefont {Estienne}\ \emph {et~al.}(2021)\citenamefont
  {Estienne}, \citenamefont {Ikhlef},\ and\ \citenamefont
  {Morin-Duchesne}}]{Estienne:2020txv}%
  \BibitemOpen
  \bibfield  {author} {\bibinfo {author} {\bibfnamefont {B.}~\bibnamefont
  {Estienne}}, \bibinfo {author} {\bibfnamefont {Y.}~\bibnamefont {Ikhlef}},\
  and\ \bibinfo {author} {\bibfnamefont {A.}~\bibnamefont {Morin-Duchesne}},\
  }\bibfield  {title} {\bibinfo {title} {{Finite-size corrections in critical
  symmetry-resolved entanglement}},\ }\href
  {https://doi.org/10.21468/SciPostPhys.10.3.054} {\bibfield  {journal}
  {\bibinfo  {journal} {SciPost Phys.}\ }\textbf {\bibinfo {volume} {10}},\
  \bibinfo {pages} {054} (\bibinfo {year} {2021})},\ \Eprint
  {https://arxiv.org/abs/2010.10515} {arXiv:2010.10515 [quant-ph]} \BibitemShut
  {NoStop}%
\bibitem [{\citenamefont {Zhao}\ \emph {et~al.}(2021)\citenamefont {Zhao},
  \citenamefont {Northe},\ and\ \citenamefont {Meyer}}]{Zhao:2020qmn}%
  \BibitemOpen
  \bibfield  {author} {\bibinfo {author} {\bibfnamefont {S.}~\bibnamefont
  {Zhao}}, \bibinfo {author} {\bibfnamefont {C.}~\bibnamefont {Northe}},\ and\
  \bibinfo {author} {\bibfnamefont {R.}~\bibnamefont {Meyer}},\ }\bibfield
  {title} {\bibinfo {title} {{Symmetry-resolved entanglement in
  AdS$_{3}$/CFT$_{2}$ coupled to U(1) Chern-Simons theory}},\ }\href
  {https://doi.org/10.1007/JHEP07(2021)030} {\bibfield  {journal} {\bibinfo
  {journal} {JHEP}\ }\textbf {\bibinfo {volume} {07}},\ \bibinfo {pages}
  {030}},\ \Eprint {https://arxiv.org/abs/2012.11274} {arXiv:2012.11274
  [hep-th]} \BibitemShut {NoStop}%
\bibitem [{\citenamefont {Horvath}\ \emph {et~al.}(2021)\citenamefont
  {Horvath}, \citenamefont {Capizzi},\ and\ \citenamefont
  {Calabrese}}]{Horvath:2021fks}%
  \BibitemOpen
  \bibfield  {author} {\bibinfo {author} {\bibfnamefont {D.~X.}\ \bibnamefont
  {Horvath}}, \bibinfo {author} {\bibfnamefont {L.}~\bibnamefont {Capizzi}},\
  and\ \bibinfo {author} {\bibfnamefont {P.}~\bibnamefont {Calabrese}},\
  }\bibfield  {title} {\bibinfo {title} {{U(1) symmetry resolved entanglement
  in free 1+1 dimensional field theories via form factor bootstrap}},\ }\href
  {https://doi.org/10.1007/JHEP05(2021)197} {\bibfield  {journal} {\bibinfo
  {journal} {JHEP}\ }\textbf {\bibinfo {volume} {05}},\ \bibinfo {pages}
  {197}},\ \Eprint {https://arxiv.org/abs/2103.03197} {arXiv:2103.03197
  [hep-th]} \BibitemShut {NoStop}%
\bibitem [{\citenamefont {Capizzi}\ \emph {et~al.}(2022)\citenamefont
  {Capizzi}, \citenamefont {Horv\'ath}, \citenamefont {Calabrese},\ and\
  \citenamefont {Castro-Alvaredo}}]{Capizzi:2021kys}%
  \BibitemOpen
  \bibfield  {author} {\bibinfo {author} {\bibfnamefont {L.}~\bibnamefont
  {Capizzi}}, \bibinfo {author} {\bibfnamefont {D.~X.}\ \bibnamefont
  {Horv\'ath}}, \bibinfo {author} {\bibfnamefont {P.}~\bibnamefont
  {Calabrese}},\ and\ \bibinfo {author} {\bibfnamefont {O.~A.}\ \bibnamefont
  {Castro-Alvaredo}},\ }\bibfield  {title} {\bibinfo {title} {{Entanglement of
  the 3-state Potts model via form factor bootstrap: total and symmetry
  resolved entropies}},\ }\href {https://doi.org/10.1007/JHEP05(2022)113}
  {\bibfield  {journal} {\bibinfo  {journal} {JHEP}\ }\textbf {\bibinfo
  {volume} {05}},\ \bibinfo {pages} {113}},\ \Eprint
  {https://arxiv.org/abs/2108.10935} {arXiv:2108.10935 [hep-th]} \BibitemShut
  {NoStop}%
\bibitem [{\citenamefont {Calabrese}\ \emph {et~al.}(2021)\citenamefont
  {Calabrese}, \citenamefont {Dubail},\ and\ \citenamefont
  {Murciano}}]{Calabrese:2021wvi}%
  \BibitemOpen
  \bibfield  {author} {\bibinfo {author} {\bibfnamefont {P.}~\bibnamefont
  {Calabrese}}, \bibinfo {author} {\bibfnamefont {J.}~\bibnamefont {Dubail}},\
  and\ \bibinfo {author} {\bibfnamefont {S.}~\bibnamefont {Murciano}},\
  }\bibfield  {title} {\bibinfo {title} {{Symmetry-resolved entanglement
  entropy in Wess-Zumino-Witten models}},\ }\href
  {https://doi.org/10.1007/JHEP10(2021)067} {\bibfield  {journal} {\bibinfo
  {journal} {JHEP}\ }\textbf {\bibinfo {volume} {10}},\ \bibinfo {pages}
  {067}},\ \Eprint {https://arxiv.org/abs/2106.15946} {arXiv:2106.15946
  [hep-th]} \BibitemShut {NoStop}%
\bibitem [{\citenamefont {Ghasemi}(2023)}]{Ghasemi:2022jxg}%
  \BibitemOpen
  \bibfield  {author} {\bibinfo {author} {\bibfnamefont {M.}~\bibnamefont
  {Ghasemi}},\ }\bibfield  {title} {\bibinfo {title} {{Universal thermal
  corrections to symmetry-resolved entanglement entropy and full counting
  statistics}},\ }\href {https://doi.org/10.1007/JHEP05(2023)209} {\bibfield
  {journal} {\bibinfo  {journal} {JHEP}\ }\textbf {\bibinfo {volume} {05}},\
  \bibinfo {pages} {209}},\ \Eprint {https://arxiv.org/abs/2203.06708}
  {arXiv:2203.06708 [hep-th]} \BibitemShut {NoStop}%
\bibitem [{\citenamefont {Foligno}\ \emph {et~al.}(2023)\citenamefont
  {Foligno}, \citenamefont {Murciano},\ and\ \citenamefont
  {Calabrese}}]{Foligno:2022ltu}%
  \BibitemOpen
  \bibfield  {author} {\bibinfo {author} {\bibfnamefont {A.}~\bibnamefont
  {Foligno}}, \bibinfo {author} {\bibfnamefont {S.}~\bibnamefont {Murciano}},\
  and\ \bibinfo {author} {\bibfnamefont {P.}~\bibnamefont {Calabrese}},\
  }\bibfield  {title} {\bibinfo {title} {{Entanglement resolution of free Dirac
  fermions on a torus}},\ }\href {https://doi.org/10.1007/JHEP03(2023)096}
  {\bibfield  {journal} {\bibinfo  {journal} {JHEP}\ }\textbf {\bibinfo
  {volume} {03}},\ \bibinfo {pages} {096}},\ \Eprint
  {https://arxiv.org/abs/2212.07261} {arXiv:2212.07261 [hep-th]} \BibitemShut
  {NoStop}%
\bibitem [{\citenamefont {Capizzi}\ \emph {et~al.}(2023)\citenamefont
  {Capizzi}, \citenamefont {Murciano},\ and\ \citenamefont
  {Calabrese}}]{Capizzi:2023bpr}%
  \BibitemOpen
  \bibfield  {author} {\bibinfo {author} {\bibfnamefont {L.}~\bibnamefont
  {Capizzi}}, \bibinfo {author} {\bibfnamefont {S.}~\bibnamefont {Murciano}},\
  and\ \bibinfo {author} {\bibfnamefont {P.}~\bibnamefont {Calabrese}},\
  }\bibfield  {title} {\bibinfo {title} {{Full counting statistics and symmetry
  resolved entanglement for free conformal theories with interface defects}},\
  }\href {https://doi.org/10.1088/1742-5468/ace3b8} {\bibfield  {journal}
  {\bibinfo  {journal} {J. Stat. Mech.}\ }\textbf {\bibinfo {volume} {2307}},\
  \bibinfo {pages} {073102} (\bibinfo {year} {2023})},\ \Eprint
  {https://arxiv.org/abs/2302.08209} {arXiv:2302.08209 [hep-th]} \BibitemShut
  {NoStop}%
\bibitem [{\citenamefont {Fossati}\ \emph {et~al.}(2023)\citenamefont
  {Fossati}, \citenamefont {Ares},\ and\ \citenamefont
  {Calabrese}}]{Fossati:2023zyz}%
  \BibitemOpen
  \bibfield  {author} {\bibinfo {author} {\bibfnamefont {M.}~\bibnamefont
  {Fossati}}, \bibinfo {author} {\bibfnamefont {F.}~\bibnamefont {Ares}},\ and\
  \bibinfo {author} {\bibfnamefont {P.}~\bibnamefont {Calabrese}},\ }\bibfield
  {title} {\bibinfo {title} {{Symmetry-resolved entanglement in critical
  non-Hermitian systems}},\ }\href
  {https://doi.org/10.1103/PhysRevB.107.205153} {\bibfield  {journal} {\bibinfo
   {journal} {Phys. Rev. B}\ }\textbf {\bibinfo {volume} {107}},\ \bibinfo
  {pages} {205153} (\bibinfo {year} {2023})},\ \Eprint
  {https://arxiv.org/abs/2303.05232} {arXiv:2303.05232 [cond-mat.stat-mech]}
  \BibitemShut {NoStop}%
\bibitem [{\citenamefont {Kusuki}\ \emph {et~al.}(2023)\citenamefont {Kusuki},
  \citenamefont {Murciano}, \citenamefont {Ooguri},\ and\ \citenamefont
  {Pal}}]{Kusuki:2023bsp}%
  \BibitemOpen
  \bibfield  {author} {\bibinfo {author} {\bibfnamefont {Y.}~\bibnamefont
  {Kusuki}}, \bibinfo {author} {\bibfnamefont {S.}~\bibnamefont {Murciano}},
  \bibinfo {author} {\bibfnamefont {H.}~\bibnamefont {Ooguri}},\ and\ \bibinfo
  {author} {\bibfnamefont {S.}~\bibnamefont {Pal}},\ }\bibfield  {title}
  {\bibinfo {title} {{Symmetry-resolved entanglement entropy, spectra \&
  boundary conformal field theory}},\ }\href
  {https://doi.org/10.1007/JHEP11(2023)216} {\bibfield  {journal} {\bibinfo
  {journal} {JHEP}\ }\textbf {\bibinfo {volume} {11}},\ \bibinfo {pages}
  {216}},\ \Eprint {https://arxiv.org/abs/2309.03287} {arXiv:2309.03287
  [hep-th]} \BibitemShut {NoStop}%
\bibitem [{\citenamefont {Murciano}\ \emph
  {et~al.}(2020{\natexlab{b}})\citenamefont {Murciano}, \citenamefont
  {Di~Giulio},\ and\ \citenamefont {Calabrese}}]{Murciano:2019wdl}%
  \BibitemOpen
  \bibfield  {author} {\bibinfo {author} {\bibfnamefont {S.}~\bibnamefont
  {Murciano}}, \bibinfo {author} {\bibfnamefont {G.}~\bibnamefont
  {Di~Giulio}},\ and\ \bibinfo {author} {\bibfnamefont {P.}~\bibnamefont
  {Calabrese}},\ }\bibfield  {title} {\bibinfo {title} {{Symmetry resolved
  entanglement in gapped integrable systems: a corner transfer matrix
  approach}},\ }\href {https://doi.org/10.21468/SciPostPhys.8.3.046} {\bibfield
   {journal} {\bibinfo  {journal} {SciPost Phys.}\ }\textbf {\bibinfo {volume}
  {8}},\ \bibinfo {pages} {046} (\bibinfo {year} {2020}{\natexlab{b}})},\
  \Eprint {https://arxiv.org/abs/1911.09588} {arXiv:1911.09588
  [cond-mat.stat-mech]} \BibitemShut {NoStop}%
\bibitem [{\citenamefont {Bonsignori}\ \emph {et~al.}(2019)\citenamefont
  {Bonsignori}, \citenamefont {Ruggiero},\ and\ \citenamefont
  {Calabrese}}]{Bonsignori:2019naz}%
  \BibitemOpen
  \bibfield  {author} {\bibinfo {author} {\bibfnamefont {R.}~\bibnamefont
  {Bonsignori}}, \bibinfo {author} {\bibfnamefont {P.}~\bibnamefont
  {Ruggiero}},\ and\ \bibinfo {author} {\bibfnamefont {P.}~\bibnamefont
  {Calabrese}},\ }\bibfield  {title} {\bibinfo {title} {{Symmetry resolved
  entanglement in free fermionic systems}},\ }\href
  {https://doi.org/10.1088/1751-8121/ab4b77} {\bibfield  {journal} {\bibinfo
  {journal} {J. Phys. A}\ }\textbf {\bibinfo {volume} {52}},\ \bibinfo {pages}
  {475302} (\bibinfo {year} {2019})},\ \Eprint
  {https://arxiv.org/abs/1907.02084} {arXiv:1907.02084 [cond-mat.stat-mech]}
  \BibitemShut {NoStop}%
\bibitem [{\citenamefont {Calabrese}\ \emph {et~al.}(2020)\citenamefont
  {Calabrese}, \citenamefont {Collura}, \citenamefont {Di~Giulio},\ and\
  \citenamefont {Murciano}}]{Calabrese:2020tci}%
  \BibitemOpen
  \bibfield  {author} {\bibinfo {author} {\bibfnamefont {P.}~\bibnamefont
  {Calabrese}}, \bibinfo {author} {\bibfnamefont {M.}~\bibnamefont {Collura}},
  \bibinfo {author} {\bibfnamefont {G.}~\bibnamefont {Di~Giulio}},\ and\
  \bibinfo {author} {\bibfnamefont {S.}~\bibnamefont {Murciano}},\ }\bibfield
  {title} {\bibinfo {title} {{Full counting statistics in the gapped XXZ spin
  chain}},\ }\href {https://doi.org/10.1209/0295-5075/129/60007} {\bibfield
  {journal} {\bibinfo  {journal} {EPL}\ }\textbf {\bibinfo {volume} {129}},\
  \bibinfo {pages} {60007} (\bibinfo {year} {2020})},\ \Eprint
  {https://arxiv.org/abs/2002.04367} {arXiv:2002.04367 [cond-mat.stat-mech]}
  \BibitemShut {NoStop}%
\bibitem [{\citenamefont {Ares}\ \emph
  {et~al.}(2022{\natexlab{a}})\citenamefont {Ares}, \citenamefont {Murciano},\
  and\ \citenamefont {Calabrese}}]{Ares:2022hdh}%
  \BibitemOpen
  \bibfield  {author} {\bibinfo {author} {\bibfnamefont {F.}~\bibnamefont
  {Ares}}, \bibinfo {author} {\bibfnamefont {S.}~\bibnamefont {Murciano}},\
  and\ \bibinfo {author} {\bibfnamefont {P.}~\bibnamefont {Calabrese}},\
  }\bibfield  {title} {\bibinfo {title} {{Symmetry-resolved entanglement in a
  long-range free-fermion chain}},\ }\href
  {https://doi.org/10.1088/1742-5468/ac7644} {\bibfield  {journal} {\bibinfo
  {journal} {J. Stat. Mech.}\ }\textbf {\bibinfo {volume} {2206}},\ \bibinfo
  {pages} {063104} (\bibinfo {year} {2022}{\natexlab{a}})},\ \Eprint
  {https://arxiv.org/abs/2202.05874} {arXiv:2202.05874 [cond-mat.stat-mech]}
  \BibitemShut {NoStop}%
\bibitem [{\citenamefont {Magan}(2021)}]{Magan:2021myk}%
  \BibitemOpen
  \bibfield  {author} {\bibinfo {author} {\bibfnamefont {J.~M.}\ \bibnamefont
  {Magan}},\ }\bibfield  {title} {\bibinfo {title} {{Proof of the universal
  density of charged states in QFT}},\ }\href
  {https://doi.org/10.1007/JHEP12(2021)100} {\bibfield  {journal} {\bibinfo
  {journal} {JHEP}\ }\textbf {\bibinfo {volume} {12}},\ \bibinfo {pages}
  {100}},\ \Eprint {https://arxiv.org/abs/2111.02418} {arXiv:2111.02418
  [hep-th]} \BibitemShut {NoStop}%
\bibitem [{\citenamefont {Weisenberger}\ \emph {et~al.}(2021)\citenamefont
  {Weisenberger}, \citenamefont {Zhao}, \citenamefont {Northe},\ and\
  \citenamefont {Meyer}}]{Weisenberger:2021eby}%
  \BibitemOpen
  \bibfield  {author} {\bibinfo {author} {\bibfnamefont {K.}~\bibnamefont
  {Weisenberger}}, \bibinfo {author} {\bibfnamefont {S.}~\bibnamefont {Zhao}},
  \bibinfo {author} {\bibfnamefont {C.}~\bibnamefont {Northe}},\ and\ \bibinfo
  {author} {\bibfnamefont {R.}~\bibnamefont {Meyer}},\ }\bibfield  {title}
  {\bibinfo {title} {{Symmetry-resolved entanglement for excited states and two
  entangling intervals in AdS$_{3}$/CFT$_{2}$}},\ }\href
  {https://doi.org/10.1007/JHEP12(2021)104} {\bibfield  {journal} {\bibinfo
  {journal} {JHEP}\ }\textbf {\bibinfo {volume} {12}},\ \bibinfo {pages}
  {104}},\ \Eprint {https://arxiv.org/abs/2108.09210} {arXiv:2108.09210
  [hep-th]} \BibitemShut {NoStop}%
\bibitem [{\citenamefont {Zhao}\ \emph {et~al.}(2022)\citenamefont {Zhao},
  \citenamefont {Northe}, \citenamefont {Weisenberger},\ and\ \citenamefont
  {Meyer}}]{Zhao:2022wnp}%
  \BibitemOpen
  \bibfield  {author} {\bibinfo {author} {\bibfnamefont {S.}~\bibnamefont
  {Zhao}}, \bibinfo {author} {\bibfnamefont {C.}~\bibnamefont {Northe}},
  \bibinfo {author} {\bibfnamefont {K.}~\bibnamefont {Weisenberger}},\ and\
  \bibinfo {author} {\bibfnamefont {R.}~\bibnamefont {Meyer}},\ }\bibfield
  {title} {\bibinfo {title} {{Charged moments in W$_{3}$ higher spin
  holography}},\ }\href {https://doi.org/10.1007/JHEP05(2022)166} {\bibfield
  {journal} {\bibinfo  {journal} {JHEP}\ }\textbf {\bibinfo {volume} {05}},\
  \bibinfo {pages} {166}},\ \Eprint {https://arxiv.org/abs/2202.11111}
  {arXiv:2202.11111 [hep-th]} \BibitemShut {NoStop}%
\bibitem [{\citenamefont {Di~Giulio}\ \emph {et~al.}(2023)\citenamefont
  {Di~Giulio}, \citenamefont {Meyer}, \citenamefont {Northe}, \citenamefont
  {Scheppach},\ and\ \citenamefont {Zhao}}]{DiGiulio:2022jjd}%
  \BibitemOpen
  \bibfield  {author} {\bibinfo {author} {\bibfnamefont {G.}~\bibnamefont
  {Di~Giulio}}, \bibinfo {author} {\bibfnamefont {R.}~\bibnamefont {Meyer}},
  \bibinfo {author} {\bibfnamefont {C.}~\bibnamefont {Northe}}, \bibinfo
  {author} {\bibfnamefont {H.}~\bibnamefont {Scheppach}},\ and\ \bibinfo
  {author} {\bibfnamefont {S.}~\bibnamefont {Zhao}},\ }\bibfield  {title}
  {\bibinfo {title} {{On the boundary conformal field theory approach to
  symmetry-resolved entanglement}},\ }\href
  {https://doi.org/10.21468/SciPostPhysCore.6.3.049} {\bibfield  {journal}
  {\bibinfo  {journal} {SciPost Phys. Core}\ }\textbf {\bibinfo {volume} {6}},\
  \bibinfo {pages} {049} (\bibinfo {year} {2023})},\ \Eprint
  {https://arxiv.org/abs/2212.09767} {arXiv:2212.09767 [hep-th]} \BibitemShut
  {NoStop}%
\bibitem [{\citenamefont {Northe}(2023)}]{Northe:2023khz}%
  \BibitemOpen
  \bibfield  {author} {\bibinfo {author} {\bibfnamefont {C.}~\bibnamefont
  {Northe}},\ }\bibfield  {title} {\bibinfo {title} {{Entanglement Resolution
  with Respect to Conformal Symmetry}},\ }\href
  {https://doi.org/10.1103/PhysRevLett.131.151601} {\bibfield  {journal}
  {\bibinfo  {journal} {Phys. Rev. Lett.}\ }\textbf {\bibinfo {volume} {131}},\
  \bibinfo {pages} {151601} (\bibinfo {year} {2023})},\ \Eprint
  {https://arxiv.org/abs/2303.07724} {arXiv:2303.07724 [hep-th]} \BibitemShut
  {NoStop}%
\bibitem [{\citenamefont {Benedetti}\ \emph {et~al.}(2024)\citenamefont
  {Benedetti}, \citenamefont {Casini}, \citenamefont {Kawahigashi},
  \citenamefont {Longo},\ and\ \citenamefont {Magan}}]{Benedetti:2024dku}%
  \BibitemOpen
  \bibfield  {author} {\bibinfo {author} {\bibfnamefont {V.}~\bibnamefont
  {Benedetti}}, \bibinfo {author} {\bibfnamefont {H.}~\bibnamefont {Casini}},
  \bibinfo {author} {\bibfnamefont {Y.}~\bibnamefont {Kawahigashi}}, \bibinfo
  {author} {\bibfnamefont {R.}~\bibnamefont {Longo}},\ and\ \bibinfo {author}
  {\bibfnamefont {J.~M.}\ \bibnamefont {Magan}},\ }\bibfield  {title} {\bibinfo
  {title} {{Modular invariance as completeness}},\ }\href
  {https://doi.org/10.1103/PhysRevD.110.125004} {\bibfield  {journal} {\bibinfo
   {journal} {Phys. Rev. D}\ }\textbf {\bibinfo {volume} {110}},\ \bibinfo
  {pages} {125004} (\bibinfo {year} {2024})},\ \Eprint
  {https://arxiv.org/abs/2408.04011} {arXiv:2408.04011 [hep-th]} \BibitemShut
  {NoStop}%
\bibitem [{\citenamefont {Casini}\ \emph {et~al.}(2020)\citenamefont {Casini},
  \citenamefont {Huerta}, \citenamefont {Mag{\'a}n},\ and\ \citenamefont
  {Pontello}}]{Casini:2019kex}%
  \BibitemOpen
  \bibfield  {author} {\bibinfo {author} {\bibfnamefont {H.}~\bibnamefont
  {Casini}}, \bibinfo {author} {\bibfnamefont {M.}~\bibnamefont {Huerta}},
  \bibinfo {author} {\bibfnamefont {J.~M.}\ \bibnamefont {Mag{\'a}n}},\ and\
  \bibinfo {author} {\bibfnamefont {D.}~\bibnamefont {Pontello}},\ }\bibfield
  {title} {\bibinfo {title} {{Entanglement entropy and superselection sectors.
  Part I. Global symmetries}},\ }\href
  {https://doi.org/10.1007/JHEP02(2020)014} {\bibfield  {journal} {\bibinfo
  {journal} {JHEP}\ }\textbf {\bibinfo {volume} {02}},\ \bibinfo {pages}
  {014}},\ \Eprint {https://arxiv.org/abs/1905.10487} {arXiv:1905.10487
  [hep-th]} \BibitemShut {NoStop}%
\bibitem [{\citenamefont {Casini}\ \emph {et~al.}(2021)\citenamefont {Casini},
  \citenamefont {Huerta}, \citenamefont {Magan},\ and\ \citenamefont
  {Pontello}}]{Casini:2020rgj}%
  \BibitemOpen
  \bibfield  {author} {\bibinfo {author} {\bibfnamefont {H.}~\bibnamefont
  {Casini}}, \bibinfo {author} {\bibfnamefont {M.}~\bibnamefont {Huerta}},
  \bibinfo {author} {\bibfnamefont {J.~M.}\ \bibnamefont {Magan}},\ and\
  \bibinfo {author} {\bibfnamefont {D.}~\bibnamefont {Pontello}},\ }\bibfield
  {title} {\bibinfo {title} {{Entropic order parameters for the phases of
  QFT}},\ }\href {https://doi.org/10.1007/JHEP04(2021)277} {\bibfield
  {journal} {\bibinfo  {journal} {JHEP}\ }\textbf {\bibinfo {volume} {04}},\
  \bibinfo {pages} {277}},\ \Eprint {https://arxiv.org/abs/2008.11748}
  {arXiv:2008.11748 [hep-th]} \BibitemShut {NoStop}%
\bibitem [{\citenamefont {Casini}\ and\ \citenamefont
  {Magan}(2021)}]{Casini:2021zgr}%
  \BibitemOpen
  \bibfield  {author} {\bibinfo {author} {\bibfnamefont {H.}~\bibnamefont
  {Casini}}\ and\ \bibinfo {author} {\bibfnamefont {J.~M.}\ \bibnamefont
  {Magan}},\ }\bibfield  {title} {\bibinfo {title} {{On completeness and
  generalized symmetries in quantum field theory}},\ }\href
  {https://doi.org/10.1142/S0217732321300251} {\bibfield  {journal} {\bibinfo
  {journal} {Mod. Phys. Lett. A}\ }\textbf {\bibinfo {volume} {36}},\ \bibinfo
  {pages} {2130025} (\bibinfo {year} {2021})},\ \Eprint
  {https://arxiv.org/abs/2110.11358} {arXiv:2110.11358 [hep-th]} \BibitemShut
  {NoStop}%
\bibitem [{\citenamefont {Ares}\ \emph {et~al.}(2023)\citenamefont {Ares},
  \citenamefont {Murciano},\ and\ \citenamefont {Calabrese}}]{Ares:2022koq}%
  \BibitemOpen
  \bibfield  {author} {\bibinfo {author} {\bibfnamefont {F.}~\bibnamefont
  {Ares}}, \bibinfo {author} {\bibfnamefont {S.}~\bibnamefont {Murciano}},\
  and\ \bibinfo {author} {\bibfnamefont {P.}~\bibnamefont {Calabrese}},\
  }\bibfield  {title} {\bibinfo {title} {{Entanglement asymmetry as a probe of
  symmetry breaking}},\ }\href {https://doi.org/10.1038/s41467-023-37747-8}
  {\bibfield  {journal} {\bibinfo  {journal} {Nature Commun.}\ }\textbf
  {\bibinfo {volume} {14}},\ \bibinfo {pages} {2036} (\bibinfo {year}
  {2023})},\ \Eprint {https://arxiv.org/abs/2207.14693} {arXiv:2207.14693
  [cond-mat.stat-mech]} \BibitemShut {NoStop}%
\bibitem [{\citenamefont {Ares}\ \emph {et~al.}(2025)\citenamefont {Ares},
  \citenamefont {Calabrese},\ and\ \citenamefont {Murciano}}]{Ares:2025onj}%
  \BibitemOpen
  \bibfield  {author} {\bibinfo {author} {\bibfnamefont {F.}~\bibnamefont
  {Ares}}, \bibinfo {author} {\bibfnamefont {P.}~\bibnamefont {Calabrese}},\
  and\ \bibinfo {author} {\bibfnamefont {S.}~\bibnamefont {Murciano}},\
  }\bibfield  {title} {\bibinfo {title} {{The quantum Mpemba effects}},\ }\href
  {https://doi.org/10.1038/s42254-025-00838-0} {\bibfield  {journal} {\bibinfo
  {journal} {Nature Rev. Phys.}\ }\textbf {\bibinfo {volume} {7}},\ \bibinfo
  {pages} {451} (\bibinfo {year} {2025})},\ \Eprint
  {https://arxiv.org/abs/2502.08087} {arXiv:2502.08087 [cond-mat.stat-mech]}
  \BibitemShut {NoStop}%
\bibitem [{\citenamefont {Castro-Alvaredo}\ and\ \citenamefont
  {Santamar\'\i{}a-Sanz}(2025)}]{Castro-Alvaredo:2024azg}%
  \BibitemOpen
  \bibfield  {author} {\bibinfo {author} {\bibfnamefont {O.~A.}\ \bibnamefont
  {Castro-Alvaredo}}\ and\ \bibinfo {author} {\bibfnamefont {L.}~\bibnamefont
  {Santamar\'\i{}a-Sanz}},\ }\bibfield  {title} {\bibinfo {title}
  {{Symmetry-resolved measures in quantum field theory: A short review}},\
  }\href {https://doi.org/10.1142/S0217984924300023} {\bibfield  {journal}
  {\bibinfo  {journal} {Mod. Phys. Lett. B}\ }\textbf {\bibinfo {volume}
  {39}},\ \bibinfo {pages} {2430002} (\bibinfo {year} {2025})},\ \Eprint
  {https://arxiv.org/abs/2403.06652} {arXiv:2403.06652 [hep-th]} \BibitemShut
  {NoStop}%
\bibitem [{\citenamefont {Cornfeld}\ \emph {et~al.}(2018)\citenamefont
  {Cornfeld}, \citenamefont {Goldstein},\ and\ \citenamefont
  {Sela}}]{Cornfeld:2018wbg}%
  \BibitemOpen
  \bibfield  {author} {\bibinfo {author} {\bibfnamefont {E.}~\bibnamefont
  {Cornfeld}}, \bibinfo {author} {\bibfnamefont {M.}~\bibnamefont
  {Goldstein}},\ and\ \bibinfo {author} {\bibfnamefont {E.}~\bibnamefont
  {Sela}},\ }\bibfield  {title} {\bibinfo {title} {{Imbalance entanglement:
  Symmetry decomposition of negativity}},\ }\href
  {https://doi.org/10.1103/PhysRevA.98.032302} {\bibfield  {journal} {\bibinfo
  {journal} {Phys. Rev. A}\ }\textbf {\bibinfo {volume} {98}},\ \bibinfo
  {pages} {032302} (\bibinfo {year} {2018})},\ \Eprint
  {https://arxiv.org/abs/1804.00632} {arXiv:1804.00632 [cond-mat.stat-mech]}
  \BibitemShut {NoStop}%
\bibitem [{\citenamefont {Murciano}\ \emph {et~al.}(2021)\citenamefont
  {Murciano}, \citenamefont {Bonsignori},\ and\ \citenamefont
  {Calabrese}}]{Murciano:2021djk}%
  \BibitemOpen
  \bibfield  {author} {\bibinfo {author} {\bibfnamefont {S.}~\bibnamefont
  {Murciano}}, \bibinfo {author} {\bibfnamefont {R.}~\bibnamefont
  {Bonsignori}},\ and\ \bibinfo {author} {\bibfnamefont {P.}~\bibnamefont
  {Calabrese}},\ }\bibfield  {title} {\bibinfo {title} {{Symmetry decomposition
  of negativity of massless free fermions}},\ }\href
  {https://doi.org/10.21468/SciPostPhys.10.5.111} {\bibfield  {journal}
  {\bibinfo  {journal} {SciPost Phys.}\ }\textbf {\bibinfo {volume} {10}},\
  \bibinfo {pages} {111} (\bibinfo {year} {2021})},\ \Eprint
  {https://arxiv.org/abs/2102.10054} {arXiv:2102.10054 [cond-mat.stat-mech]}
  \BibitemShut {NoStop}%
\bibitem [{\citenamefont {Capizzi}\ and\ \citenamefont
  {Calabrese}(2021)}]{Capizzi:2021zga}%
  \BibitemOpen
  \bibfield  {author} {\bibinfo {author} {\bibfnamefont {L.}~\bibnamefont
  {Capizzi}}\ and\ \bibinfo {author} {\bibfnamefont {P.}~\bibnamefont
  {Calabrese}},\ }\bibfield  {title} {\bibinfo {title} {{Symmetry resolved
  relative entropies and distances in conformal field theory}},\ }\href
  {https://doi.org/10.1007/JHEP10(2021)195} {\bibfield  {journal} {\bibinfo
  {journal} {JHEP}\ }\textbf {\bibinfo {volume} {10}},\ \bibinfo {pages}
  {195}},\ \Eprint {https://arxiv.org/abs/2105.08596} {arXiv:2105.08596
  [hep-th]} \BibitemShut {NoStop}%
\bibitem [{\citenamefont {Chen}(2022{\natexlab{a}})}]{Chen:2021nma}%
  \BibitemOpen
  \bibfield  {author} {\bibinfo {author} {\bibfnamefont {H.-H.}\ \bibnamefont
  {Chen}},\ }\bibfield  {title} {\bibinfo {title} {{Charged R\'enyi negativity
  of massless free bosons}},\ }\href {https://doi.org/10.1007/JHEP02(2022)117}
  {\bibfield  {journal} {\bibinfo  {journal} {JHEP}\ }\textbf {\bibinfo
  {volume} {02}},\ \bibinfo {pages} {117}},\ \Eprint
  {https://arxiv.org/abs/2111.11028} {arXiv:2111.11028 [hep-th]} \BibitemShut
  {NoStop}%
\bibitem [{\citenamefont {Chen}(2021)}]{Chen:2021pls}%
  \BibitemOpen
  \bibfield  {author} {\bibinfo {author} {\bibfnamefont {H.-H.}\ \bibnamefont
  {Chen}},\ }\bibfield  {title} {\bibinfo {title} {{Symmetry decomposition of
  relative entropies in conformal field theory}},\ }\href
  {https://doi.org/10.1007/JHEP07(2021)084} {\bibfield  {journal} {\bibinfo
  {journal} {JHEP}\ }\textbf {\bibinfo {volume} {07}},\ \bibinfo {pages}
  {084}},\ \Eprint {https://arxiv.org/abs/2104.03102} {arXiv:2104.03102
  [hep-th]} \BibitemShut {NoStop}%
\bibitem [{\citenamefont {Ares}\ \emph
  {et~al.}(2022{\natexlab{b}})\citenamefont {Ares}, \citenamefont {Calabrese},
  \citenamefont {Di~Giulio},\ and\ \citenamefont {Murciano}}]{Ares:2022gjb}%
  \BibitemOpen
  \bibfield  {author} {\bibinfo {author} {\bibfnamefont {F.}~\bibnamefont
  {Ares}}, \bibinfo {author} {\bibfnamefont {P.}~\bibnamefont {Calabrese}},
  \bibinfo {author} {\bibfnamefont {G.}~\bibnamefont {Di~Giulio}},\ and\
  \bibinfo {author} {\bibfnamefont {S.}~\bibnamefont {Murciano}},\ }\bibfield
  {title} {\bibinfo {title} {{Multi-charged moments of two intervals in
  conformal field theory}},\ }\href {https://doi.org/10.1007/JHEP09(2022)051}
  {\bibfield  {journal} {\bibinfo  {journal} {JHEP}\ }\textbf {\bibinfo
  {volume} {09}},\ \bibinfo {pages} {051}},\ \Eprint
  {https://arxiv.org/abs/2206.01534} {arXiv:2206.01534 [hep-th]} \BibitemShut
  {NoStop}%
\bibitem [{\citenamefont {Chen}(2022{\natexlab{b}})}]{Chen:2022gyy}%
  \BibitemOpen
  \bibfield  {author} {\bibinfo {author} {\bibfnamefont {H.-H.}\ \bibnamefont
  {Chen}},\ }\bibfield  {title} {\bibinfo {title} {{Dynamics of charge
  imbalance resolved negativity after a global quench in free scalar field
  theory}},\ }\href {https://doi.org/10.1007/JHEP08(2022)146} {\bibfield
  {journal} {\bibinfo  {journal} {JHEP}\ }\textbf {\bibinfo {volume} {08}},\
  \bibinfo {pages} {146}},\ \bibinfo {note} {[Erratum: JHEP 10, 157 (2022)]},\
  \Eprint {https://arxiv.org/abs/2205.09532} {arXiv:2205.09532 [hep-th]}
  \BibitemShut {NoStop}%
\bibitem [{\citenamefont {Rath}\ \emph {et~al.}(2023)\citenamefont {Rath},
  \citenamefont {Vitale}, \citenamefont {Murciano}, \citenamefont {Votto},
  \citenamefont {Dubail}, \citenamefont {Kueng}, \citenamefont {Branciard},
  \citenamefont {Calabrese},\ and\ \citenamefont {Vermersch}}]{Rath:2022qif}%
  \BibitemOpen
  \bibfield  {author} {\bibinfo {author} {\bibfnamefont {A.}~\bibnamefont
  {Rath}}, \bibinfo {author} {\bibfnamefont {V.}~\bibnamefont {Vitale}},
  \bibinfo {author} {\bibfnamefont {S.}~\bibnamefont {Murciano}}, \bibinfo
  {author} {\bibfnamefont {M.}~\bibnamefont {Votto}}, \bibinfo {author}
  {\bibfnamefont {J.}~\bibnamefont {Dubail}}, \bibinfo {author} {\bibfnamefont
  {R.}~\bibnamefont {Kueng}}, \bibinfo {author} {\bibfnamefont
  {C.}~\bibnamefont {Branciard}}, \bibinfo {author} {\bibfnamefont
  {P.}~\bibnamefont {Calabrese}},\ and\ \bibinfo {author} {\bibfnamefont
  {B.}~\bibnamefont {Vermersch}},\ }\bibfield  {title} {\bibinfo {title}
  {{Entanglement Barrier and its Symmetry Resolution: Theory and Experimental
  Observation}},\ }\href {https://doi.org/10.1103/PRXQuantum.4.010318}
  {\bibfield  {journal} {\bibinfo  {journal} {PRX Quantum}\ }\textbf {\bibinfo
  {volume} {4}},\ \bibinfo {pages} {010318} (\bibinfo {year} {2023})},\ \Eprint
  {https://arxiv.org/abs/2209.04393} {arXiv:2209.04393 [quant-ph]} \BibitemShut
  {NoStop}%
\bibitem [{\citenamefont {Parez}(2022)}]{Parez:2022sgc}%
  \BibitemOpen
  \bibfield  {author} {\bibinfo {author} {\bibfnamefont {G.}~\bibnamefont
  {Parez}},\ }\bibfield  {title} {\bibinfo {title} {{Symmetry-resolved R\'enyi
  fidelities and quantum phase transitions}},\ }\href
  {https://doi.org/10.1103/PhysRevB.106.235101} {\bibfield  {journal} {\bibinfo
   {journal} {Phys. Rev. B}\ }\textbf {\bibinfo {volume} {106}},\ \bibinfo
  {pages} {235101} (\bibinfo {year} {2022})},\ \Eprint
  {https://arxiv.org/abs/2208.09457} {arXiv:2208.09457 [cond-mat.stat-mech]}
  \BibitemShut {NoStop}%
\bibitem [{\citenamefont {Parez}\ \emph {et~al.}(2022)\citenamefont {Parez},
  \citenamefont {Bonsignori},\ and\ \citenamefont {Calabrese}}]{Parez:2022xur}%
  \BibitemOpen
  \bibfield  {author} {\bibinfo {author} {\bibfnamefont {G.}~\bibnamefont
  {Parez}}, \bibinfo {author} {\bibfnamefont {R.}~\bibnamefont {Bonsignori}},\
  and\ \bibinfo {author} {\bibfnamefont {P.}~\bibnamefont {Calabrese}},\
  }\bibfield  {title} {\bibinfo {title} {{Dynamics of charge-imbalance-resolved
  entanglement negativity after a quench in a free-fermion model}},\ }\href
  {https://doi.org/10.1088/1742-5468/ac666c} {\bibfield  {journal} {\bibinfo
  {journal} {J. Stat. Mech.}\ }\textbf {\bibinfo {volume} {2205}},\ \bibinfo
  {pages} {053103} (\bibinfo {year} {2022})},\ \Eprint
  {https://arxiv.org/abs/2202.05309} {arXiv:2202.05309 [cond-mat.stat-mech]}
  \BibitemShut {NoStop}%
\bibitem [{\citenamefont {Murciano}\ \emph {et~al.}(2024)\citenamefont
  {Murciano}, \citenamefont {Dubail},\ and\ \citenamefont
  {Calabrese}}]{Murciano:2023ofp}%
  \BibitemOpen
  \bibfield  {author} {\bibinfo {author} {\bibfnamefont {S.}~\bibnamefont
  {Murciano}}, \bibinfo {author} {\bibfnamefont {J.}~\bibnamefont {Dubail}},\
  and\ \bibinfo {author} {\bibfnamefont {P.}~\bibnamefont {Calabrese}},\
  }\bibfield  {title} {\bibinfo {title} {{More on symmetry resolved operator
  entanglement}},\ }\href {https://doi.org/10.1088/1751-8121/ad30d1} {\bibfield
   {journal} {\bibinfo  {journal} {J. Phys. A}\ }\textbf {\bibinfo {volume}
  {57}},\ \bibinfo {pages} {145002} (\bibinfo {year} {2024})},\ \Eprint
  {https://arxiv.org/abs/2309.04032} {arXiv:2309.04032 [cond-mat.stat-mech]}
  \BibitemShut {NoStop}%
\bibitem [{\citenamefont {Di~Giulio}\ and\ \citenamefont
  {Erdmenger}(2023)}]{DiGiulio:2023nvz}%
  \BibitemOpen
  \bibfield  {author} {\bibinfo {author} {\bibfnamefont {G.}~\bibnamefont
  {Di~Giulio}}\ and\ \bibinfo {author} {\bibfnamefont {J.}~\bibnamefont
  {Erdmenger}},\ }\bibfield  {title} {\bibinfo {title} {{Symmetry-resolved
  modular correlation functions in free fermionic theories}},\ }\href
  {https://doi.org/10.1007/JHEP07(2023)058} {\bibfield  {journal} {\bibinfo
  {journal} {JHEP}\ }\textbf {\bibinfo {volume} {07}},\ \bibinfo {pages}
  {058}},\ \Eprint {https://arxiv.org/abs/2305.02343} {arXiv:2305.02343
  [hep-th]} \BibitemShut {NoStop}%
\bibitem [{\citenamefont {Yan}\ \emph {et~al.}(2025)\citenamefont {Yan},
  \citenamefont {Murciano}, \citenamefont {Calabrese},\ and\ \citenamefont
  {Konik}}]{Yan:2024rcl}%
  \BibitemOpen
  \bibfield  {author} {\bibinfo {author} {\bibfnamefont {F.}~\bibnamefont
  {Yan}}, \bibinfo {author} {\bibfnamefont {S.}~\bibnamefont {Murciano}},
  \bibinfo {author} {\bibfnamefont {P.}~\bibnamefont {Calabrese}},\ and\
  \bibinfo {author} {\bibfnamefont {R.~M.}\ \bibnamefont {Konik}},\ }\bibfield
  {title} {\bibinfo {title} {{On symmetry-resolved generalized entropies}},\
  }\href {https://doi.org/10.21468/SciPostPhys.18.6.211} {\bibfield  {journal}
  {\bibinfo  {journal} {SciPost Phys.}\ }\textbf {\bibinfo {volume} {18}},\
  \bibinfo {pages} {211} (\bibinfo {year} {2025})},\ \Eprint
  {https://arxiv.org/abs/2412.14165} {arXiv:2412.14165 [quant-ph]} \BibitemShut
  {NoStop}%
\bibitem [{\citenamefont {Nie}(2021)}]{Nie:2021ond}%
  \BibitemOpen
  \bibfield  {author} {\bibinfo {author} {\bibfnamefont {L.}~\bibnamefont
  {Nie}},\ }\bibfield  {title} {\bibinfo {title} {{Operator Growth and
  Symmetry-Resolved Coefficient Entropy in Quantum Maps}},\ }\href@noop {} {\
  (\bibinfo {year} {2021})},\ \Eprint {https://arxiv.org/abs/2111.08729}
  {arXiv:2111.08729 [cond-mat.stat-mech]} \BibitemShut {NoStop}%
\bibitem [{\citenamefont {Parker}\ \emph {et~al.}(2019)\citenamefont {Parker},
  \citenamefont {Cao}, \citenamefont {Avdoshkin}, \citenamefont {Scaffidi},\
  and\ \citenamefont {Altman}}]{Parker:2018yvk}%
  \BibitemOpen
  \bibfield  {author} {\bibinfo {author} {\bibfnamefont {D.~E.}\ \bibnamefont
  {Parker}}, \bibinfo {author} {\bibfnamefont {X.}~\bibnamefont {Cao}},
  \bibinfo {author} {\bibfnamefont {A.}~\bibnamefont {Avdoshkin}}, \bibinfo
  {author} {\bibfnamefont {T.}~\bibnamefont {Scaffidi}},\ and\ \bibinfo
  {author} {\bibfnamefont {E.}~\bibnamefont {Altman}},\ }\bibfield  {title}
  {\bibinfo {title} {{A Universal Operator Growth Hypothesis}},\ }\href
  {https://doi.org/10.1103/PhysRevX.9.041017} {\bibfield  {journal} {\bibinfo
  {journal} {Phys. Rev. X}\ }\textbf {\bibinfo {volume} {9}},\ \bibinfo {pages}
  {041017} (\bibinfo {year} {2019})},\ \Eprint
  {https://arxiv.org/abs/1812.08657} {arXiv:1812.08657 [cond-mat.stat-mech]}
  \BibitemShut {NoStop}%
\bibitem [{\citenamefont {Balasubramanian}\ \emph {et~al.}(2022)\citenamefont
  {Balasubramanian}, \citenamefont {Caputa}, \citenamefont {Magan},\ and\
  \citenamefont {Wu}}]{Balasubramanian:2022tpr}%
  \BibitemOpen
  \bibfield  {author} {\bibinfo {author} {\bibfnamefont {V.}~\bibnamefont
  {Balasubramanian}}, \bibinfo {author} {\bibfnamefont {P.}~\bibnamefont
  {Caputa}}, \bibinfo {author} {\bibfnamefont {J.~M.}\ \bibnamefont {Magan}},\
  and\ \bibinfo {author} {\bibfnamefont {Q.}~\bibnamefont {Wu}},\ }\bibfield
  {title} {\bibinfo {title} {{Quantum chaos and the complexity of spread of
  states}},\ }\href {https://doi.org/10.1103/PhysRevD.106.046007} {\bibfield
  {journal} {\bibinfo  {journal} {Phys. Rev. D}\ }\textbf {\bibinfo {volume}
  {106}},\ \bibinfo {pages} {046007} (\bibinfo {year} {2022})},\ \Eprint
  {https://arxiv.org/abs/2202.06957} {arXiv:2202.06957 [hep-th]} \BibitemShut
  {NoStop}%
\bibitem [{\citenamefont {Roberts}\ \emph {et~al.}(2018)\citenamefont
  {Roberts}, \citenamefont {Stanford},\ and\ \citenamefont
  {Streicher}}]{Roberts:2018mnp}%
  \BibitemOpen
  \bibfield  {author} {\bibinfo {author} {\bibfnamefont {D.~A.}\ \bibnamefont
  {Roberts}}, \bibinfo {author} {\bibfnamefont {D.}~\bibnamefont {Stanford}},\
  and\ \bibinfo {author} {\bibfnamefont {A.}~\bibnamefont {Streicher}},\
  }\bibfield  {title} {\bibinfo {title} {{Operator growth in the SYK model}},\
  }\href {https://doi.org/10.1007/JHEP06(2018)122} {\bibfield  {journal}
  {\bibinfo  {journal} {JHEP}\ }\textbf {\bibinfo {volume} {06}},\ \bibinfo
  {pages} {122}},\ \Eprint {https://arxiv.org/abs/1802.02633} {arXiv:1802.02633
  [hep-th]} \BibitemShut {NoStop}%
\bibitem [{\citenamefont {Rabinovici}\ \emph {et~al.}(2022)\citenamefont
  {Rabinovici}, \citenamefont {S\'anchez-Garrido}, \citenamefont {Shir},\ and\
  \citenamefont {Sonner}}]{Rabinovici:2022beu}%
  \BibitemOpen
  \bibfield  {author} {\bibinfo {author} {\bibfnamefont {E.}~\bibnamefont
  {Rabinovici}}, \bibinfo {author} {\bibfnamefont {A.}~\bibnamefont
  {S\'anchez-Garrido}}, \bibinfo {author} {\bibfnamefont {R.}~\bibnamefont
  {Shir}},\ and\ \bibinfo {author} {\bibfnamefont {J.}~\bibnamefont {Sonner}},\
  }\bibfield  {title} {\bibinfo {title} {{Krylov complexity from integrability
  to chaos}},\ }\href {https://doi.org/10.1007/JHEP07(2022)151} {\bibfield
  {journal} {\bibinfo  {journal} {JHEP}\ }\textbf {\bibinfo {volume} {07}},\
  \bibinfo {pages} {151}},\ \Eprint {https://arxiv.org/abs/2207.07701}
  {arXiv:2207.07701 [hep-th]} \BibitemShut {NoStop}%
\bibitem [{\citenamefont {Balasubramanian}\ \emph {et~al.}(2023)\citenamefont
  {Balasubramanian}, \citenamefont {Magan},\ and\ \citenamefont
  {Wu}}]{Balasubramanian:2022dnj}%
  \BibitemOpen
  \bibfield  {author} {\bibinfo {author} {\bibfnamefont {V.}~\bibnamefont
  {Balasubramanian}}, \bibinfo {author} {\bibfnamefont {J.~M.}\ \bibnamefont
  {Magan}},\ and\ \bibinfo {author} {\bibfnamefont {Q.}~\bibnamefont {Wu}},\
  }\bibfield  {title} {\bibinfo {title} {{Tridiagonalizing random matrices}},\
  }\href {https://doi.org/10.1103/PhysRevD.107.126001} {\bibfield  {journal}
  {\bibinfo  {journal} {Phys. Rev. D}\ }\textbf {\bibinfo {volume} {107}},\
  \bibinfo {pages} {126001} (\bibinfo {year} {2023})},\ \Eprint
  {https://arxiv.org/abs/2208.08452} {arXiv:2208.08452 [hep-th]} \BibitemShut
  {NoStop}%
\bibitem [{\citenamefont {Erdmenger}\ \emph {et~al.}(2023)\citenamefont
  {Erdmenger}, \citenamefont {Jian},\ and\ \citenamefont
  {Xian}}]{Erdmenger:2023wjg}%
  \BibitemOpen
  \bibfield  {author} {\bibinfo {author} {\bibfnamefont {J.}~\bibnamefont
  {Erdmenger}}, \bibinfo {author} {\bibfnamefont {S.-K.}\ \bibnamefont
  {Jian}},\ and\ \bibinfo {author} {\bibfnamefont {Z.-Y.}\ \bibnamefont
  {Xian}},\ }\bibfield  {title} {\bibinfo {title} {{Universal chaotic dynamics
  from Krylov space}},\ }\href {https://doi.org/10.1007/JHEP08(2023)176}
  {\bibfield  {journal} {\bibinfo  {journal} {JHEP}\ }\textbf {\bibinfo
  {volume} {08}},\ \bibinfo {pages} {176}},\ \Eprint
  {https://arxiv.org/abs/2303.12151} {arXiv:2303.12151 [hep-th]} \BibitemShut
  {NoStop}%
\bibitem [{\citenamefont {Loc}(2024)}]{Loc:2024oen}%
  \BibitemOpen
  \bibfield  {author} {\bibinfo {author} {\bibfnamefont {T.~Q.}\ \bibnamefont
  {Loc}},\ }\bibfield  {title} {\bibinfo {title} {{Lanczos spectrum for random
  operator growth}},\ }\href@noop {} {\  (\bibinfo {year} {2024})},\ \Eprint
  {https://arxiv.org/abs/2402.07980} {arXiv:2402.07980 [hep-th]} \BibitemShut
  {NoStop}%
\bibitem [{\citenamefont {Caputa}\ and\ \citenamefont
  {Liu}(2022)}]{Caputa:2022eye}%
  \BibitemOpen
  \bibfield  {author} {\bibinfo {author} {\bibfnamefont {P.}~\bibnamefont
  {Caputa}}\ and\ \bibinfo {author} {\bibfnamefont {S.}~\bibnamefont {Liu}},\
  }\bibfield  {title} {\bibinfo {title} {{Quantum complexity and topological
  phases of matter}},\ }\href {https://doi.org/10.1103/PhysRevB.106.195125}
  {\bibfield  {journal} {\bibinfo  {journal} {Phys. Rev. B}\ }\textbf {\bibinfo
  {volume} {106}},\ \bibinfo {pages} {195125} (\bibinfo {year} {2022})},\
  \Eprint {https://arxiv.org/abs/2205.05688} {arXiv:2205.05688 [hep-th]}
  \BibitemShut {NoStop}%
\bibitem [{\citenamefont {Caputa}\ \emph {et~al.}(2023)\citenamefont {Caputa},
  \citenamefont {Gupta}, \citenamefont {Haque}, \citenamefont {Liu},
  \citenamefont {Murugan},\ and\ \citenamefont {Van~Zyl}}]{Caputa:2022yju}%
  \BibitemOpen
  \bibfield  {author} {\bibinfo {author} {\bibfnamefont {P.}~\bibnamefont
  {Caputa}}, \bibinfo {author} {\bibfnamefont {N.}~\bibnamefont {Gupta}},
  \bibinfo {author} {\bibfnamefont {S.~S.}\ \bibnamefont {Haque}}, \bibinfo
  {author} {\bibfnamefont {S.}~\bibnamefont {Liu}}, \bibinfo {author}
  {\bibfnamefont {J.}~\bibnamefont {Murugan}},\ and\ \bibinfo {author}
  {\bibfnamefont {H.~J.~R.}\ \bibnamefont {Van~Zyl}},\ }\bibfield  {title}
  {\bibinfo {title} {{Spread complexity and topological transitions in the
  Kitaev chain}},\ }\href {https://doi.org/10.1007/JHEP01(2023)120} {\bibfield
  {journal} {\bibinfo  {journal} {JHEP}\ }\textbf {\bibinfo {volume} {01}},\
  \bibinfo {pages} {120}},\ \Eprint {https://arxiv.org/abs/2208.06311}
  {arXiv:2208.06311 [hep-th]} \BibitemShut {NoStop}%
\bibitem [{\citenamefont {Dymarsky}\ and\ \citenamefont
  {Smolkin}(2021)}]{Dymarsky:2021bjq}%
  \BibitemOpen
  \bibfield  {author} {\bibinfo {author} {\bibfnamefont {A.}~\bibnamefont
  {Dymarsky}}\ and\ \bibinfo {author} {\bibfnamefont {M.}~\bibnamefont
  {Smolkin}},\ }\bibfield  {title} {\bibinfo {title} {{Krylov complexity in
  conformal field theory}},\ }\href
  {https://doi.org/10.1103/PhysRevD.104.L081702} {\bibfield  {journal}
  {\bibinfo  {journal} {Phys. Rev. D}\ }\textbf {\bibinfo {volume} {104}},\
  \bibinfo {pages} {L081702} (\bibinfo {year} {2021})},\ \Eprint
  {https://arxiv.org/abs/2104.09514} {arXiv:2104.09514 [hep-th]} \BibitemShut
  {NoStop}%
\bibitem [{\citenamefont {Avdoshkin}\ \emph {et~al.}(2024)\citenamefont
  {Avdoshkin}, \citenamefont {Dymarsky},\ and\ \citenamefont
  {Smolkin}}]{Avdoshkin:2022xuw}%
  \BibitemOpen
  \bibfield  {author} {\bibinfo {author} {\bibfnamefont {A.}~\bibnamefont
  {Avdoshkin}}, \bibinfo {author} {\bibfnamefont {A.}~\bibnamefont
  {Dymarsky}},\ and\ \bibinfo {author} {\bibfnamefont {M.}~\bibnamefont
  {Smolkin}},\ }\bibfield  {title} {\bibinfo {title} {{Krylov complexity in
  quantum field theory, and beyond}},\ }\href
  {https://doi.org/10.1007/JHEP06(2024)066} {\bibfield  {journal} {\bibinfo
  {journal} {JHEP}\ }\textbf {\bibinfo {volume} {06}},\ \bibinfo {pages}
  {066}},\ \Eprint {https://arxiv.org/abs/2212.14429} {arXiv:2212.14429
  [hep-th]} \BibitemShut {NoStop}%
\bibitem [{\citenamefont {Camargo}\ \emph {et~al.}(2023)\citenamefont
  {Camargo}, \citenamefont {Jahnke}, \citenamefont {Kim},\ and\ \citenamefont
  {Nishida}}]{Camargo:2022rnt}%
  \BibitemOpen
  \bibfield  {author} {\bibinfo {author} {\bibfnamefont {H.~A.}\ \bibnamefont
  {Camargo}}, \bibinfo {author} {\bibfnamefont {V.}~\bibnamefont {Jahnke}},
  \bibinfo {author} {\bibfnamefont {K.-Y.}\ \bibnamefont {Kim}},\ and\ \bibinfo
  {author} {\bibfnamefont {M.}~\bibnamefont {Nishida}},\ }\bibfield  {title}
  {\bibinfo {title} {{Krylov complexity in free and interacting scalar field
  theories with bounded power spectrum}},\ }\href
  {https://doi.org/10.1007/JHEP05(2023)226} {\bibfield  {journal} {\bibinfo
  {journal} {JHEP}\ }\textbf {\bibinfo {volume} {05}},\ \bibinfo {pages}
  {226}},\ \Eprint {https://arxiv.org/abs/2212.14702} {arXiv:2212.14702
  [hep-th]} \BibitemShut {NoStop}%
\bibitem [{\citenamefont {Caputa}\ and\ \citenamefont
  {Datta}(2021)}]{Caputa:2021ori}%
  \BibitemOpen
  \bibfield  {author} {\bibinfo {author} {\bibfnamefont {P.}~\bibnamefont
  {Caputa}}\ and\ \bibinfo {author} {\bibfnamefont {S.}~\bibnamefont {Datta}},\
  }\bibfield  {title} {\bibinfo {title} {{Operator growth in 2d CFT}},\ }\href
  {https://doi.org/10.1007/JHEP12(2021)188} {\bibfield  {journal} {\bibinfo
  {journal} {JHEP}\ }\textbf {\bibinfo {volume} {12}},\ \bibinfo {pages}
  {188}},\ \bibinfo {note} {[Erratum: JHEP 09, 113 (2022)]},\ \Eprint
  {https://arxiv.org/abs/2110.10519} {arXiv:2110.10519 [hep-th]} \BibitemShut
  {NoStop}%
\bibitem [{\citenamefont {Caputa}\ \emph
  {et~al.}(2024{\natexlab{a}})\citenamefont {Caputa}, \citenamefont {Magan},
  \citenamefont {Patramanis},\ and\ \citenamefont {Tonni}}]{Caputa:2023vyr}%
  \BibitemOpen
  \bibfield  {author} {\bibinfo {author} {\bibfnamefont {P.}~\bibnamefont
  {Caputa}}, \bibinfo {author} {\bibfnamefont {J.~M.}\ \bibnamefont {Magan}},
  \bibinfo {author} {\bibfnamefont {D.}~\bibnamefont {Patramanis}},\ and\
  \bibinfo {author} {\bibfnamefont {E.}~\bibnamefont {Tonni}},\ }\bibfield
  {title} {\bibinfo {title} {{Krylov complexity of modular Hamiltonian
  evolution}},\ }\href {https://doi.org/10.1103/PhysRevD.109.086004} {\bibfield
   {journal} {\bibinfo  {journal} {Phys. Rev. D}\ }\textbf {\bibinfo {volume}
  {109}},\ \bibinfo {pages} {086004} (\bibinfo {year} {2024}{\natexlab{a}})},\
  \Eprint {https://arxiv.org/abs/2306.14732} {arXiv:2306.14732 [hep-th]}
  \BibitemShut {NoStop}%
\bibitem [{\citenamefont {Malvimat}\ \emph {et~al.}(2025)\citenamefont
  {Malvimat}, \citenamefont {Porey},\ and\ \citenamefont
  {Roy}}]{Malvimat:2024vhr}%
  \BibitemOpen
  \bibfield  {author} {\bibinfo {author} {\bibfnamefont {V.}~\bibnamefont
  {Malvimat}}, \bibinfo {author} {\bibfnamefont {S.}~\bibnamefont {Porey}},\
  and\ \bibinfo {author} {\bibfnamefont {B.}~\bibnamefont {Roy}},\ }\bibfield
  {title} {\bibinfo {title} {{Krylov complexity in 2d CFTs with SL(2,
  \ensuremath{\mathbb{R}}) deformed Hamiltonians}},\ }\href
  {https://doi.org/10.1007/JHEP02(2025)035} {\bibfield  {journal} {\bibinfo
  {journal} {JHEP}\ }\textbf {\bibinfo {volume} {02}},\ \bibinfo {pages}
  {035}},\ \Eprint {https://arxiv.org/abs/2402.15835} {arXiv:2402.15835
  [hep-th]} \BibitemShut {NoStop}%
\bibitem [{\citenamefont {Caputa}\ and\ \citenamefont
  {Di~Giulio}(2025)}]{Caputa:2025dep}%
  \BibitemOpen
  \bibfield  {author} {\bibinfo {author} {\bibfnamefont {P.}~\bibnamefont
  {Caputa}}\ and\ \bibinfo {author} {\bibfnamefont {G.}~\bibnamefont
  {Di~Giulio}},\ }\bibfield  {title} {\bibinfo {title} {{Local quenches from a
  Krylov perspective}},\ }\href {https://doi.org/10.1007/JHEP07(2025)164}
  {\bibfield  {journal} {\bibinfo  {journal} {JHEP}\ }\textbf {\bibinfo
  {volume} {07}},\ \bibinfo {pages} {164}},\ \Eprint
  {https://arxiv.org/abs/2502.19485} {arXiv:2502.19485 [hep-th]} \BibitemShut
  {NoStop}%
\bibitem [{\citenamefont {Liu}\ \emph {et~al.}(2023)\citenamefont {Liu},
  \citenamefont {Tang},\ and\ \citenamefont {Zhai}}]{Liu:2022god}%
  \BibitemOpen
  \bibfield  {author} {\bibinfo {author} {\bibfnamefont {C.}~\bibnamefont
  {Liu}}, \bibinfo {author} {\bibfnamefont {H.}~\bibnamefont {Tang}},\ and\
  \bibinfo {author} {\bibfnamefont {H.}~\bibnamefont {Zhai}},\ }\bibfield
  {title} {\bibinfo {title} {{Krylov complexity in open quantum systems}},\
  }\href {https://doi.org/10.1103/PhysRevResearch.5.033085} {\bibfield
  {journal} {\bibinfo  {journal} {Phys. Rev. Res.}\ }\textbf {\bibinfo {volume}
  {5}},\ \bibinfo {pages} {033085} (\bibinfo {year} {2023})},\ \Eprint
  {https://arxiv.org/abs/2207.13603} {arXiv:2207.13603 [cond-mat.str-el]}
  \BibitemShut {NoStop}%
\bibitem [{\citenamefont {Bhattacharya}\ \emph {et~al.}(2022)\citenamefont
  {Bhattacharya}, \citenamefont {Nandy}, \citenamefont {Nath},\ and\
  \citenamefont {Sahu}}]{Bhattacharya:2022gbz}%
  \BibitemOpen
  \bibfield  {author} {\bibinfo {author} {\bibfnamefont {A.}~\bibnamefont
  {Bhattacharya}}, \bibinfo {author} {\bibfnamefont {P.}~\bibnamefont {Nandy}},
  \bibinfo {author} {\bibfnamefont {P.~P.}\ \bibnamefont {Nath}},\ and\
  \bibinfo {author} {\bibfnamefont {H.}~\bibnamefont {Sahu}},\ }\bibfield
  {title} {\bibinfo {title} {{Operator growth and Krylov construction in
  dissipative open quantum systems}},\ }\href
  {https://doi.org/10.1007/JHEP12(2022)081} {\bibfield  {journal} {\bibinfo
  {journal} {JHEP}\ }\textbf {\bibinfo {volume} {12}},\ \bibinfo {pages}
  {081}},\ \Eprint {https://arxiv.org/abs/2207.05347} {arXiv:2207.05347
  [quant-ph]} \BibitemShut {NoStop}%
\bibitem [{\citenamefont {Medina-Guerra}\ \emph
  {et~al.}(2025{\natexlab{a}})\citenamefont {Medina-Guerra}, \citenamefont
  {Gornyi},\ and\ \citenamefont {Gefen}}]{Medina-Guerra:2025wxg}%
  \BibitemOpen
  \bibfield  {author} {\bibinfo {author} {\bibfnamefont {E.}~\bibnamefont
  {Medina-Guerra}}, \bibinfo {author} {\bibfnamefont {I.~V.}\ \bibnamefont
  {Gornyi}},\ and\ \bibinfo {author} {\bibfnamefont {Y.}~\bibnamefont
  {Gefen}},\ }\bibfield  {title} {\bibinfo {title} {{Phase transitions in a
  non-Hermitian Su-Schrieffer-Heeger model via Krylov spread complexity}},\
  }\href {https://doi.org/10.1103/6lvg-7qdn} {\bibfield  {journal} {\bibinfo
  {journal} {Phys. Rev. B}\ }\textbf {\bibinfo {volume} {112}},\ \bibinfo
  {pages} {035427} (\bibinfo {year} {2025}{\natexlab{a}})},\ \Eprint
  {https://arxiv.org/abs/2503.18936} {arXiv:2503.18936 [cond-mat.str-el]}
  \BibitemShut {NoStop}%
\bibitem [{\citenamefont {Medina-Guerra}\ \emph
  {et~al.}(2025{\natexlab{b}})\citenamefont {Medina-Guerra}, \citenamefont
  {Gornyi},\ and\ \citenamefont {Gefen}}]{Medina-Guerra:2025rwa}%
  \BibitemOpen
  \bibfield  {author} {\bibinfo {author} {\bibfnamefont {E.}~\bibnamefont
  {Medina-Guerra}}, \bibinfo {author} {\bibfnamefont {I.~V.}\ \bibnamefont
  {Gornyi}},\ and\ \bibinfo {author} {\bibfnamefont {Y.}~\bibnamefont
  {Gefen}},\ }\bibfield  {title} {\bibinfo {title} {{Correlations and Krylov
  spread for a non-Hermitian Hamiltonian: Ising chain with a complex-valued
  transverse magnetic field}},\ }\href
  {https://doi.org/10.1103/PhysRevB.111.174207} {\bibfield  {journal} {\bibinfo
   {journal} {Phys. Rev. B}\ }\textbf {\bibinfo {volume} {111}},\ \bibinfo
  {pages} {174207} (\bibinfo {year} {2025}{\natexlab{b}})},\ \Eprint
  {https://arxiv.org/abs/2502.07775} {arXiv:2502.07775 [quant-ph]} \BibitemShut
  {NoStop}%
\bibitem [{\citenamefont {Nandy}\ \emph {et~al.}(2025)\citenamefont {Nandy},
  \citenamefont {Matsoukas-Roubeas}, \citenamefont {Mart{\'\i}nez-Azcona},
  \citenamefont {Dymarsky},\ and\ \citenamefont {del Campo}}]{Nandy:2024htc}%
  \BibitemOpen
  \bibfield  {author} {\bibinfo {author} {\bibfnamefont {P.}~\bibnamefont
  {Nandy}}, \bibinfo {author} {\bibfnamefont {A.~S.}\ \bibnamefont
  {Matsoukas-Roubeas}}, \bibinfo {author} {\bibfnamefont {P.}~\bibnamefont
  {Mart{\'\i}nez-Azcona}}, \bibinfo {author} {\bibfnamefont {A.}~\bibnamefont
  {Dymarsky}},\ and\ \bibinfo {author} {\bibfnamefont {A.}~\bibnamefont {del
  Campo}},\ }\bibfield  {title} {\bibinfo {title} {{Quantum dynamics in Krylov
  space: Methods and applications}},\ }\href
  {https://doi.org/10.1016/j.physrep.2025.05.001} {\bibfield  {journal}
  {\bibinfo  {journal} {Phys. Rept.}\ }\textbf {\bibinfo {volume}
  {1125-1128}},\ \bibinfo {pages} {1} (\bibinfo {year} {2025})},\ \Eprint
  {https://arxiv.org/abs/2405.09628} {arXiv:2405.09628 [quant-ph]} \BibitemShut
  {NoStop}%
\bibitem [{\citenamefont {Rabinovici}\ \emph {et~al.}(2023)\citenamefont
  {Rabinovici}, \citenamefont {S\'anchez-Garrido}, \citenamefont {Shir},\ and\
  \citenamefont {Sonner}}]{Rabinovici:2023yex}%
  \BibitemOpen
  \bibfield  {author} {\bibinfo {author} {\bibfnamefont {E.}~\bibnamefont
  {Rabinovici}}, \bibinfo {author} {\bibfnamefont {A.}~\bibnamefont
  {S\'anchez-Garrido}}, \bibinfo {author} {\bibfnamefont {R.}~\bibnamefont
  {Shir}},\ and\ \bibinfo {author} {\bibfnamefont {J.}~\bibnamefont {Sonner}},\
  }\bibfield  {title} {\bibinfo {title} {{A bulk manifestation of Krylov
  complexity}},\ }\href {https://doi.org/10.1007/JHEP08(2023)213} {\bibfield
  {journal} {\bibinfo  {journal} {JHEP}\ }\textbf {\bibinfo {volume} {08}},\
  \bibinfo {pages} {213}},\ \Eprint {https://arxiv.org/abs/2305.04355}
  {arXiv:2305.04355 [hep-th]} \BibitemShut {NoStop}%
\bibitem [{\citenamefont {Balasubramanian}\ \emph {et~al.}(2024)\citenamefont
  {Balasubramanian}, \citenamefont {Magan}, \citenamefont {Nandi},\ and\
  \citenamefont {Wu}}]{Balasubramanian:2024lqk}%
  \BibitemOpen
  \bibfield  {author} {\bibinfo {author} {\bibfnamefont {V.}~\bibnamefont
  {Balasubramanian}}, \bibinfo {author} {\bibfnamefont {J.~M.}\ \bibnamefont
  {Magan}}, \bibinfo {author} {\bibfnamefont {P.}~\bibnamefont {Nandi}},\ and\
  \bibinfo {author} {\bibfnamefont {Q.}~\bibnamefont {Wu}},\ }\bibfield
  {title} {\bibinfo {title} {{Spread complexity and the saturation of wormhole
  size}},\ }\href@noop {} {\  (\bibinfo {year} {2024})},\ \Eprint
  {https://arxiv.org/abs/2412.02038} {arXiv:2412.02038 [hep-th]} \BibitemShut
  {NoStop}%
\bibitem [{\citenamefont {Miyaji}\ \emph {et~al.}(2025)\citenamefont {Miyaji},
  \citenamefont {Ruan}, \citenamefont {Shibuya},\ and\ \citenamefont
  {Yano}}]{Miyaji:2025yvm}%
  \BibitemOpen
  \bibfield  {author} {\bibinfo {author} {\bibfnamefont {M.}~\bibnamefont
  {Miyaji}}, \bibinfo {author} {\bibfnamefont {S.-M.}\ \bibnamefont {Ruan}},
  \bibinfo {author} {\bibfnamefont {S.}~\bibnamefont {Shibuya}},\ and\ \bibinfo
  {author} {\bibfnamefont {K.}~\bibnamefont {Yano}},\ }\bibfield  {title}
  {\bibinfo {title} {{Non-perturbative overlaps in JT gravity: from spectral
  form factor to generating functions of complexity}},\ }\href
  {https://doi.org/10.1007/JHEP06(2025)251} {\bibfield  {journal} {\bibinfo
  {journal} {JHEP}\ }\textbf {\bibinfo {volume} {06}},\ \bibinfo {pages}
  {251}},\ \Eprint {https://arxiv.org/abs/2502.12266} {arXiv:2502.12266
  [hep-th]} \BibitemShut {NoStop}%
\bibitem [{\citenamefont {Caputa}\ \emph
  {et~al.}(2024{\natexlab{b}})\citenamefont {Caputa}, \citenamefont {Chen},
  \citenamefont {McDonald}, \citenamefont {Sim\'on},\ and\ \citenamefont
  {Strittmatter}}]{Caputa:2024sux}%
  \BibitemOpen
  \bibfield  {author} {\bibinfo {author} {\bibfnamefont {P.}~\bibnamefont
  {Caputa}}, \bibinfo {author} {\bibfnamefont {B.}~\bibnamefont {Chen}},
  \bibinfo {author} {\bibfnamefont {R.~W.}\ \bibnamefont {McDonald}}, \bibinfo
  {author} {\bibfnamefont {J.}~\bibnamefont {Sim\'on}},\ and\ \bibinfo {author}
  {\bibfnamefont {B.}~\bibnamefont {Strittmatter}},\ }\bibfield  {title}
  {\bibinfo {title} {{Spread Complexity Rate as Proper Momentum}},\ }\href@noop
  {} {\  (\bibinfo {year} {2024}{\natexlab{b}})},\ \Eprint
  {https://arxiv.org/abs/2410.23334} {arXiv:2410.23334 [hep-th]} \BibitemShut
  {NoStop}%
\bibitem [{\citenamefont {Maldacena}(1998)}]{Maldacena:1997re}%
  \BibitemOpen
  \bibfield  {author} {\bibinfo {author} {\bibfnamefont {J.~M.}\ \bibnamefont
  {Maldacena}},\ }\bibfield  {title} {\bibinfo {title} {{The Large N limit of
  superconformal field theories and supergravity}},\ }\href
  {https://doi.org/10.1023/A:1026654312961} {\bibfield  {journal} {\bibinfo
  {journal} {Adv. Theor. Math. Phys.}\ }\textbf {\bibinfo {volume} {2}},\
  \bibinfo {pages} {231} (\bibinfo {year} {1998})},\ \Eprint
  {https://arxiv.org/abs/hep-th/9711200} {arXiv:hep-th/9711200} \BibitemShut
  {NoStop}%
\bibitem [{\citenamefont {Berges}\ \emph {et~al.}(2004)\citenamefont {Berges},
  \citenamefont {Borsanyi},\ and\ \citenamefont {Wetterich}}]{Berges:2004ce}%
  \BibitemOpen
  \bibfield  {author} {\bibinfo {author} {\bibfnamefont {J.}~\bibnamefont
  {Berges}}, \bibinfo {author} {\bibfnamefont {S.}~\bibnamefont {Borsanyi}},\
  and\ \bibinfo {author} {\bibfnamefont {C.}~\bibnamefont {Wetterich}},\
  }\bibfield  {title} {\bibinfo {title} {{Prethermalization}},\ }\href
  {https://doi.org/10.1103/PhysRevLett.93.142002} {\bibfield  {journal}
  {\bibinfo  {journal} {Phys. Rev. Lett.}\ }\textbf {\bibinfo {volume} {93}},\
  \bibinfo {pages} {142002} (\bibinfo {year} {2004})},\ \Eprint
  {https://arxiv.org/abs/hep-ph/0403234} {arXiv:hep-ph/0403234} \BibitemShut
  {NoStop}%
\bibitem [{\citenamefont {Berges}\ \emph {et~al.}(2011)\citenamefont {Berges},
  \citenamefont {Borsanyi},\ and\ \citenamefont {Wetterich}}]{Kollar11}%
  \BibitemOpen
  \bibfield  {author} {\bibinfo {author} {\bibfnamefont {J.}~\bibnamefont
  {Berges}}, \bibinfo {author} {\bibfnamefont {S.}~\bibnamefont {Borsanyi}},\
  and\ \bibinfo {author} {\bibfnamefont {C.}~\bibnamefont {Wetterich}},\
  }\bibfield  {title} {\bibinfo {title} {{Kollar, Marcus and Wolf, F. Alexander
  and Eckstein, Martin}},\ }\href {https://doi.org/10.1103/PhysRevB.84.054304}
  {\bibfield  {journal} {\bibinfo  {journal} {Phys. Rev. B}\ }\textbf {\bibinfo
  {volume} {84}},\ \bibinfo {pages} {054304} (\bibinfo {year} {2011})},\
  \Eprint {https://arxiv.org/abs/1102.2117} {arXiv:1102.2117 [cond-mat.str-el]}
  \BibitemShut {NoStop}%
\bibitem [{\citenamefont {Langen}\ \emph {et~al.}(2016)\citenamefont {Langen},
  \citenamefont {Gasenzer},\ and\ \citenamefont
  {Schmiedmayer}}]{Langen:2016vdb}%
  \BibitemOpen
  \bibfield  {author} {\bibinfo {author} {\bibfnamefont {T.}~\bibnamefont
  {Langen}}, \bibinfo {author} {\bibfnamefont {T.}~\bibnamefont {Gasenzer}},\
  and\ \bibinfo {author} {\bibfnamefont {J.}~\bibnamefont {Schmiedmayer}},\
  }\bibfield  {title} {\bibinfo {title} {{Prethermalization and universal
  dynamics in near-integrable quantum systems}},\ }\href
  {https://doi.org/10.1088/1742-5468/2016/06/064009} {\bibfield  {journal}
  {\bibinfo  {journal} {J. Stat. Mech.}\ }\textbf {\bibinfo {volume} {1606}},\
  \bibinfo {pages} {064009} (\bibinfo {year} {2016})},\ \Eprint
  {https://arxiv.org/abs/1603.09385} {arXiv:1603.09385 [cond-mat.quant-gas]}
  \BibitemShut {NoStop}%
\bibitem [{\citenamefont {Choi}\ \emph {et~al.}(2023)\citenamefont {Choi} \emph
  {et~al.}}]{Choi:2021npc}%
  \BibitemOpen
  \bibfield  {author} {\bibinfo {author} {\bibfnamefont {J.}~\bibnamefont
  {Choi}} \emph {et~al.},\ }\bibfield  {title} {\bibinfo {title} {{Preparing
  random states and benchmarking with many-body quantum chaos}},\ }\href
  {https://doi.org/10.1038/s41586-022-05442-1} {\bibfield  {journal} {\bibinfo
  {journal} {Nature}\ }\textbf {\bibinfo {volume} {613}},\ \bibinfo {pages}
  {468} (\bibinfo {year} {2023})},\ \Eprint {https://arxiv.org/abs/2103.03535}
  {arXiv:2103.03535 [quant-ph]} \BibitemShut {NoStop}%
\bibitem [{\citenamefont {Cotler}\ \emph {et~al.}(2023)\citenamefont {Cotler},
  \citenamefont {Mark}, \citenamefont {Huang}, \citenamefont {Hernandez},
  \citenamefont {Choi}, \citenamefont {Shaw}, \citenamefont {Endres},\ and\
  \citenamefont {Choi}}]{Cotler:2021pbc}%
  \BibitemOpen
  \bibfield  {author} {\bibinfo {author} {\bibfnamefont {J.~S.}\ \bibnamefont
  {Cotler}}, \bibinfo {author} {\bibfnamefont {D.~K.}\ \bibnamefont {Mark}},
  \bibinfo {author} {\bibfnamefont {H.-Y.}\ \bibnamefont {Huang}}, \bibinfo
  {author} {\bibfnamefont {F.}~\bibnamefont {Hernandez}}, \bibinfo {author}
  {\bibfnamefont {J.}~\bibnamefont {Choi}}, \bibinfo {author} {\bibfnamefont
  {A.~L.}\ \bibnamefont {Shaw}}, \bibinfo {author} {\bibfnamefont
  {M.}~\bibnamefont {Endres}},\ and\ \bibinfo {author} {\bibfnamefont
  {S.}~\bibnamefont {Choi}},\ }\bibfield  {title} {\bibinfo {title} {{Emergent
  Quantum State Designs from Individual Many-Body Wave Functions}},\ }\href
  {https://doi.org/10.1103/PRXQuantum.4.010311} {\bibfield  {journal} {\bibinfo
   {journal} {PRX Quantum}\ }\textbf {\bibinfo {volume} {4}},\ \bibinfo {pages}
  {010311} (\bibinfo {year} {2023})},\ \Eprint
  {https://arxiv.org/abs/2103.03536} {arXiv:2103.03536 [quant-ph]} \BibitemShut
  {NoStop}%
\bibitem [{\citenamefont {Chang}\ \emph {et~al.}(2025)\citenamefont {Chang},
  \citenamefont {Shrotriya}, \citenamefont {Ho},\ and\ \citenamefont
  {Ippoliti}}]{Chang:2024cic}%
  \BibitemOpen
  \bibfield  {author} {\bibinfo {author} {\bibfnamefont {R.-A.}\ \bibnamefont
  {Chang}}, \bibinfo {author} {\bibfnamefont {H.}~\bibnamefont {Shrotriya}},
  \bibinfo {author} {\bibfnamefont {W.~W.}\ \bibnamefont {Ho}},\ and\ \bibinfo
  {author} {\bibfnamefont {M.}~\bibnamefont {Ippoliti}},\ }\bibfield  {title}
  {\bibinfo {title} {{Deep Thermalization under Charge-Conserving Quantum
  Dynamics}},\ }\href {https://doi.org/10.1103/PRXQuantum.6.020343} {\bibfield
  {journal} {\bibinfo  {journal} {PRX Quantum}\ }\textbf {\bibinfo {volume}
  {6}},\ \bibinfo {pages} {020343} (\bibinfo {year} {2025})},\ \Eprint
  {https://arxiv.org/abs/2408.15325} {arXiv:2408.15325 [quant-ph]} \BibitemShut
  {NoStop}%
\bibitem [{\citenamefont {Turner}\ \emph {et~al.}(2018)\citenamefont {Turner},
  \citenamefont {Michailidis}, \citenamefont {Abanin}, \citenamefont {Serbyn},\
  and\ \citenamefont {Papi\'c}}]{Turner:2018kjz}%
  \BibitemOpen
  \bibfield  {author} {\bibinfo {author} {\bibfnamefont {C.~J.}\ \bibnamefont
  {Turner}}, \bibinfo {author} {\bibfnamefont {A.~A.}\ \bibnamefont
  {Michailidis}}, \bibinfo {author} {\bibfnamefont {D.~A.}\ \bibnamefont
  {Abanin}}, \bibinfo {author} {\bibfnamefont {M.}~\bibnamefont {Serbyn}},\
  and\ \bibinfo {author} {\bibfnamefont {Z.}~\bibnamefont {Papi\'c}},\
  }\bibfield  {title} {\bibinfo {title} {{Weak ergodicity breaking from quantum
  many-body scars}},\ }\href {https://doi.org/10.1038/s41567-018-0137-5}
  {\bibfield  {journal} {\bibinfo  {journal} {Nature Phys.}\ }\textbf {\bibinfo
  {volume} {14}},\ \bibinfo {pages} {745} (\bibinfo {year} {2018})}\BibitemShut
  {NoStop}%
\bibitem [{\citenamefont {Choi}\ \emph {et~al.}(2019)\citenamefont {Choi},
  \citenamefont {Turner}, \citenamefont {Pichler}, \citenamefont {Ho},
  \citenamefont {Michailidis}, \citenamefont {Papi\'c}, \citenamefont {Serbyn},
  \citenamefont {Lukin},\ and\ \citenamefont {Abanin}}]{Choi:2019wqq}%
  \BibitemOpen
  \bibfield  {author} {\bibinfo {author} {\bibfnamefont {S.}~\bibnamefont
  {Choi}}, \bibinfo {author} {\bibfnamefont {C.~J.}\ \bibnamefont {Turner}},
  \bibinfo {author} {\bibfnamefont {H.}~\bibnamefont {Pichler}}, \bibinfo
  {author} {\bibfnamefont {W.~W.}\ \bibnamefont {Ho}}, \bibinfo {author}
  {\bibfnamefont {A.~A.}\ \bibnamefont {Michailidis}}, \bibinfo {author}
  {\bibfnamefont {Z.}~\bibnamefont {Papi\'c}}, \bibinfo {author} {\bibfnamefont
  {M.}~\bibnamefont {Serbyn}}, \bibinfo {author} {\bibfnamefont {M.~D.}\
  \bibnamefont {Lukin}},\ and\ \bibinfo {author} {\bibfnamefont {D.~A.}\
  \bibnamefont {Abanin}},\ }\bibfield  {title} {\bibinfo {title} {{Emergent
  SU(2) Dynamics and Perfect Quantum Many-Body Scars}},\ }\href
  {https://doi.org/10.1103/PhysRevLett.122.220603} {\bibfield  {journal}
  {\bibinfo  {journal} {Phys. Rev. Lett.}\ }\textbf {\bibinfo {volume} {122}},\
  \bibinfo {pages} {220603} (\bibinfo {year} {2019})}\BibitemShut {NoStop}%
\bibitem [{\citenamefont {Viswanath}\ and\ \citenamefont
  {M{\"u}ller}(1994)}]{LanczosBook}%
  \BibitemOpen
  \bibfield  {author} {\bibinfo {author} {\bibfnamefont {V.~S.}\ \bibnamefont
  {Viswanath}}\ and\ \bibinfo {author} {\bibfnamefont {G.}~\bibnamefont
  {M{\"u}ller}},\ }\href@noop {} {\emph {\bibinfo {title} {The Recursion
  Method: Application to Many-Body Dynamics}}}\ (\bibinfo  {publisher}
  {Springer Berlin},\ \bibinfo {address} {Heidelberg, Germany},\ \bibinfo
  {year} {1994})\BibitemShut {NoStop}%
\bibitem [{\citenamefont {Lanczos}(1950)}]{Lanczos:1950zz}%
  \BibitemOpen
  \bibfield  {author} {\bibinfo {author} {\bibfnamefont {C.}~\bibnamefont
  {Lanczos}},\ }\bibfield  {title} {\bibinfo {title} {{An iteration method for
  the solution of the eigenvalue problem of linear differential and integral
  operators}},\ }\href {https://doi.org/10.6028/jres.045.026} {\bibfield
  {journal} {\bibinfo  {journal} {J. Res. Natl. Bur. Stand. B}\ }\textbf
  {\bibinfo {volume} {45}},\ \bibinfo {pages} {255} (\bibinfo {year}
  {1950})}\BibitemShut {NoStop}%
\bibitem [{\citenamefont {Craps}\ \emph {et~al.}(2025)\citenamefont {Craps},
  \citenamefont {Evnin},\ and\ \citenamefont {Pascuzzi}}]{Craps:2024suj}%
  \BibitemOpen
  \bibfield  {author} {\bibinfo {author} {\bibfnamefont {B.}~\bibnamefont
  {Craps}}, \bibinfo {author} {\bibfnamefont {O.}~\bibnamefont {Evnin}},\ and\
  \bibinfo {author} {\bibfnamefont {G.}~\bibnamefont {Pascuzzi}},\ }\bibfield
  {title} {\bibinfo {title} {{Multiseed Krylov Complexity}},\ }\href
  {https://doi.org/10.1103/PhysRevLett.134.050402} {\bibfield  {journal}
  {\bibinfo  {journal} {Phys. Rev. Lett.}\ }\textbf {\bibinfo {volume} {134}},\
  \bibinfo {pages} {050402} (\bibinfo {year} {2025})},\ \Eprint
  {https://arxiv.org/abs/2409.15666} {arXiv:2409.15666 [quant-ph]} \BibitemShut
  {NoStop}%
\bibitem [{\citenamefont {Caputa}\ \emph
  {et~al.}(2024{\natexlab{c}})\citenamefont {Caputa}, \citenamefont {Jeong},
  \citenamefont {Liu}, \citenamefont {Pedraza},\ and\ \citenamefont
  {Qu}}]{Caputa:2024vrn}%
  \BibitemOpen
  \bibfield  {author} {\bibinfo {author} {\bibfnamefont {P.}~\bibnamefont
  {Caputa}}, \bibinfo {author} {\bibfnamefont {H.-S.}\ \bibnamefont {Jeong}},
  \bibinfo {author} {\bibfnamefont {S.}~\bibnamefont {Liu}}, \bibinfo {author}
  {\bibfnamefont {J.~F.}\ \bibnamefont {Pedraza}},\ and\ \bibinfo {author}
  {\bibfnamefont {L.-C.}\ \bibnamefont {Qu}},\ }\bibfield  {title} {\bibinfo
  {title} {{Krylov complexity of density matrix operators}},\ }\href
  {https://doi.org/10.1007/JHEP05(2024)337} {\bibfield  {journal} {\bibinfo
  {journal} {JHEP}\ }\textbf {\bibinfo {volume} {05}},\ \bibinfo {pages}
  {337}},\ \Eprint {https://arxiv.org/abs/2402.09522} {arXiv:2402.09522
  [hep-th]} \BibitemShut {NoStop}%
\bibitem [{\citenamefont {Serafini}(2017)}]{Serafini17book}%
  \BibitemOpen
  \bibfield  {author} {\bibinfo {author} {\bibfnamefont {A.}~\bibnamefont
  {Serafini}},\ }\href@noop {} {\emph {\bibinfo {title} {{Quantum Continuous
  Variables: A Primer of Theoretical Methods}}}}\ (\bibinfo  {publisher} {CRC
  Press},\ \bibinfo {year} {2017})\BibitemShut {NoStop}%
\bibitem [{\citenamefont {Lukin}\ \emph
  {et~al.}(2019{\natexlab{b}})\citenamefont {Lukin}, \citenamefont {Rispoli},
  \citenamefont {Schittko}, \citenamefont {Tai}, \citenamefont {Kaufman},
  \citenamefont {Choi}, \citenamefont {Khemani}, \citenamefont {L{\'e}onard},\
  and\ \citenamefont {Greiner}}]{Lukin:2019tkq}%
  \BibitemOpen
  \bibfield  {author} {\bibinfo {author} {\bibfnamefont {A.}~\bibnamefont
  {Lukin}}, \bibinfo {author} {\bibfnamefont {M.}~\bibnamefont {Rispoli}},
  \bibinfo {author} {\bibfnamefont {R.}~\bibnamefont {Schittko}}, \bibinfo
  {author} {\bibfnamefont {M.~E.}\ \bibnamefont {Tai}}, \bibinfo {author}
  {\bibfnamefont {A.~M.}\ \bibnamefont {Kaufman}}, \bibinfo {author}
  {\bibfnamefont {S.}~\bibnamefont {Choi}}, \bibinfo {author} {\bibfnamefont
  {V.}~\bibnamefont {Khemani}}, \bibinfo {author} {\bibfnamefont
  {J.}~\bibnamefont {L{\'e}onard}},\ and\ \bibinfo {author} {\bibfnamefont
  {M.}~\bibnamefont {Greiner}},\ }\bibfield  {title} {\bibinfo {title}
  {{Probing entanglement in a many-body{\textendash}localized system}},\ }\href
  {https://doi.org/10.1126/science.aau0818} {\bibfield  {journal} {\bibinfo
  {journal} {Science}\ }\textbf {\bibinfo {volume} {364}},\ \bibinfo {pages}
  {aau0818} (\bibinfo {year} {2019}{\natexlab{b}})}\BibitemShut {NoStop}%
\bibitem [{\citenamefont {Caputa}\ \emph {et~al.}(2025)\citenamefont {Caputa},
  \citenamefont {Di~Giulio},\ and\ \citenamefont {Loc}}]{Caputa:2025ozd}%
  \BibitemOpen
  \bibfield  {author} {\bibinfo {author} {\bibfnamefont {P.}~\bibnamefont
  {Caputa}}, \bibinfo {author} {\bibfnamefont {G.}~\bibnamefont {Di~Giulio}},\
  and\ \bibinfo {author} {\bibfnamefont {T.~Q.}\ \bibnamefont {Loc}},\
  }\bibfield  {title} {\bibinfo {title} {{Symmetry-Resolved Spread
  Complexity}},\ }\href@noop {} {\  (\bibinfo {year} {2025})},\ \Eprint
  {https://arxiv.org/abs/2509.12992} {arXiv:2509.12992 [hep-th]} \BibitemShut
  {NoStop}%
\bibitem [{\citenamefont {Kar}\ \emph {et~al.}(2022)\citenamefont {Kar},
  \citenamefont {Lamprou}, \citenamefont {Rozali},\ and\ \citenamefont
  {Sully}}]{Kar:2021nbm}%
  \BibitemOpen
  \bibfield  {author} {\bibinfo {author} {\bibfnamefont {A.}~\bibnamefont
  {Kar}}, \bibinfo {author} {\bibfnamefont {L.}~\bibnamefont {Lamprou}},
  \bibinfo {author} {\bibfnamefont {M.}~\bibnamefont {Rozali}},\ and\ \bibinfo
  {author} {\bibfnamefont {J.}~\bibnamefont {Sully}},\ }\bibfield  {title}
  {\bibinfo {title} {{Random matrix theory for complexity growth and black hole
  interiors}},\ }\href {https://doi.org/10.1007/JHEP01(2022)016} {\bibfield
  {journal} {\bibinfo  {journal} {JHEP}\ }\textbf {\bibinfo {volume} {01}},\
  \bibinfo {pages} {016}},\ \Eprint {https://arxiv.org/abs/2106.02046}
  {arXiv:2106.02046 [hep-th]} \BibitemShut {NoStop}%
\bibitem [{\citenamefont {Dymarsky}\ and\ \citenamefont
  {Gorsky}(2020)}]{Dymarsky:2019elm}%
  \BibitemOpen
  \bibfield  {author} {\bibinfo {author} {\bibfnamefont {A.}~\bibnamefont
  {Dymarsky}}\ and\ \bibinfo {author} {\bibfnamefont {A.}~\bibnamefont
  {Gorsky}},\ }\bibfield  {title} {\bibinfo {title} {{Quantum chaos as
  delocalization in Krylov space}},\ }\href
  {https://doi.org/10.1103/PhysRevB.102.085137} {\bibfield  {journal} {\bibinfo
   {journal} {Phys. Rev. B}\ }\textbf {\bibinfo {volume} {102}},\ \bibinfo
  {pages} {085137} (\bibinfo {year} {2020})},\ \Eprint
  {https://arxiv.org/abs/1912.12227} {arXiv:1912.12227 [cond-mat.stat-mech]}
  \BibitemShut {NoStop}%
\bibitem [{\citenamefont {Caputa}\ \emph {et~al.}(2022)\citenamefont {Caputa},
  \citenamefont {Magan},\ and\ \citenamefont {Patramanis}}]{Caputa:2021sib}%
  \BibitemOpen
  \bibfield  {author} {\bibinfo {author} {\bibfnamefont {P.}~\bibnamefont
  {Caputa}}, \bibinfo {author} {\bibfnamefont {J.~M.}\ \bibnamefont {Magan}},\
  and\ \bibinfo {author} {\bibfnamefont {D.}~\bibnamefont {Patramanis}},\
  }\bibfield  {title} {\bibinfo {title} {{Geometry of Krylov complexity}},\
  }\href {https://doi.org/10.1103/PhysRevResearch.4.013041} {\bibfield
  {journal} {\bibinfo  {journal} {Phys. Rev. Res.}\ }\textbf {\bibinfo {volume}
  {4}},\ \bibinfo {pages} {013041} (\bibinfo {year} {2022})},\ \Eprint
  {https://arxiv.org/abs/2109.03824} {arXiv:2109.03824 [hep-th]} \BibitemShut
  {NoStop}%
\bibitem [{\citenamefont {Baiguera}\ \emph {et~al.}(2025)\citenamefont
  {Baiguera}, \citenamefont {Balasubramanian}, \citenamefont {Caputa},
  \citenamefont {Chapman}, \citenamefont {Haferkamp}, \citenamefont {Heller},\
  and\ \citenamefont {Halpern}}]{Baiguera:2025dkc}%
  \BibitemOpen
  \bibfield  {author} {\bibinfo {author} {\bibfnamefont {S.}~\bibnamefont
  {Baiguera}}, \bibinfo {author} {\bibfnamefont {V.}~\bibnamefont
  {Balasubramanian}}, \bibinfo {author} {\bibfnamefont {P.}~\bibnamefont
  {Caputa}}, \bibinfo {author} {\bibfnamefont {S.}~\bibnamefont {Chapman}},
  \bibinfo {author} {\bibfnamefont {J.}~\bibnamefont {Haferkamp}}, \bibinfo
  {author} {\bibfnamefont {M.~P.}\ \bibnamefont {Heller}},\ and\ \bibinfo
  {author} {\bibfnamefont {N.~Y.}\ \bibnamefont {Halpern}},\ }\bibfield
  {title} {\bibinfo {title} {{Quantum complexity in gravity, quantum field
  theory, and quantum information science}},\ }\href@noop {} {\  (\bibinfo
  {year} {2025})},\ \Eprint {https://arxiv.org/abs/2503.10753}
  {arXiv:2503.10753 [hep-th]} \BibitemShut {NoStop}%
\bibitem [{\citenamefont {Zanardi}(2001)}]{Zanardi:2001zza}%
  \BibitemOpen
  \bibfield  {author} {\bibinfo {author} {\bibfnamefont {P.}~\bibnamefont
  {Zanardi}},\ }\bibfield  {title} {\bibinfo {title} {{Entanglement of quantum
  evolutions}},\ }\href {https://doi.org/10.1103/PhysRevA.63.040304} {\bibfield
   {journal} {\bibinfo  {journal} {Phys. Rev. A}\ }\textbf {\bibinfo {volume}
  {63}},\ \bibinfo {pages} {040304} (\bibinfo {year} {2001})},\ \Eprint
  {https://arxiv.org/abs/quant-ph/0010074} {arXiv:quant-ph/0010074}
  \BibitemShut {NoStop}%
\bibitem [{\citenamefont {Prosen}\ and\ \citenamefont
  {Pi{\v{z}}orn}(2007)}]{Prosen:2007gfp}%
  \BibitemOpen
  \bibfield  {author} {\bibinfo {author} {\bibfnamefont {T.}~\bibnamefont
  {Prosen}}\ and\ \bibinfo {author} {\bibfnamefont {I.}~\bibnamefont
  {Pi{\v{z}}orn}},\ }\bibfield  {title} {\bibinfo {title} {{Operator space
  entanglement entropy in a transverse Ising chain}},\ }\href
  {https://doi.org/10.1103/PhysRevA.76.032316} {\bibfield  {journal} {\bibinfo
  {journal} {Phys. Rev. A}\ }\textbf {\bibinfo {volume} {76}},\ \bibinfo
  {pages} {032316} (\bibinfo {year} {2007})}\BibitemShut {NoStop}%
\bibitem [{\citenamefont {Dubail}(2017)}]{Dubail:2016xht}%
  \BibitemOpen
  \bibfield  {author} {\bibinfo {author} {\bibfnamefont {J.}~\bibnamefont
  {Dubail}},\ }\bibfield  {title} {\bibinfo {title} {{Entanglement scaling of
  operators: a conformal field theory approach, with a glimpse of simulability
  of long-time dynamics in 1 + 1d}},\ }\href
  {https://doi.org/10.1088/1751-8121/aa6f38} {\bibfield  {journal} {\bibinfo
  {journal} {J. Phys. A}\ }\textbf {\bibinfo {volume} {50}},\ \bibinfo {pages}
  {234001} (\bibinfo {year} {2017})},\ \Eprint
  {https://arxiv.org/abs/1612.08630} {arXiv:1612.08630 [cond-mat.str-el]}
  \BibitemShut {NoStop}%
\bibitem [{\citenamefont {Fan}(2022)}]{Fan:2022xaa}%
  \BibitemOpen
  \bibfield  {author} {\bibinfo {author} {\bibfnamefont {Z.-Y.}\ \bibnamefont
  {Fan}},\ }\bibfield  {title} {\bibinfo {title} {{Universal relation for
  operator complexity}},\ }\href {https://doi.org/10.1103/PhysRevA.105.062210}
  {\bibfield  {journal} {\bibinfo  {journal} {Phys. Rev. A}\ }\textbf {\bibinfo
  {volume} {105}},\ \bibinfo {pages} {062210} (\bibinfo {year} {2022})},\
  \Eprint {https://arxiv.org/abs/2202.07220} {arXiv:2202.07220 [quant-ph]}
  \BibitemShut {NoStop}%
\end{thebibliography}%

\end{document}